# Virtual Data Cosmos

## Information design in Modern Astronomy

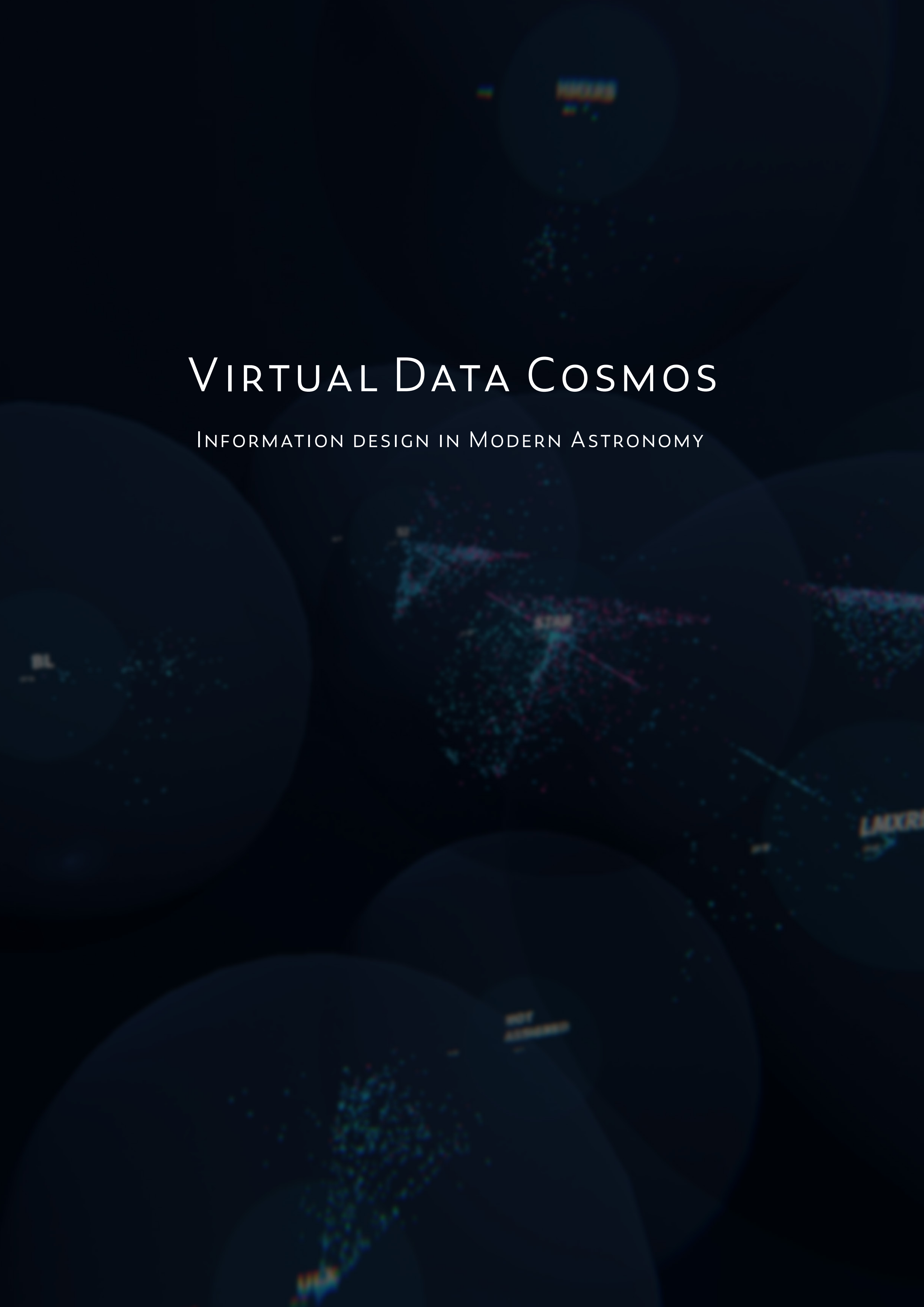

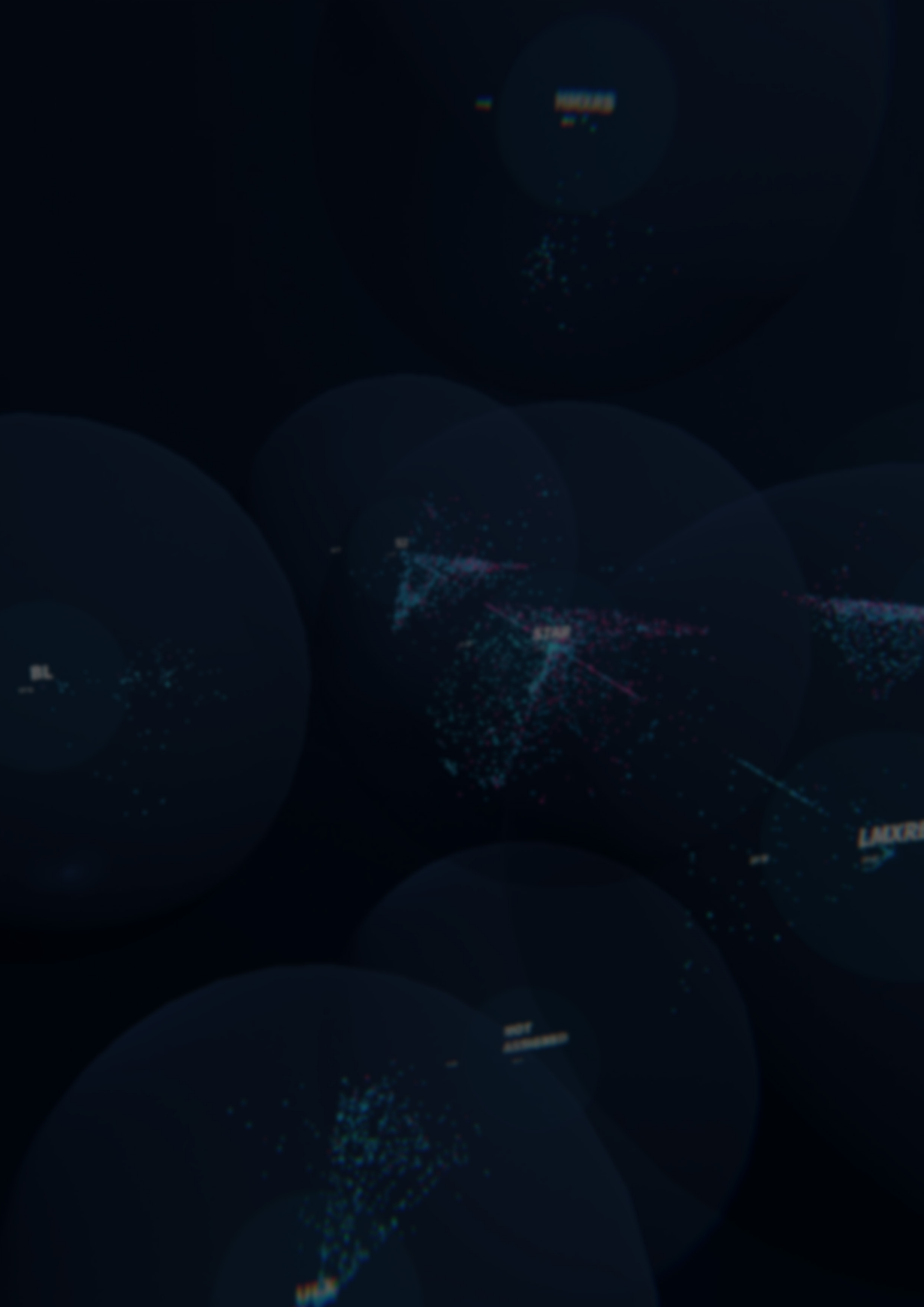

# Virtual Data Cosmos

## Information design in Modern Astronomy

University of Applied Sciences
Würzburg-Schweinfurt
Faculty of Design

Annika Kreikenbohm

**Information Design in modern Astronomy**

**„The greatest value of a picture is when it forces us to notice what we never expected to see“**

– John Tukey, founder of Exploratory Data Analysis, Tuckey (1977)

# Content

# 1 Introduction

**What is the history of our universe? What are the originis of galaxies like our Milky Way? Where does dark matter come from, which apparently determines the basic structure of the universe?**

These are some of the big questions that astronomy deals with. As one of the oldest natural sciences, astronomy explores the basic connections of our universe. And it has always been a data driven discipline. Theory and experiments (in terms of observations) are in a close and constant interplay. However, in contrast to experiments on Earth, experiments in the universe cannot usually be carried out under laboratory conditions or repeated as often as required. Because the phenomena in question are dynamic processes in distant stars, galaxies or even the big bang itself. What astronomers take advantage of is the unimaginable size of the universe and the infinite number of objects within. For example, while the death of a star cannot be observed multiple times on the same star, statistical surveys allow for the detection and comparison of many different dying stars. In this way, the similarities and differences of the objects can be examined. In other cases, e.g. the exploraion of solar flares, it helps to observe the same object in different states (i.e., before, during and after a solar flare) to examine the conditions of the process. Very large data sets are therefore not uncommon in astronomy and the size of the data sets grows with the development of increasingly efficient observation methods (e.g., Hassan & Fluke, 2011). With individual observations, more and more data are being collected, which not only document the condition of the telescope during an observation, but also collect more and more properties of the observed object. This creates new challenges for data processing and its interpretation (cf., Longo et al., 2019; Maarten et al., 2018). On the one hand, the measured data is compared to a concept of reality using a theory. On the other hand, the observation data construct an abstract image of nature, based on which new theories can be formed.

Information designers are also concerned with the question of how data is transformed into information and ultimately knowledge. Cairo (2013) describes the translation of data into information and knowledge through various processes of coding, decryption and comparison with experience (see Fig. 1.1): Reality is understood here as unstructured information. Data represents a first encoding of reality. By means of a suitable arrangement and linkage, this data becomes structured information in a second encoding phase. This can finally be transformed into knowledge by decoding it and by comparing it with ones own experiences and memories. The first coding can be understood here with the design and conception of an experiment by the scientist. It decides which statistical parameters are detected and how. Decoding the data is a crucial step in the chain of knowledge acquisition. This step involves recognizing patterns in the data and building a cognitive model. In order to obtain information and statements from the observation data, astronomers examine the underlying patterns and structures of the data sets. In the natural sciences, these are usually expressed concisely by abstract mathematical relations and correlations. For humans, however, visual representation is above all an effective method to understand complex data structures (see e.g., Ware, 2004; Cairo, 2013)

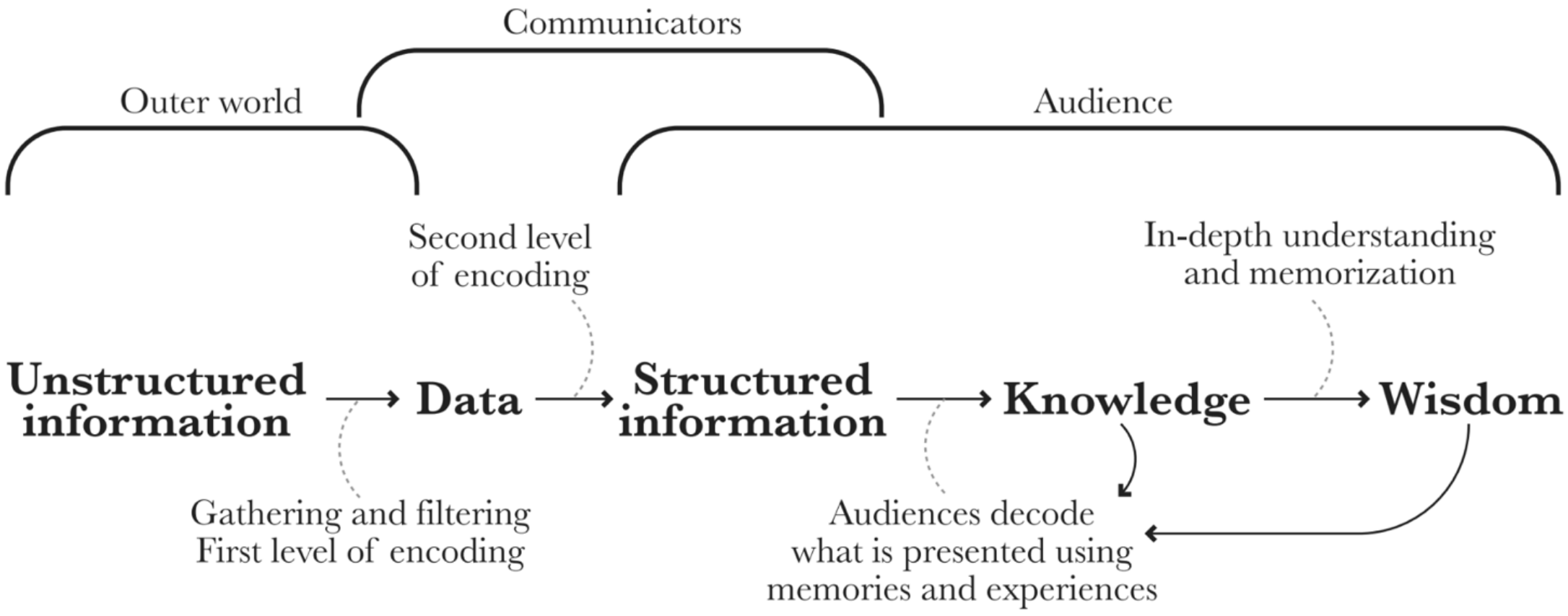

**Fig. 1.1:** Diagram by Cairo (2013) explaining the transformation of data into knowledge as a series of encoding and description.

As explained in the following section, data visualization in the sense of a graphical representation of data structures is an important tool for understanding and communicating data and theoretical concepts in astronomy and science in general. Although the role of visualization has changed over the course of history and the method of analysis, from an outside perspective, information design plays a key role in knowledge acquisition and transfer, as does data analysis itself result from large, complex data sets, not only requiring new information technology, but also creative approaches. How can big data sets with many correlated parameters be visualized? The human brain has an outstanding ability in pattern recognition (e.g., Damasio, 2010). Due to the complexity of today's data sets, however, intelligent algorithms, such as artificial neural networks, are increasingly being used to extract the knowledge from data sets and their internal correlations. Can data visualization itself become a tool to understand data more intuitively?

In this article, I deal with the question of the extent to which data visualizations in virtual reality can be a tool in processing multidimensional data in order to understand their patterns and to derive information from them.

## Plan of the thesis

In Chapter 2, I first explain to what extent astronomy can be understood as visual science and identify various challenges of modern astronomy. Chapter 3 looks at data visualization in virtual reality. The hypothesis derived from this is set out in Chapter 4. The data set to which this work is exemplary is presented in Chapter 5. The main focus is on the context in which the data was collected and the questions addressed to it by the astronomer. The development of a visualization concept and its implementation are then documented in Chapter 6 and Chapter 7. The latter primarily deals with the parametric representation of the data in virtual space. Finally, Chapter 8 summarizes the findings from this work and provides an outlook on possible further developments.

# 2 Astronomy – A visual science

Astronomy is naturally dependent on visualizations. Humans are simply not able to perceive most astronomical objects and phenomena sensually. Almost every popular scientific publication on an astronomical topic contains at least one artistic graphic representation or schematic drawing of an astronomical phenomenon. Joan Costa summarizes the task of visualization in the sense of information design, as follows (Joan Costa, 1998):

*„[The goal] is to make certain phenomena and portions of reality visible and understandable; many of these phenomena are not naturally accessible to the bare eye and many of them are not even of visual nature."*

Not only the general public, but researchers as well want to get an idea of the reality "out there". For example, consider Fig. 2.1, which shows "the first image" (of the shadow) of a black hole. Technically speaking, it is a 2D color diagram that shows the brightness distribution of radio radiation in the galaxy M87. The fascination with this data visualization is based on the one hand on the context that the dark spot in the blurry cloud is supposed to be a black hole. On the other hand, it is fascinating to know that this is real data that conveys an image of something that is not visible to the eye. Black holes, stars or planets per se are not natural objects that you can encounter on Earth. In order to still be able to imagine them, one uses visualization. The fact that some astronomers are masters of coding reality is demonstrated by the famous Golden Record of the Voyager missions (cf., Fig. 2.2), which provides those who are able to decrypt it with fundamental information about our location in the universe.

**Fig. 2.1:** The display of radio radiation in the center of the Galaxy M87 is the first image of a black hole. (Quelle: EHT, 2019)

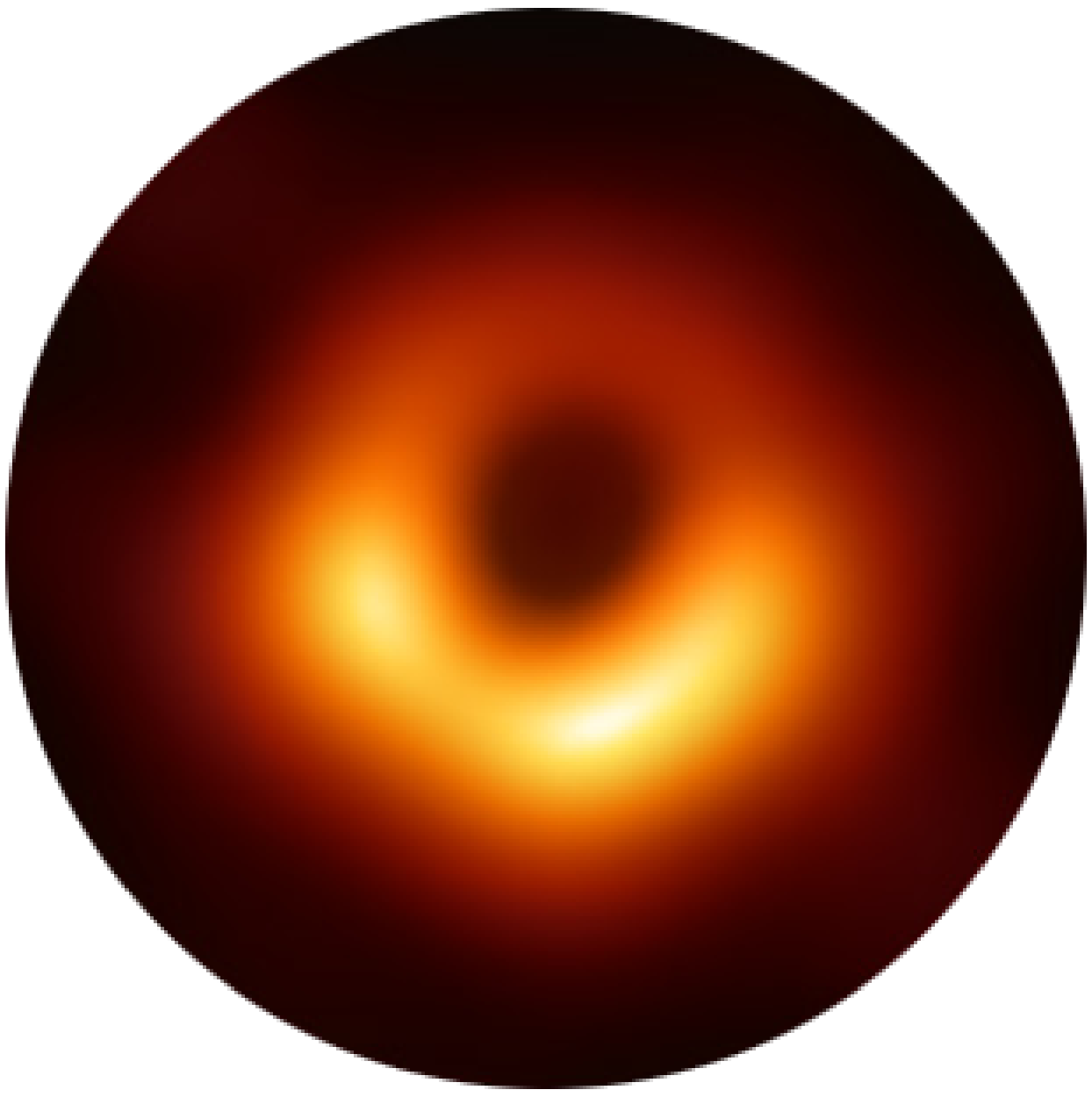

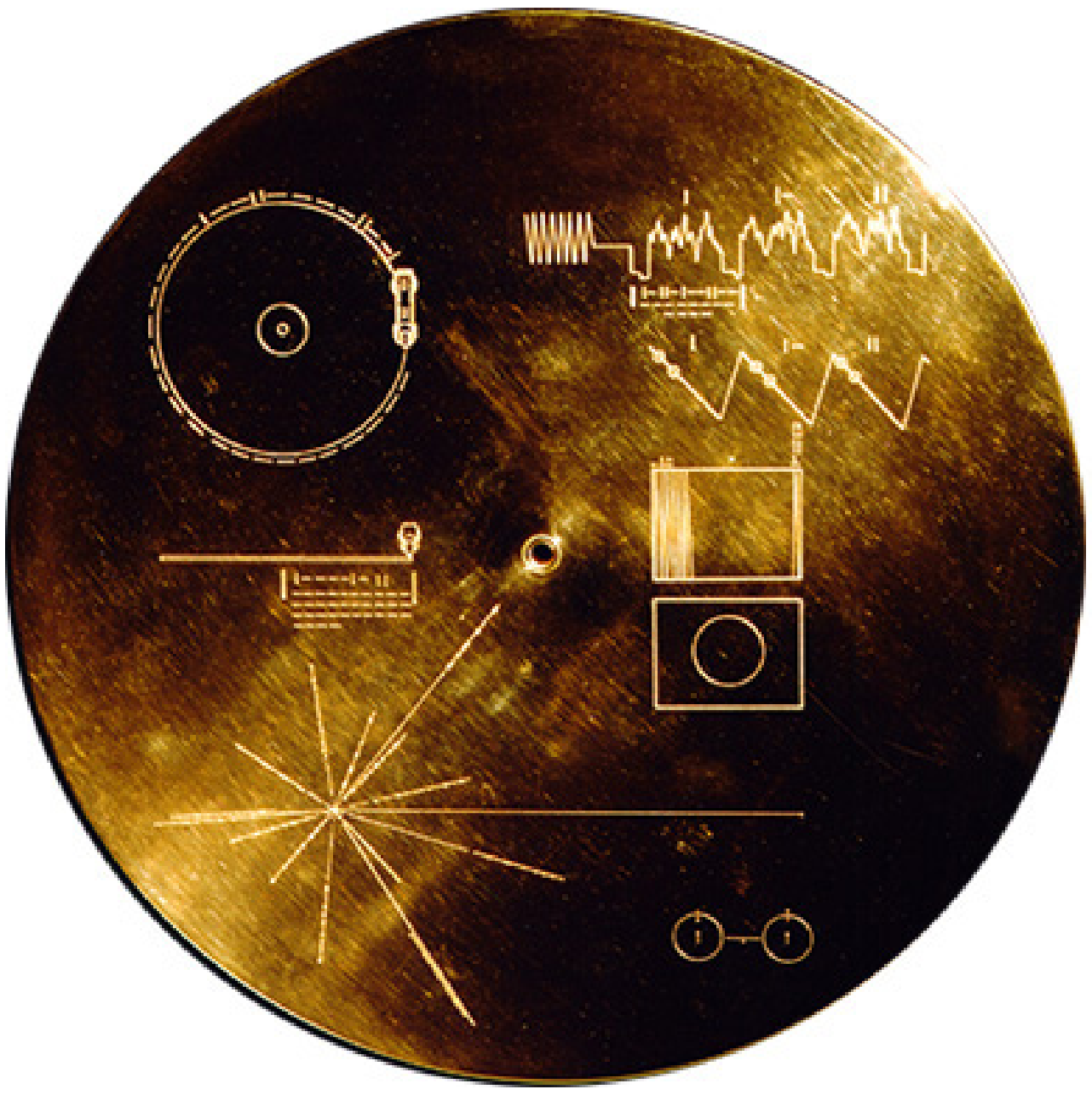

**Fig. 2.2:** The Golden Record on the interstellar space mission Voyager 1 and 2. The infographics are intended to convey information about Earth to extraterrestrial life forms. (Quelle: NASA/JPL)

## 2.1 Visualization as a tool

The entire history of astronomy testifies to the ability of humans to form a visual image of invisible phenomena based on information. In this respect, world views themselves are of a visual nature. First of all, they do not correspond to a direct observation of reality. Instead, based on observations of the night sky, a mental space is created and the sky objects are located in it. Figure 2.3 shows a historical visualization of the worldviews of Tycho Brahe and René Descartes, which make it clear how different this location in space can be and how it changes with time and the level of knowledge.
The order of celestial objects also plays an important role in epicyclic theory (see Fig.2.4). The epicyclic theory is a geometric method with which, among other things, the Greek scientist Claudius Ptolemy described the planetary movements in the sky. He placed the planets on epicycles (circles whose centers move on another circle), which were on the main circles of the planetary orbits. This resulted in loop tracks with different distances to the center. With the description Ptolemy searched for an explanation for the observed changes in brightness of the planets (cf. Spektrum der Wissenschaft, 1998).

For thousands of years, the human eye has been the primary instrument for observing and detecting astronomical sources. In order to be able to archive, interpret and communicate what was seen, the observations were drawn by hand. Figure 2.5 shows one of the first historical drawings of the planet Mars, made by Christiaan Huygens around 1659. The drawing shows a large dark structure, which today probably corresponds to the Syrtis Major, the largest extensive plateau on the surface of Mars. Huygens made several of these drawings and determined the rotational speed of the planet from the recurring position of the structure.

The drawing as an archiving tool lost its importance with the invention of light-sensitive materials (photo plates). Observations of the sky were archived directly on the photo plates. The analysis was initially based on visual comparisons and the handy transfer to statistical graphics. The blink comparator is an example of a direct visual method as a tool (cf., Gutekunst, 2015). Due to the rapidly changing illumination of two superimposed photoplates of the same sky region, the stars

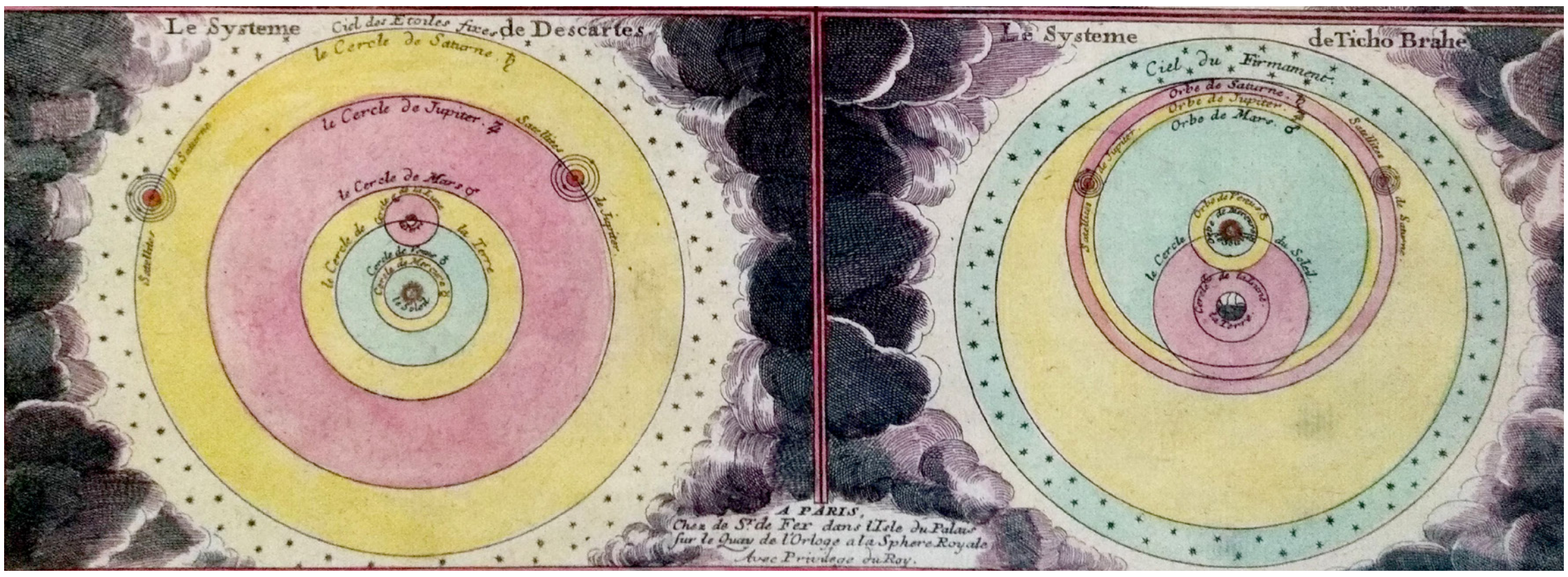

**Fig. 2.3:** Historical visualizations of the geocentric and heliocentric worldviews by René Descartes (left) and Tycho Brahe (right). (Quelle: Rendgen, 2019)

whose brightness has changed between observations seem to flash. So new changeable stars could be discovered. In 1930, the dwarf planet Pluto was discovered with this visualization technique.

Visual comparisons were quickly translated into statistical data. As a result, information graphics and diagrams increasingly developed into statistical representations. The development of the butterfly diagram is based, for example, on the same principle as Huygen's observations of Mars: repeated observations of a phenomenon are recorded at different times. While Huygens graphically depicted his observations, the observation by the sun observer Edward Maunder already shows an abstraction and thus a coding of the observation. By simultaneously recording the position of a sunspot and the time of observation in a diagram, he was able to recognize the pattern of a periodic geographic shift of the sunspots. The diagram thus represents an extension of the simple local order of data by visualizing both local and temporal parameters. The consideration of additional dimensions enables the viewer to place the observation in a large context and makes correlations visible.

The examples clearly show that graphic visualization can be understood as a tool for analyzing information and data from astronomical observations. Be it in the representation of concepts, through the direct geographic order and mapping of observations or in the abstract representation in the form of statistical diagrams. The role of visualization finally changed due to the development of computer-aided analyzes. The digital detection of light by semiconductor technology triggered a revolutionary change. Measurements spread across the entire electromagnetic spectrum and due to the direct conversion of light into electronic signals, digital and algorithmic analysis methods became more important. The examples above, in which the visual analysis was used, are data sets with few parameters that could be sensibly plotted and perceived in simple two-dimensional, statistical diagrams. With the development of computer-assisted generation and storage of data, the amount and complexity of data sets grew and computers took over their exploratory analysis. According to Goodman (2012), data analysis and its visualizations are now viewed as separate aspects of research. The visualization serves primarily to as-sess and communicate results and only follow after the computer-based analysis and processing of the data.

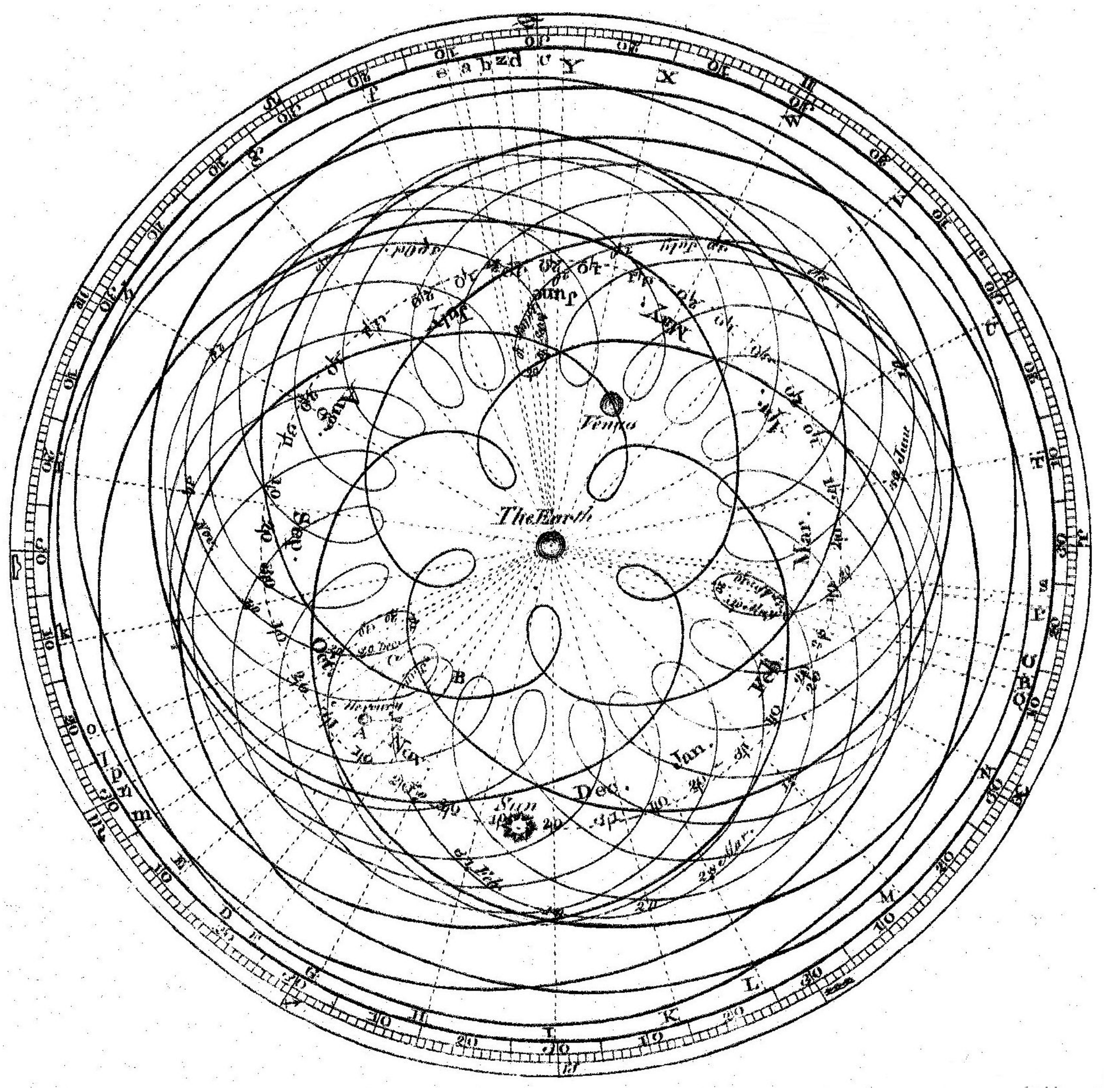

**Fig. 2.4:** Epicyclic diagram of the orbits of the Sun, Mercury and Venus by James Ferguso (1710–1776). (Source: Encyclopaedia Britannica)

In general, demands for scientific data visualizations depend heavily on the analysis goal and the data sets. Is it e.g. Visualization can continue to serve as a tool for precise mapping, such as when researching planet surfaces. Visual examinations are sometimes also used at the beginning of analysis processes to filter databases according to areas of interest and objects (see Fig. 2.7 and 2.8). Statistical diagrams are then used to classify and evaluate the results. In this sense, data visualization serves more as a control tool.

*„Oddly though, as astronomy's wavelength coverage increased, the value of the ‚visual' to astronomers seems to have declined – not as a wavelength, but as a tool."* (Goodman, 2012)

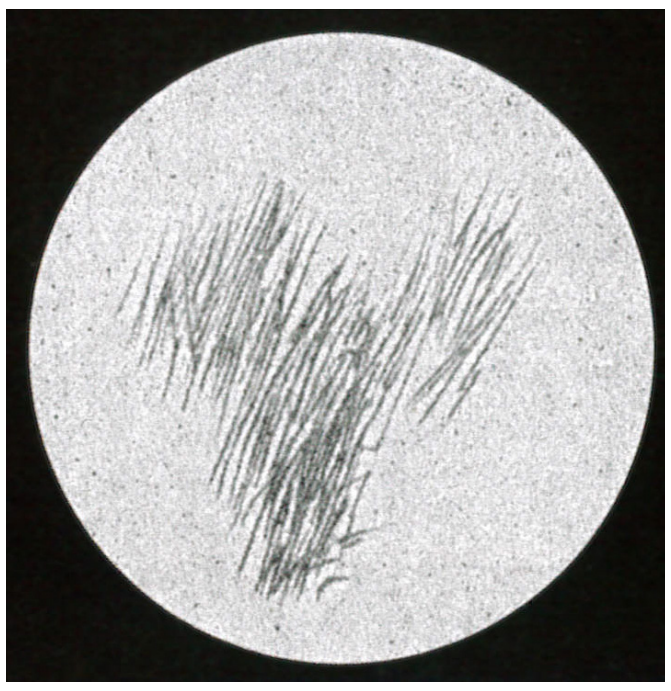

**Fig. 2.5:** First drawing of the Martian surface by Christiaan Huygens from 1659. (Quelle: Archiv Uwe Reicher, SuW, 2016)

Contrary to Goodman's hypothesis, there are still areas in which visual analysis comes into play. In these cases, it is often a matter of examining the structures and distributions of astronomical objects in so-called parameter spaces. For the Cosmic Flows Project (see Fig. 2.9, Courtois et al., 2013) an entire series of publications was based on a film. This represented an animated data visualization and showed the discovery of the super galaxy cluster Laniakea, in which our own Milky Way is located. Another class is physical simulations, which convey images of causal relationships and provide clues for new analyzes. A prominent example is the millennium simulation (see Fig. 2.10) of the structure distribution in the universe based on cosmological models.

The more efficient and prominent statistical analyzes were made by computers, the more visual representations moved into the background as analysis tools. With the increasing dimensionality and complexity of the data sets, it became more and more difficult for the astronomer to get an overview of the data situation and its context. Instead, they are forced to trust the results of the computer.

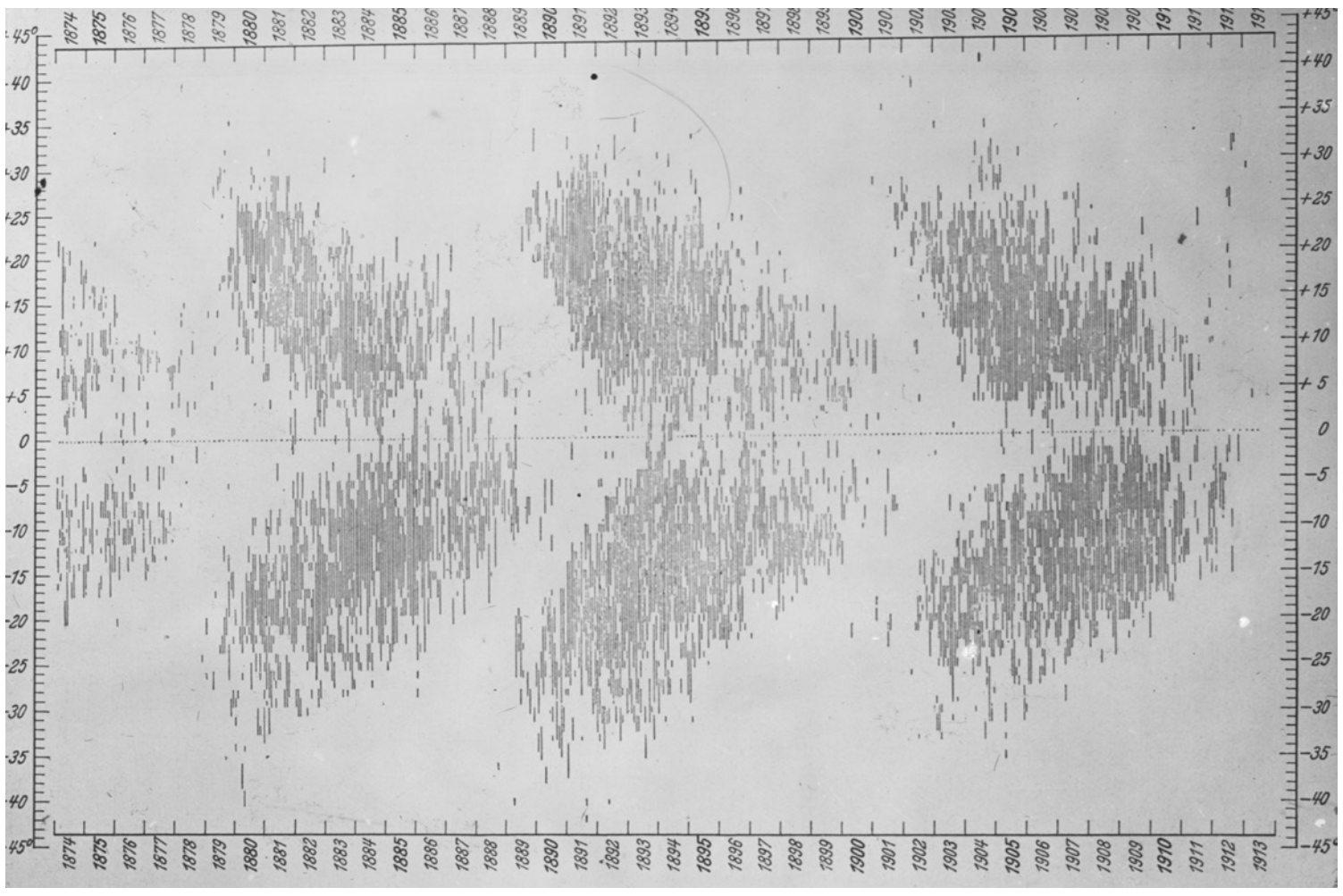

**Fig. 2.6:** Butterfly chart by Edward Maunder. (Source: Hamburg Observatory Digital Plate Archives)

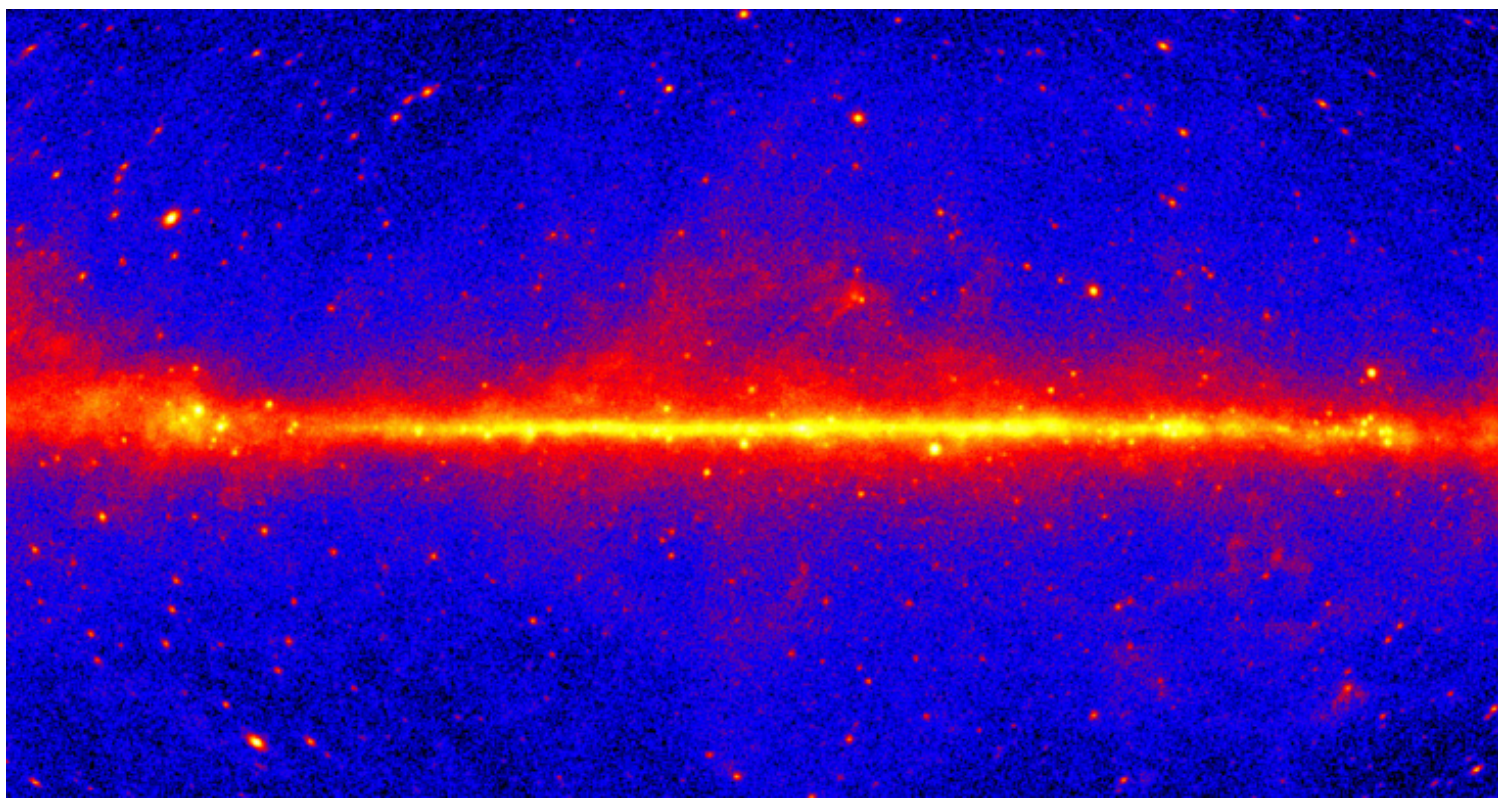

**Fig. 2.7:** Mapping the high-energy sky as a two-dimensional color diagram. The intensity of gamma radiation is coded as a color gradient from dark blue (weak), through red, to yellow (strong). (Source: NASA / DOE / Fermi LAT Collaboration)

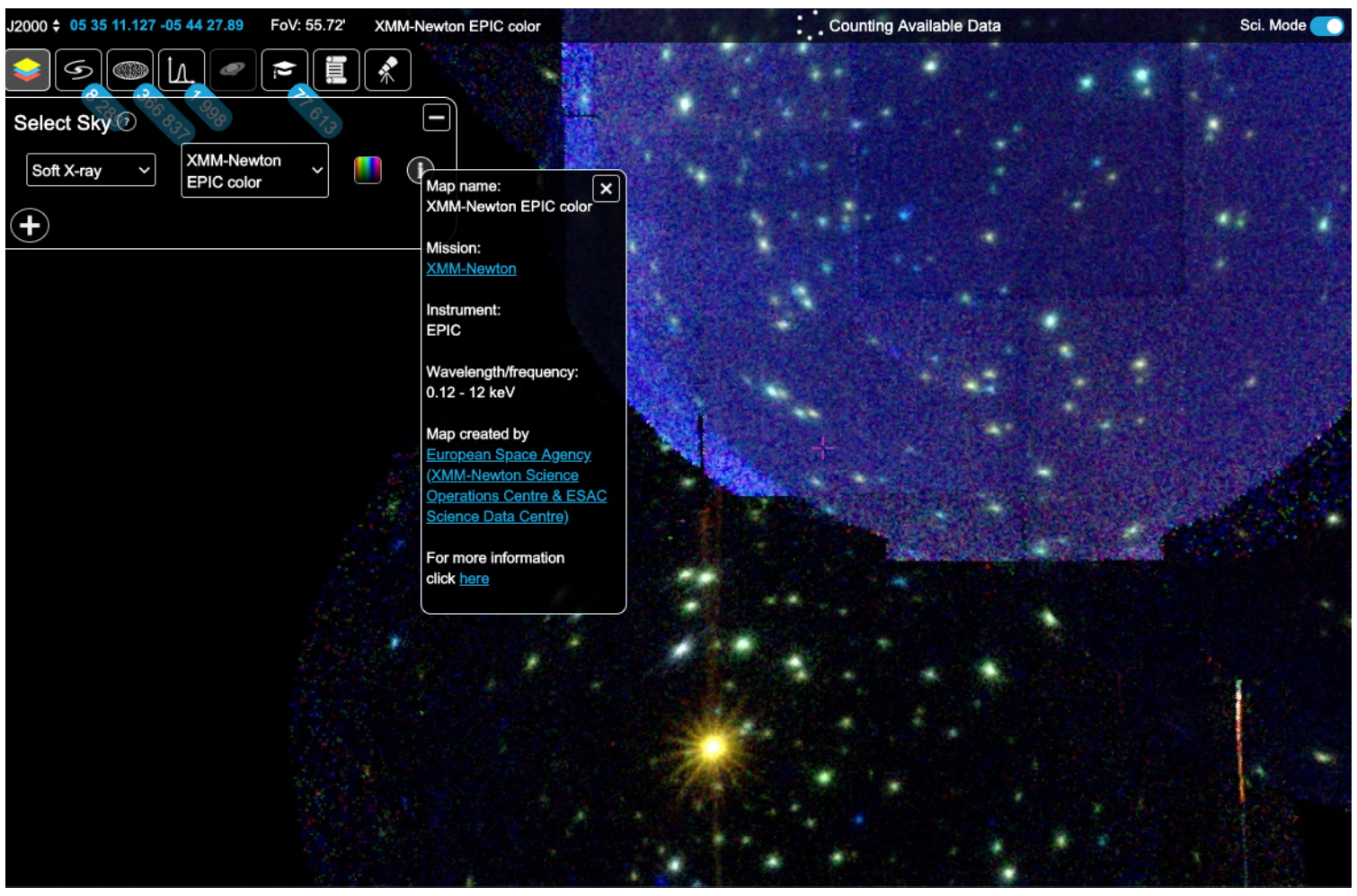

**Fig. 2.8:** Interface of the ESASky application of the European Space Agency. With the visual search for data sets, ESA responded to requests from astronomers to make the surface more visual. (Source: ESASky)

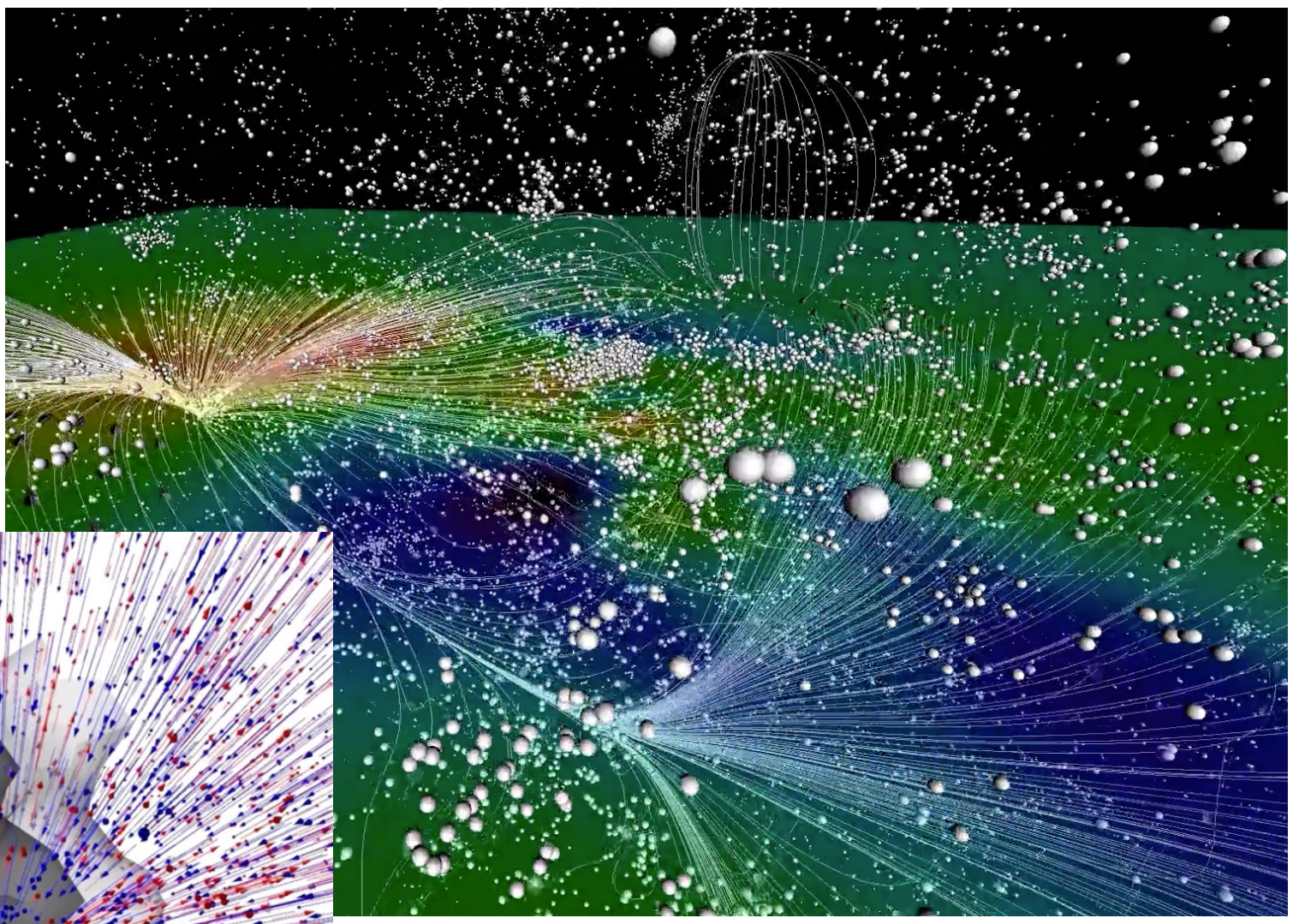

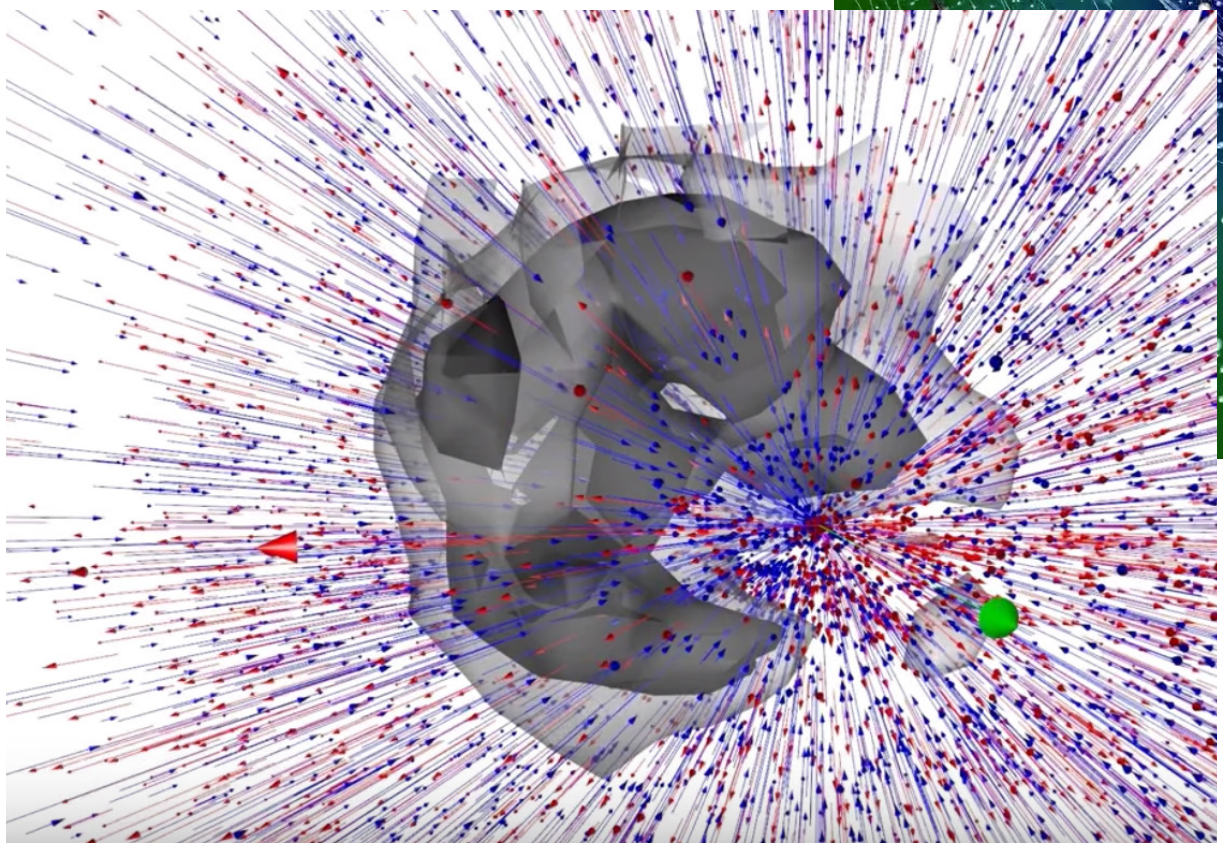

**Fig. 2.9:** Animated visualizations of galaxy movements that were used to analyze and discover the super cluster Laneakaia. (Source: Cosmic Flows Project)

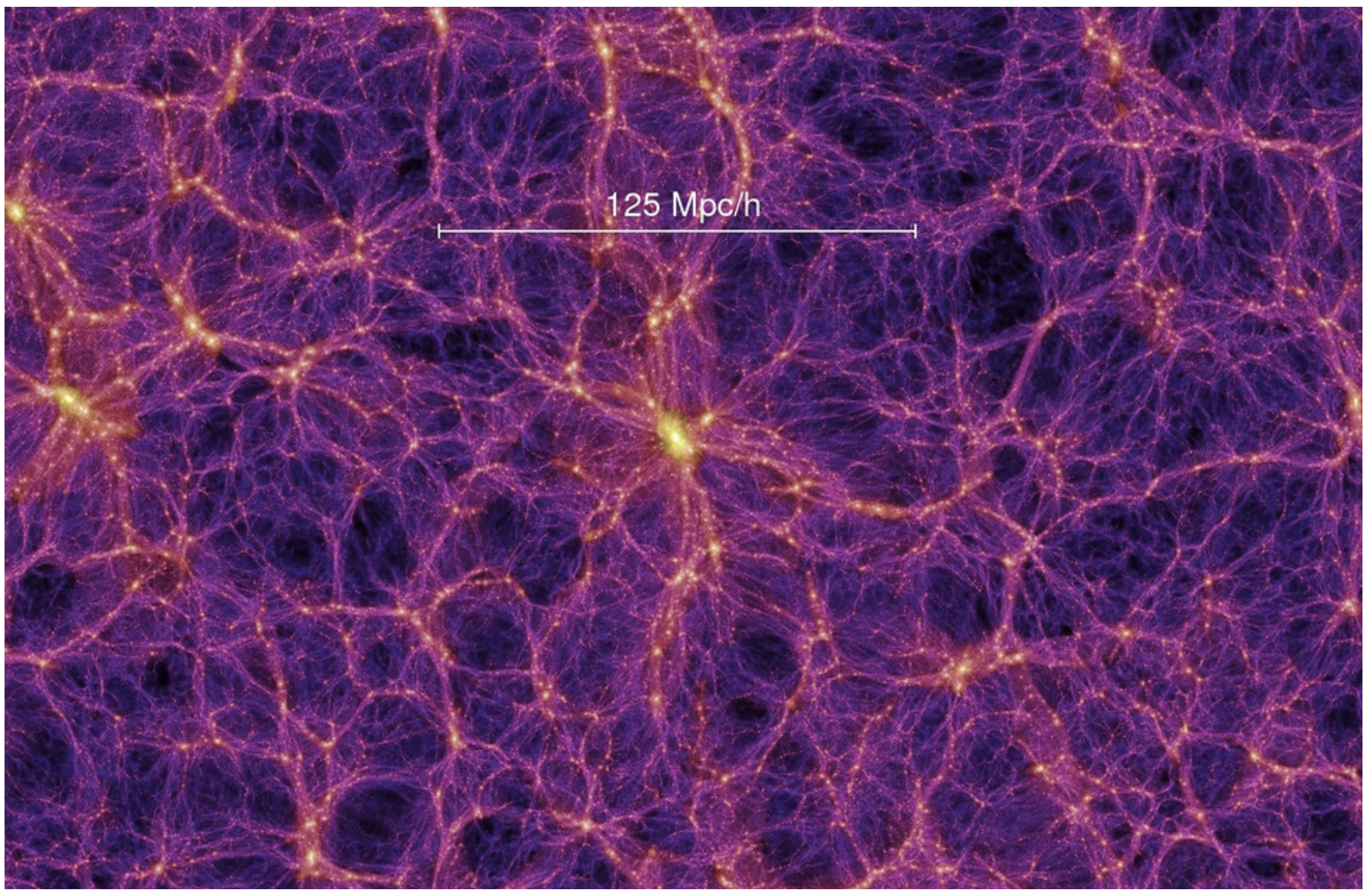

**Fig. 2.10:** Visualization of simulated data. The Millennium Project is currently one of the most accurate simulations for the structure formation of the universe. (Source: Millenium Project)

## 2.2 Visual challenges in astronomy

In modern astronomy, individual measurements consist of images, simulations, catalog information and multidimensional data sets. In the analysis, different sizes must be considered at the same time, which take into account both the condition of the telescope and that of the sky region to be observed. Due to the increasingly complex preparation and analysis of data, especially since the opening up of newer observation channels such as cosmic particles or gravitational waves, the analysis focused on statistical methods. A problem with focusing on statistical methods is the lack of or difficult access to contextual information, or the limited overview of the content of the data set and the relationships of the data dimensions to one another (see e.g., Hassan & Fluke, 2011; Goodman, 2012). Many astronomical data centers and archives, such as the European Space Agency, ESA, are therefore looking for new interactive, visual concepts to enable scientists to access their databases more intuitively (see Fig. 2.8).
The increasing importance of intelligent, machine-learning algorithms, the analysis steps of which are no longer easy to understand, also increases the need for transparency during data analysis (cf., Hassan & Fluke, 2011). Although many steps of data preparation and analysis can be processed by intelligent algorithms, the conversion of data into information and ultimately knowledge is only possible through humans (see e.g. Cairo, 2013). In order to be able to fulfill this position, as a scientist it is necessary to have an overview of the existing data set, its connection to research and its importance for the following steps. This overview is currently very difficult to obtain, depending on the data situation. In the case of highly complex data sets in particular, the necessary transparency is often lacking, among other things due to the lack of visualization concepts. In his article on the visualization of complex data, Goodman (2012) identified four future challenges for (explorative) data visualization: Big Data, the variety and individual challenges in analysis, interface design, and the 3D selection of subsets. In the last two points, the design of information can obviously contribute to improvement. But the first two challenges offer the opportunity to find new design approaches that are not only used in astronomy.

As early as the 1970s, the American statistician John Tukey defined a methodology with his work Exploratory Data Analysis (cf. Tukey, 1977) to support the ability of humans to recognize patterns. It highlights the visual exploration of data sets. The visualization should be met with skepticism and openness: skepticism about potential errors in the display, and openness to new unexpected data patterns.

*„In the case of high-dimensional data sets, though, interactive exploratory data visualization can give far more insight than an approach where data processing and statistical analysis are followed, rather than accompanied, by visualization“, (cf. Goodman, 2012).*

While researching the data set, hypotheses can then be formed and tested on the basis of the data (Hartwig & Dearing, 1979). In exploratory data analysis (EDA), new patterns and unexpected results are seen as a starting point for generating new hypotheses and future studies, rather than statistical conclusions. Ideally, the result would be a set of different plausible models that will be evaluated in further analysis steps. In the EDA, the possibilities can above all be exploited through modern visualization systems (cf. Behrens, 1997). These allow the viewer greater interactivity with the data in order to be able to assess certain working hypotheses quickly and fluently by viewing data from different perspectives or highlighting different parameters of a problem. The challenge for EDA in data-driven modern astronomy, however, is to display data sets with a large number of parameters (multi-dimensional data sets). These must be clearly translated into visual parameters, such as position, shape, color, etc., so that the viewer can recognize possible patterns. Today, statistical data mining methods are used which easily capture multidimensional parameter spaces. The results of these analyzes are often incomprehensible to the astronomer and cause a certain amount of distrust/doubt. Nevertheless, especially due to the steadily growing amount of data, the need for new presentation concepts has been recognized and has been investigated in various individual projects for several years (see Kent, 2013; Baines, 2017; Taylor, 2017).

A common approach is the principle of linked views (cf. Goodman, 2012). Different displays and representations are linked to each other so that the effect of data processing, e.g. the filtering is visualized simultaneously in all representations. This gives researchers a direct overview of what an action does not only result in one parameter space, but in the entire data set. There are already various applications, such as Scientific Toolkit for Astrophysical Research (STAR, Mickus et al., 1990) or the LSSTGALPY (Argudo-Fernandez et al., 2017), which were either adapted for use in astronomy or newly developed. However, the displays are still based on traditional display in the form of histograms and curves on two-dimensional displays. In order to break out of two-dimensionality, some studies are already testing to what extent the new media, i.e., virtual and augmented reality, can be used for data visualization (cf., Fluke & Barnes, 2018).

# 3 Data Visualisation in Virtual Reality

The research results of the last chapter show a common problem of information design in modern astronomy: many applications are still limited to the visualization of large multidimensional data sets in 2D and quickly reach the limits of clarity due to the abundance of different display windows on the screen. Three-dimensional representations are also usually reproduced on two-dimensional displays. Size relationships and spatial structures are therefore more difficult to understand. In astronomy, the possibilities of 3D data visualization in combination with 3D modeling and game development software (e.g. Unity or Houdini) have so far been used only in the context of science communication (see Kent, 2013; Taylor, 2017; Naiman, 2017). The intention here is the pure visualization and explanation of data without interactive analysis. Combining generative information design and statistical analysis would be possible with today's programs. Many programs for generating these virtual data worlds allow the use of programming languages so that the parameters of a visualization can be influenced and changed by the data using program code.

The problem of visualizing complex data interactively and dynamically is by no means new or limited to research. Examples from digital information design, such as e.g...., show how visualization can support the understanding of structures within data sets in addition to mathematical methods. Genome Valence by Ben Fry (Fry, 2002), or Meviatis by Ricarda Schuhmann (Schumann, 2018). In the projects, the possibilities of generative design are used to visualize relationships in different data sets from biology and medicine. The parameterization of the representation through the different dimensions of the data set creates a dynamic and aesthetic data sculpture and enables quick interpretation and comparison of the data.

Die neuen digitalen Medien erlauben uns heute einen Schritt weiter zu The new digital media allow us today to go one step further than to design interactive three-dimensional applications for the screen. Both augmented reality (augmented reality, AR) and completely artificially generated reality (virtual reality, VR) make it possible to experience the geometry of data rooms in three-dimensional spaces anew. Thanks to the data-based coding of color, shape, transparency and animation, far more than three dimensions can be represented. As a user, you can also easily move and navigate around the room. An interesting aspect of the location of data in three-dimensional space is the immersive moment of perceiving a data structure from the inside and not, as in traditional visualization, from the outside in (see, Djorgovski, 2018).

Data visualization in VR and AR is a highly topical subject, the applications of which are not only aimed at science, but also at computer science and industry. This is also shown by the large number of start-ups (e.g. 3data1, badVR2, Virtualitics3, nanome4) that specialize in this topic. However, the visual design of the data worlds continues to be based on the two-dimensional, statistical diagrams. One aspect is certainly that scientists are used to these representations and have learned to read and interpret them. The question is, however, what advantage VR and AR actually have and whether this results in an improved perception of multidimensional data sets. The potential of these technologies is certainly not yet exhausted in the field of information design.

[1] 3data: https://3data.io/
[2] badVR: https://badVR.com
[3] Virtualitics: https://www.virtualitcs.com/
[4] nanome: https://nanome.ai/

## The benefit of data visualization in virtual reality

There are various studies that demonstrate the benefits of data visualizations in comparison of VR with conventional methods on the screen and discuss their advantages (e.g., Chandler, 2015). An obvious difference is the increased field of view in virtual reality compared to a computer screen. By immersing yourself in a complete data room, various information can be placed around the user in all directions. This enables additional information to be displayed in the same room as the data itself. Such an arrangement could improve the overview in the linked view graphics mentioned.
Another advantage of three-dimensional information graphics in VR is that the user can move freely around the data. This makes it easy to take different viewing angles on the data. Recognizing three-dimensional structures and patterns is a natural process in the human brain. A graphic does not have to be rotated in different directions in order to recognize its three-dimensional structure. This speeds up cognition and enables structures to be perceived quickly (cf., Warfel, 2016).

In VR, flat structures on screens or print media can literally be pulled apart. The viewer can go through data and objects, immerse them in them and experience them in a spatial context. The scalability of virtual space is extremely advantageous (cf., Olshannikova, 2015). For these reasons, physical representations and data installations are often shown in public relations. Since the perception of data structures in real or virtual three-dimensional space happens more intuitively and faster than e.g. on posters. However, the experience of extreme scalability and manipulation of space is only possible in virtual space.

In addition to the additional space dimension that can be assigned as a parameter, sound and especially haptics also provide other parameters for the data display (Chandler, 2015). VR also makes size ratios easier to perceive. It is easy to get an impression of the relative size of an object or room in virtual space. Perspective distortions in VR glasses make direct comparisons or quantitative statements about the size of objects difficult (Kurbatov, 2017). Here, for example, two-dimensional bar charts have an advantage.

Millais et al. (2018) investigated to what extent students process scientific data on screen and in VR. The subjects were divided into two groups and, depending on the group, had the same scatter and bar charts either in VR or as interactive infographic on the computer. During the exploration of the data set with the given representations, the test subjects should make statements and hypotheses about the data. In comparison, the authors found no difference in the workload during exploration. The number of correct statements regarding the correlations of the data between the groups was also indistinguishable. A significant difference, however, was found in the number of incorrect statements, which was significantly lower for the VR group. The VR application also enabled a more natural and faster handling of the data. The VR subjects felt much more satisfied after the examination than the control group and felt that they understood the data set well. The authors suspect that the reason for this feeling was the more immersive experience of VR Exploration.

The advantages of data visualization in virtual reality can be summarized as follows:

More space!
VR rooms offer a larger field of view, allow multiple views to be arranged in the room, making it easier to draw cross-references and connections.

More dimensions!
Compared to two-dimensional graphics, VR representations offer additional parameters that can represent data (sound, haptics, light, movement, position, etc.).

More structure!
The perception of space and depth is more direct, shapes and volumes can be recognized more quickly.

More fun!
Immersing yourself in the data and the ability to go from overview to detail by scaling the space is a powerful immersive moment.

# 4 Hypothesis und Objective

On the basis of the research described, this work hypothesizes that the presentation of scientific data with new digital media, especially VR, offers great potential for data analysis in science. Currently, information design in the scientific context is often too immature to meet the requirements of multidimensional data sets. For many scientists, exploratory data analysis cannot be used to get a quick overview of their data.

The aim of my work should be to use the example of a data set from astronomy to find concepts and methods for scientific visualization with new digital technologies that make exploratory data analysis more usable.

# 5 The Data Set

The elaborated problems of modern information graphics in astronomy and the advantages of data visualization in virtual reality, defined concrete demands on the example data set for the analysis of this work: the data set should contain a high number of parameters and astronomical objects. Since the subject of the investigation was whether a visualization of the data set could become an active tool for analyzing the data set, it was also important for the selection that the individual analysis steps and questions to be answered by the analysis were known. This fact led the research to the projects that I worked on during my doctorate in astronomy (Kreikenbohm, 2019a). A data analysis proved to be the most suitable, the result of which was already visualized in a reduced form for the general public in VR in the summer semester project 2019, The Tangible Universe (TTU, see Kreikenbohm, 2019b). The entire data set of the analysis should now be considered in this work. While the visualization in the semester project had an educational character in the sense of science communication, the goal of this work was to find a scientific visualization that enables the astronomer to explore the data set in the sense of exploratory data analysis. The diagram in Fig. 5.1 summarizes the analysis and content of the data set. The analysis took place within the framework of the EXTraS (Exploring the X-ray Transient Sky) project5 and dealt with the classification of unknown X-ray sources in the visible universe.

The ESA XMM-Newton X-ray satellite detects X-rays emitted by different sources in the universe. The starting point of the analysis was the XMM-Newton archive, which contains the raw data of all observed sources. Not every source knows what an astronomical object it is. The top map in Fig. 5.1 shows the positions of all radiation sources in the galactic coordinate system. The positions of the sources were mapped onto the inner surface of an imaginary celestial sphere. Like an earthly world map, the two-dimensional representation corresponds to an unfolded sphere. The equator corresponds to the galactic disk of the Milky Way.

In order to classify the sources, statistical and physical properties such as intensity or energy distribution were determined for each individual source in spectral analyzes. This resulted in a data set of 144 parameters for 137.212 sources each. Each parameter can be understood as the dimension of the data set. Together, they span a 144-dimensional parameter space in which each source has a unique position.
The model for the classification was then created. That is, an intelligent algorithm learned the structure of the parameter space based on a number of known sources that were described with the same parameters and differentiated into eight classes (star, galaxy, etc.). In particular, he learned the patterns according to which the sources could be differentiated into the respective classes based on a previously defined selection of parameters. The result was a ranking of the relevance of the selected parameters, as well as an assessment of how high the error rate of the classification was. Four different parameter selections were made during the analysis. This resulted in four classification models with their own ranking lists and error rates.

The respective classification models were then applied to the unknown sources. The algorithm calculated for each source with what

[5] EXTraS Projekt Homepage: http://www.extras-fp7.eu/

**Fig. 5.1:** Scheme of data analysis in the EXTraS project.

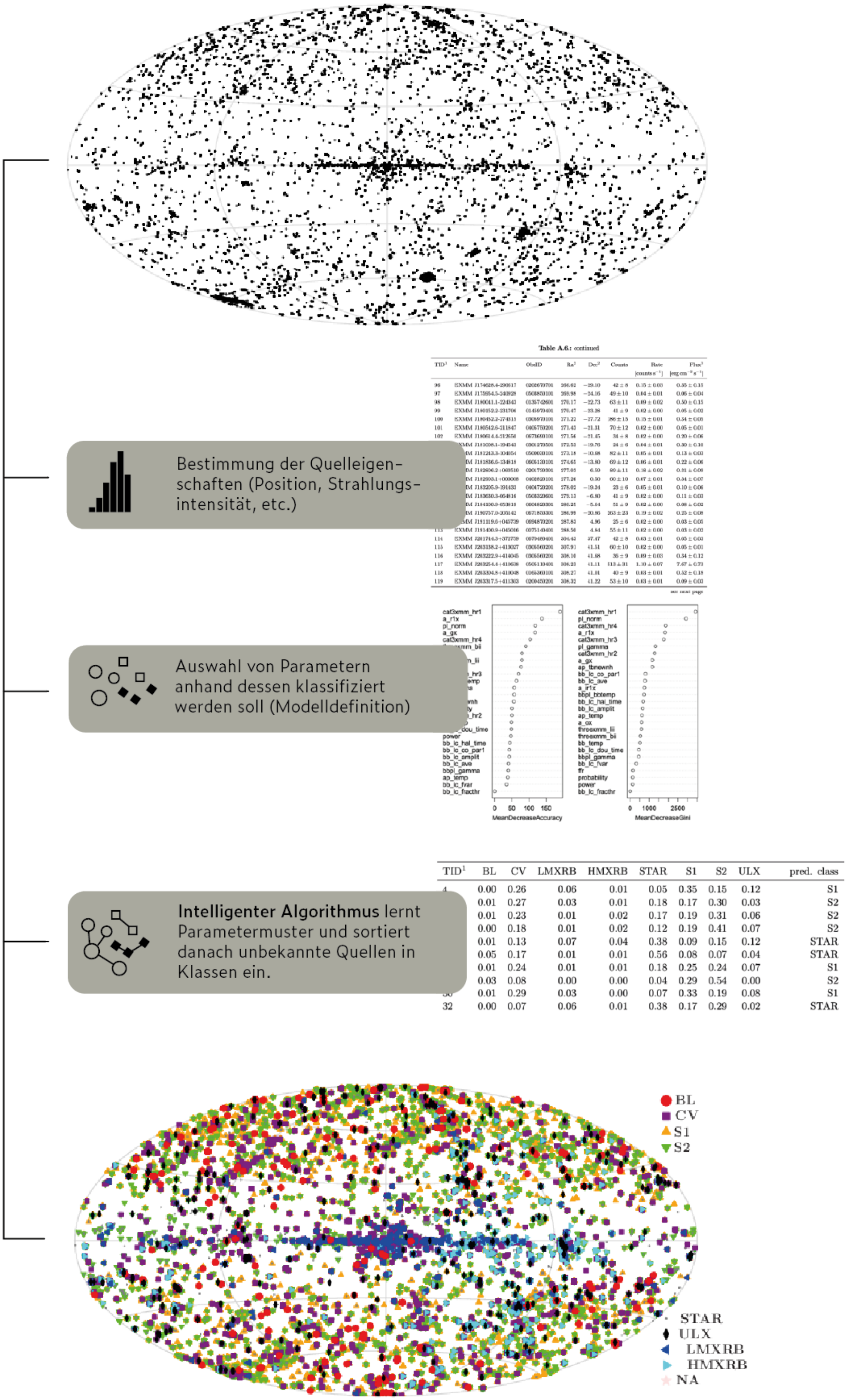

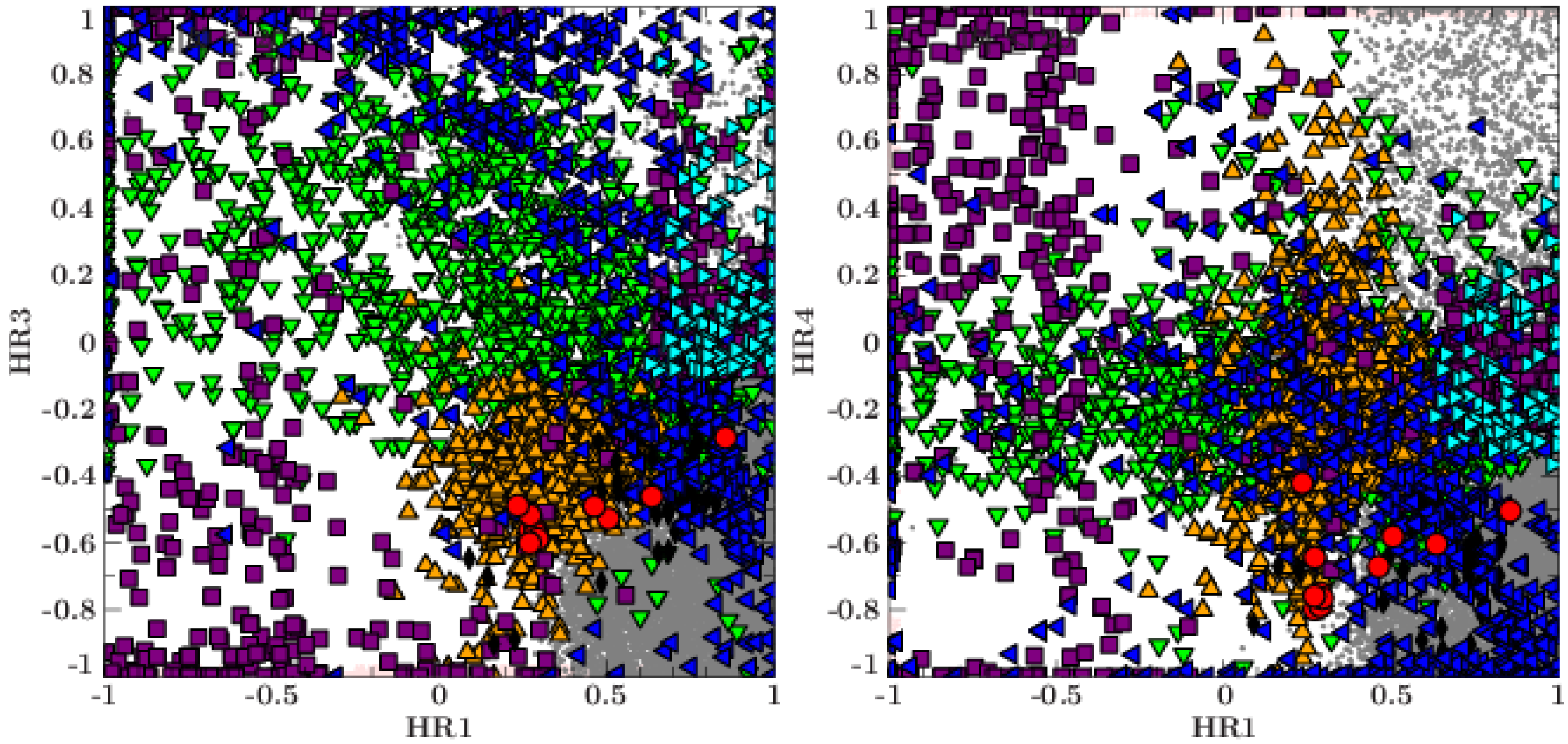

**Fig. 5.2:** Scatterplots are used to search for correlations between classification and parameter values. The diagrams show the distribution of sources, the probability of which is more than 60%, for only two parameters. Despite the limitation, the zones overlap several classes. (Kreikenbohm, 2019a)

probability it could be assigned to a certain class. This resulted in a data table with the probabilities calculated per source and class. The four defined models result in four different classifications of the unknown sources. One of them can be seen as a galactic map in Fig.5.1 (below). The different source types were represented by different colors and symbols. However, since the position is very close, many of the sources overlap. The map is unable to show the full picture of the class distribution.
Eight classes were distinguished in the analysis. That is, for each unknown source, there are four sets of probabilities in addition to the parameters. Since a maximum of 27 parameters were selected for the model definition in the analysis, each of the 137,212 sources is described by 59 dimensions (27 parameters and 32 probabilities).

After the classification, the question arose which model most sensibly classified the sources? To answer this question, one first looked for the differences in classifications or class distribution between the models. Associated with this is the question of where the differences come from. What are the patterns on which the algorithm acted? What correlations are there between the parameters of the sources with each other and with the respective classes?

The answers to these questions are ultimately the basis for new hypotheses as to how and why x-ray sources differ. When looking for the answers, histograms and scatter plots are often used, as in Fig. 5.2. But the amount of data makes two-dimensional representation very difficult. The data points overlap and as a viewer you have the feeling that you cannot capture the entire data set. **How can the 8,095,508 entries be sensibly visualized to find the first signs of parameter correlations and differences in the classifications?**

# 6 Concept

*„Excellence in statistical graphics consists of complex ideas communicated with clarity, and efficiency“, (Tufte, 1983).*

Edward Tufte coined the term of graphic excellence in data visualization (Tufte, 1983). He postulated various properties that successful statistical graphics require. This includes that data should be displayed directly without distracting the user through the design and method. Furthermore, statistical graphics should serve a clear purpose (description, exploration, tabulation or decoration) and show several levels of details of the data (from the rough overview to the fine structure). Similar claims were made by Olshannikova et al. (2015) a study on the visualization of big data in VR and AR. In order for the data visualization to serve as an analysis tool, it requires the exact representation of information that is present in the data records. In addition, the graphics must effectively use the cognitive abilities of the human visual system.

These conditions of good information graphics described by Tufte and Olshannikova should be considered in the conception of the data visualization of this work. It was also based on two aspects that were already being pursued in the The Tangible Universe project and which proved to be very effective in being able to perceive many dimensions of a data set at the same time: the parameterization of the visual appearance through the data itself and the positioning of the individual data objects in space by their parameters. The former meant that the visual appearance should mainly be parameterized by the data and the aesthetics of the virtual space resulted from the data itself. This affected both the design of the data objects and their position in space. Regardless of how the data was organized, the basic approach was to pull the data cloud (see e.g. in Fig.5.2) apart and display it in three-dimensional space. This should enable the viewer to dive deep into the data. The following diagram shows how the data records for one X-ray source are linked:

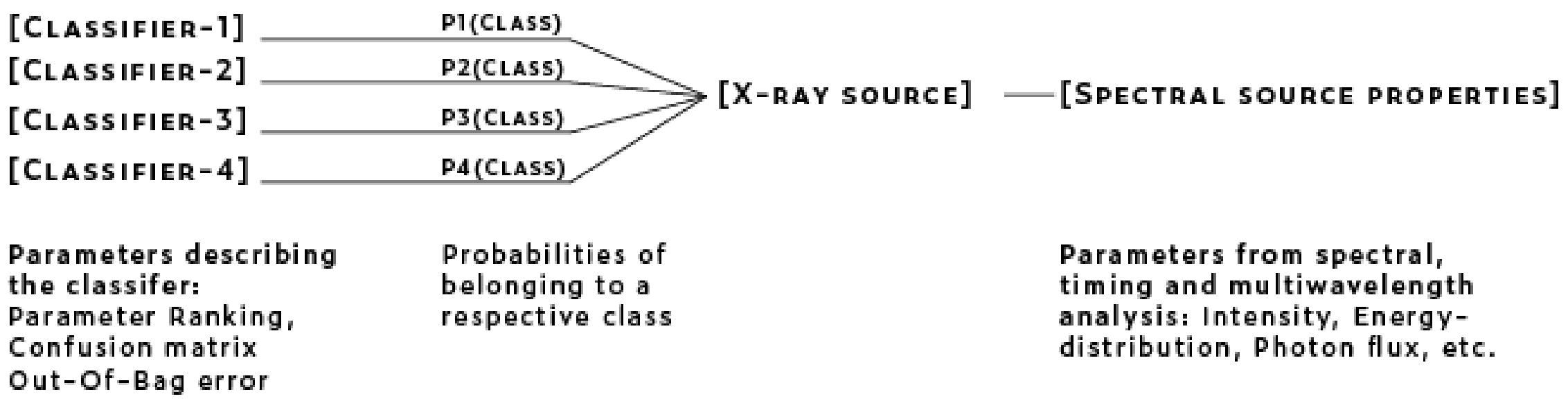

On the one hand, each source has an entry in the database of the radiation properties that result from the satellite data (hereinafter referred to as spectral properties). On the other hand, each source was assigned eight probabilities for a certain class affiliation by four classification models. These models are in turn described with their own parameters that describe the quality of the classification. One difficulty that arose at the beginning was the construction of a narrative. How should the different data sets be linked? In what order would an astronomer explore them? Should the viewer first start with the spectral properties, get an overview of the correlations there and then, based on this, examine the models? Or should, conversely, be based on the models and, after choosing a classification, explore their result data set (as in TTU)?

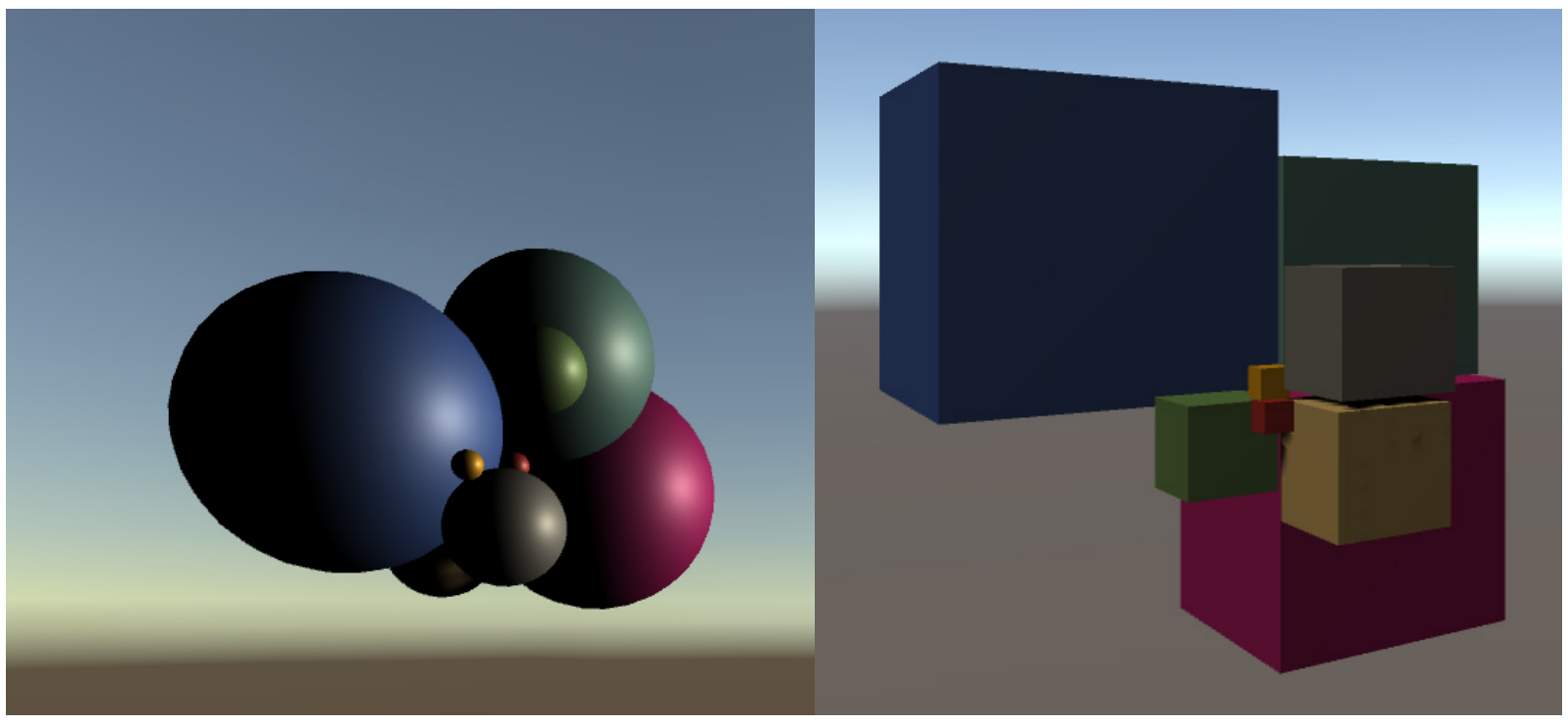

**Fig. 6.1:** Construction of the data objects by single ball (left) or cubes (right), which represent individual classes. The size and color of the components is determined by the probabilities of the source to belong to the respective classes.

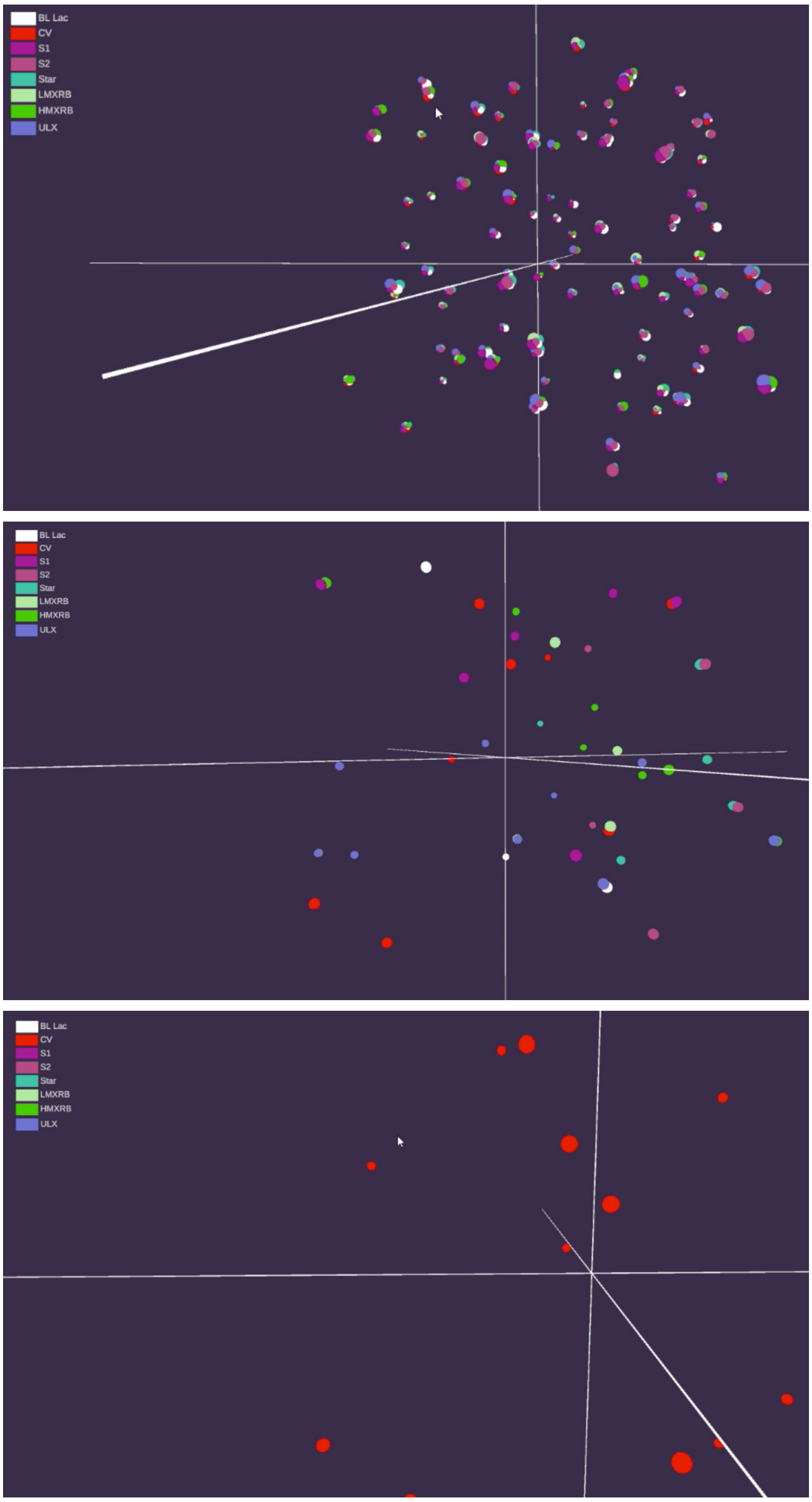

**Fig. 6.2:** Representation of simulated data objects in a 3D coordinate system based on three parameters. Above: The classification of the individual objects coded by color and size is clearly recognizable. Middle: Filtering by probability value (minimum & maximum) and selection of objects belonging to a certain class.

In order to approach the solution to this problem, different experiments were first carried out, which tested how different information from the data sets could be visualized. In a first approach, each X-ray source was constructed using several smaller components, such as spheres or cuboids, which were each identified with one of the possible classes (see Fig. 6.1). The probability of a class was shown by scaling the component. The respective class was differentiated by color coding. According to this rule, the assigned classes of a source were directly visible on the data object as part of a model, while other source parameters could be used to locate the object in space. Using this data object, the viewer could directly see which was the most likely class of the object and what the relationships between the values were.

Based on this representation of data objects, the first selection options were also tested (cf., Fig.6.2): e.g., the selection of a minimum and maximum threshold value for probabilities, which decides whether a class component is shown or not. This filter meant that the relationships between the components were easier to classify, which meant that the most probable classes could also be displayed directly. The distribution of certain classes in the coordinate space was easier to perceive. The disadvantage, however, was that only one classification model was displayed due to this representation. The different class assignments of the models for one X-ray source were not visible.

Figure 6.3 outlines a narrative based on these experiments, which starts with the choice of a model and then transfers the user to the results in a defined parameter space. This means that the user moves from the detailed view of the model objects to an overview of the result data record: First of all, the user is in a room with four model objects facing him. The parameters of the model should be used to characterize the appearance of the model objects. By interacting with the model objects, the user would now have the opportunity to get details about the respective classifications and then decide on one that should be examined more closely. The model objects should then serve as a kind of portal and bring the user into the result space. This contains the data objects that were constructed as in the experiments and positioned in the coordinate system. The question that arose here was how to choose the spectral parameters according to which the objects are located. Could the viewer choose this in a previous step or change it in real time? And how are the remaining source properties displayed if the objects are already built up by the classes?

In the course of these considerations, the first experiments to visualize various parameters of the models were also tested. Here it turned out that it would not have made sense to translate all parameters and tables into a three-dimensional representation. One example is the confusion matrix (see Table 6.1) that the intelligent algorithm automatically creates during the learning process. The confusion matrix describes how often in the model a known object is assigned to a wrong class, i.e. was confused. The matrices are usually presented in tabular form, where each row corresponds to a true class and each column corresponds to an assigned class. All control objects that have been correctly classified are located on the diagonal. All entries on the secondary diagonals suggest confusion.

**Fig.6.3:** Concept sketches

1. Explore model properties and select a desired classification

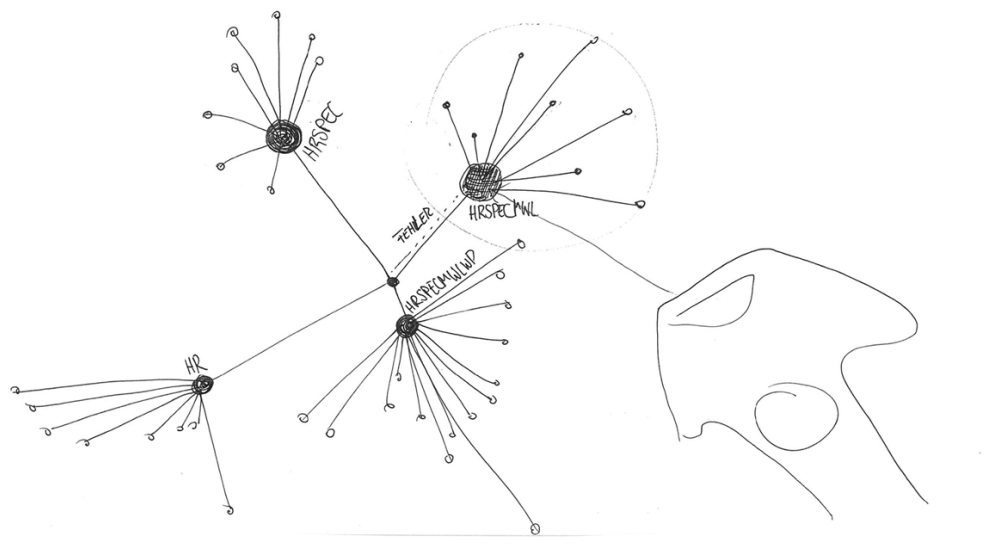

2. Objects are portals to the parameter space.

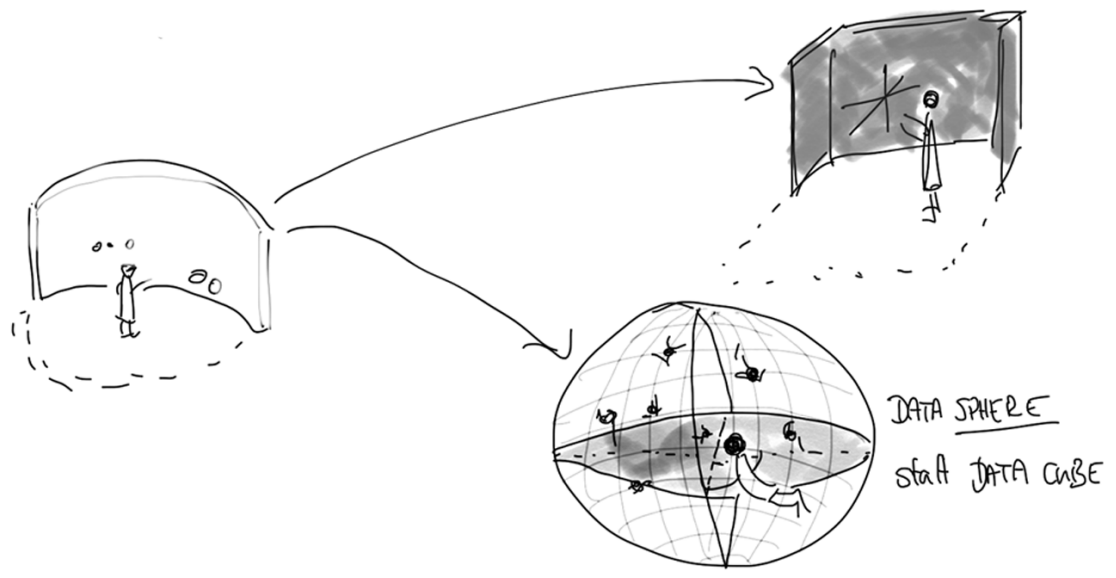

3. Explore the results in the coordinate space by "immersing" and "turning them around". Definition of the coordinate axes by the user; Interact with data objects for details.

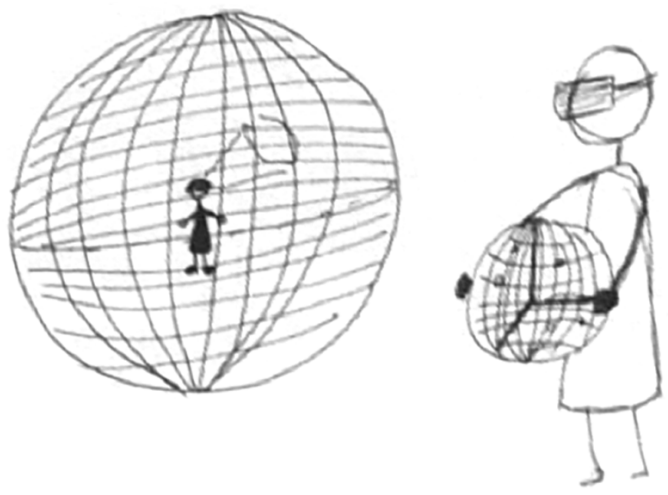

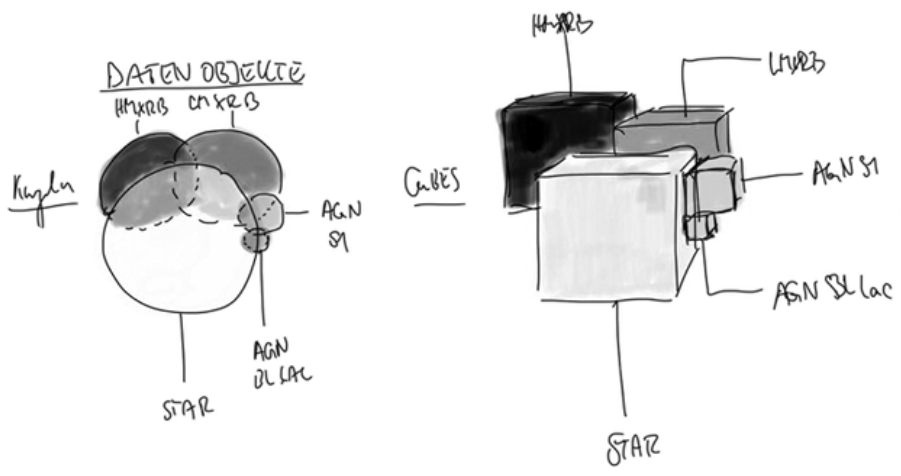

Figures 6.4 & 6.5 shows an attempt to translate this table into VR so that it can be attached to a model object if necessary. In fact, the transmission in VR was only partially successful. As already described in Chapter 3, it is easier to make relative size comparisons in VR than concrete numerical assignments. In this case, the use of matrices and tables is much more efficient in order to make statistical comparisons about the confusion. Ultimately, the confusion matrix says something about which classes were difficult to distinguish for the algorithm. As was shown later, this information could also be conveyed in a different way (see chapter 7.1).

One advantage of VR is that the positions and distributions of objects in space can be perceived much better than on two-dimensional screens (see chapter 3). Re-sorting can also reveal other dimensions of a data set (Cairo, 2013). In a further experiment, the data objects were therefore rearranged in the coordinate system. Now their highest, second and third highest probability values defined their position along the axes of the coordinate system (see, Fig.6.7). The sorting allowed the viewer to perceive the class distribution differently than from the object components and to formulate possible hypothesis. Through the clear assignment and reference of the axes, it quickly became clear that, ideally, all objects would lie on the axis of the primary class. The structure in the room shows the deviations from the ideal case. This meant that the arrangement of the objects according to classes not only made it easier to find significant classifications and outliers. She helped perceive the distribution of probabilities per source and across the data set.

Table 6.1: Example of a confusion matrix

| **A: Hardness ratios only** | | | | | | | | | **(OBB error rate 8.3%)** |
|---|---|---|---|---|---|---|---|---|---|
| | BL | CV | HMXRB | LMXRB | S1 | S2 | STAR | ULX | Class. Error |
| BL | 3380 | 34 | 0 | 2 | 62 | 17 | 9 | 32 | 0.04 |
| CV | 80 | 3055 | 23 | 52 | 110 | 189 | 26 | 29 | 0.14 |
| HMXRB | 0 | 4 | 3729 | 1 | 0 | 15 | 1 | 0 | $\leq 0.01$ |
| LMXRB | 1 | 25 | 8 | 3375 | 3 | 9 | 34 | 47 | 0.04 |
| S1 | 115 | 126 | 3 | 12 | 3204 | 264 | 18 | 31 | 0.15 |
| S2 | 28 | 204 | 34 | 52 | 297 | 3384 | 55 | 50 | 0.18 |
| STAR | 15 | 8 | 2 | 57 | 15 | 52 | 3009 | 68 | 0.08 |
| ULX | 19 | 8 | 0 | 33 | 7 | 17 | 34 | 3792 | 0.03 |

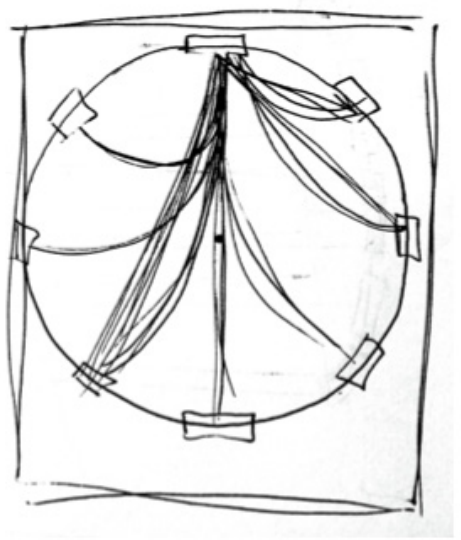

Fig. 6.4: A type of information graphic, which is often used to represent transactions, served as inspiration for the confusion matrix.

The final concept was derived from this knowledge. Instead of going from the detail to the overview, the visualization should first give the user an overview of the data distribution and only then go to the detailed view as an option. The scalability of the virtual space should be exploited. So by going further and further from the big one (the distribution of the classification), one should finally reach the DNA of the X-ray source (its spectral parameters) and therefore the reason why the algorithm assigned the source to a certain class. Figure 6.7 summarizes the final concept:
The starting point is the classroom, a room in which each class has its own volume. The number of classes can be arbitrary here. The presentation is modular and scalable. X-ray sources are positioned in the volume according to their classification. Their position results from the different probability values (cf. chapter 7.1). In the classroom, the user should get an overview of what the distribution of the X-ray sources looks like through the models used. One data point is created for each X-ray source for each model. Any number of models can be displayed at the same time. A uniform, parameterized assignment results in different positions for different classifications of the sources. This means that an X-ray source in this room is described by four data objects.

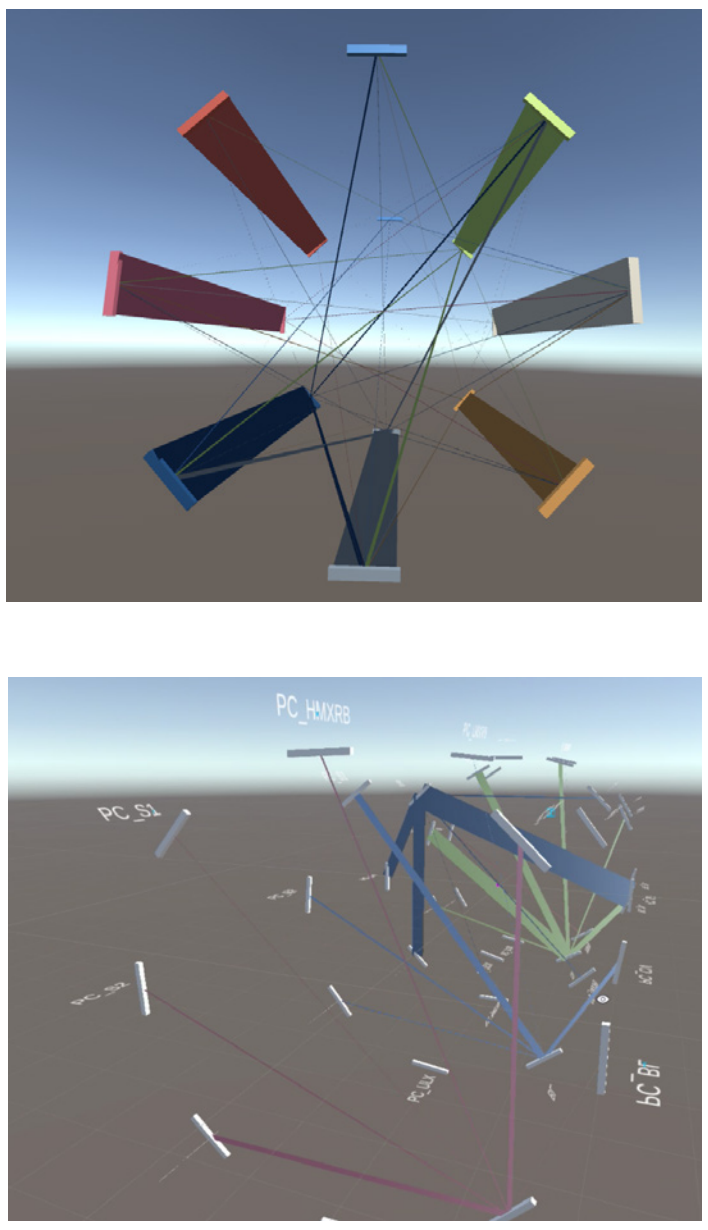

Fig. 6.5: Realisation of the confusion matrix in VR.

Since the classroom is about comparing models, the data objects are determined by the parameters of the underlying model. If you take a proverbial look at a model object, the user should gradually get more detailed information and finally be able to zoom into the parameter space. In this room, the user receives detailed information on the spectral properties of the selected X-ray source. According to the principle equivalent to the classroom, the spectral properties of the sources define volumes in which data objects are located. By connecting the individual data objects, traces develop in the parameter space that characterize a source. By comparing

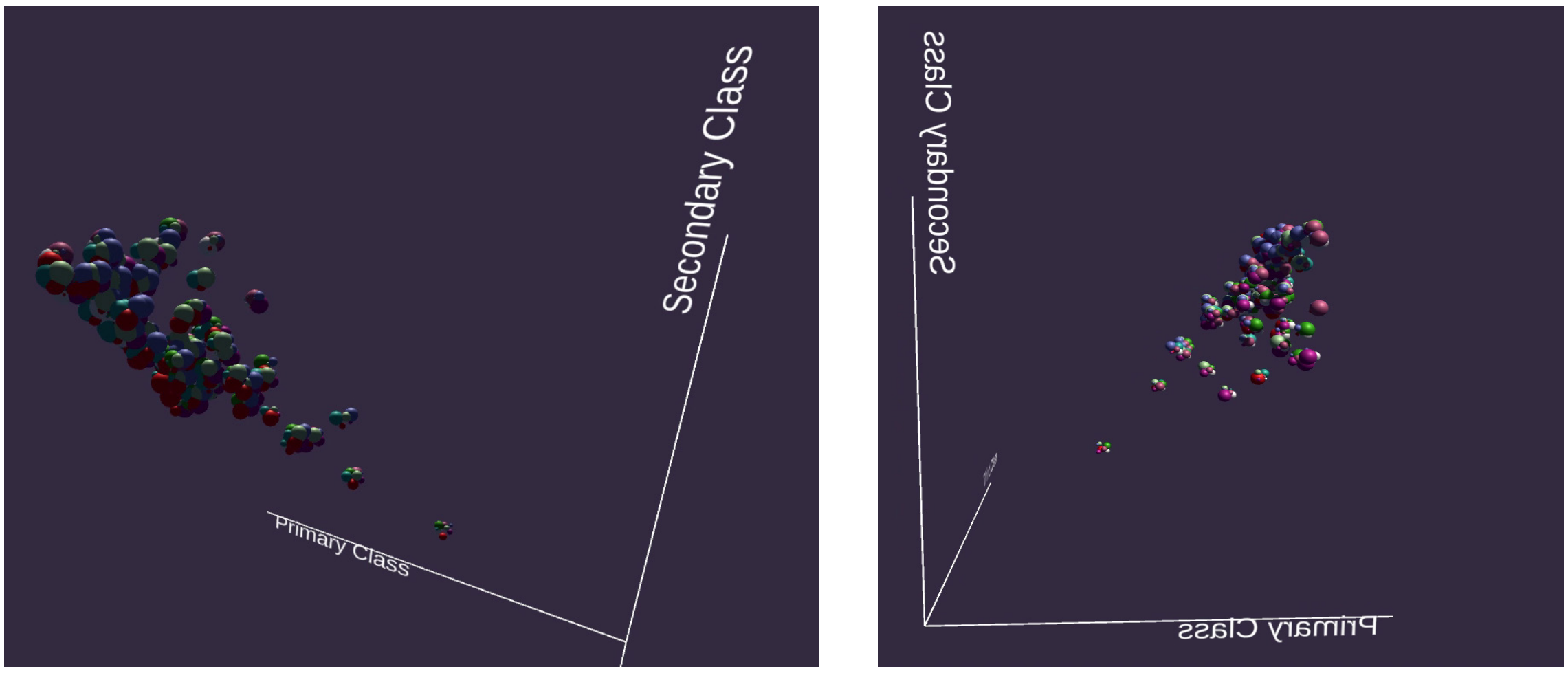

**Fig. 6.6:** Positioning of the data objects in the coordinate system, based on their probabilities.

**Fig. 6.7:** Final conceptual sketches

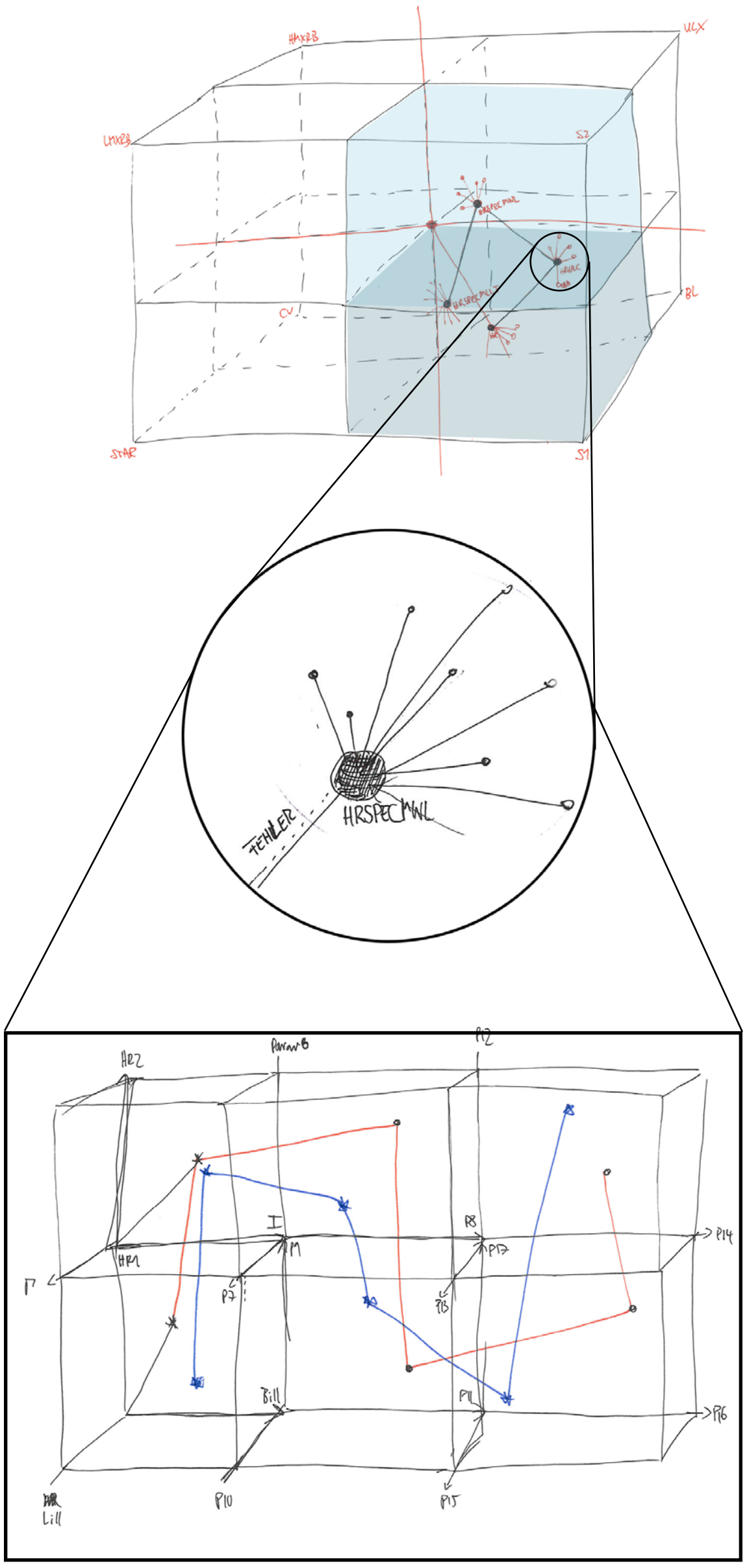

**Classroom:**
Parent room maps X-ray sources to the classes; shows overview of class distribution.

Objects in the classroom show model properties.

Transition between rooms via zoom-in / zoom-out

**Parameter space:**
Properties of an X-ray source are shown in modular coordinate spaces and draw a characterizing path.

the traces of different X-ray sources, the user should be able to understand and understand the parameter correlations according to which the intelligent algorithm classified the sources. Based on their experience, the user could then place this knowledge in a larger context and potentially find new correlations.

In the narrative of this work, the user corresponds to a researcher who can move between the macroscopic world of classes and the microscopic world of spectral traces of the X-ray sources of our universe.

# 7 Realisation

The implementation of the visualization or creation of the virtual space was done with the game engine Unity (version 2019.3.0b12). Unity allows the visual and functional design of the room to be controlled via scripts in the programming languages C # or Java. By designing with code, it is possible to define and parameterize all data objects uniformly in their appearance and position, so that they are completely determined by their own parameter values. The game engine was therefore already used for the development of the data visualization TTU.

A limitation of the program is that Unity can only render around 1000 – 1500 3D objects („GameObjects") at the same time. This corresponds to approximately ~ 3 per mille of the number of data points that are to be displayed in the classroom, for example. In order to still be able to display the large amount of data, it was exploited that a 3D object consists of a mesh, with vertices and 2D triangles drawn in between. The number of grid points increases depending on the complexity of an object and can amount to several thousand. The trick was not to define the data points as individual objects, but as a vertex in a single grid structure. Each model then only corresponds to a single 3D object, the grid points of which represent the classifications of the X-ray sources. The assignment meant that the data points as a whole could not be represented as 3D objects. Instead of the 2D triangles, a 2D circle was therefore drawn on each vertex. Fig. 7.1 on the left shows a data cloud with 150,000 randomly generated data points (orange). The rendering of 3D objects was then regulated by a query asking which data points are close to the user. In the figure, the green dots represent a thousand vertices that are in a fixed radius around the user, whose view is shown in Fig.6.1 on the right. The user can move freely through the clouds and his 3D environment is generated in real time. The higher-level structure is only visible through the data points. The improved perception of dimensions in VR helps to recognize the 3D structure of the data clouds, although only 2D points were used. To distinguish the data points of several models, each model was marked with its own color. Since the models build on each other, the color palette is chosen from a gradient to signal the development of the models.

## 7.1 The Classroom

### 7.1.1 The mapping of probabilities

In the classroom, X-ray sources are mapped into the volumes of the different classes. Various forms (cf., Fig. 7.1) were tried out. Cubes have the advantage that their edges provide direct orientation for the user. However, the choice of a spherical shape resulted directly from the calculation of the positions of the data points. The underlying design principle is the order according to comparability. Objects with similar probabilities should be collected. Since a position in a sphere is clearly defined by its distance to the center point and a direction (see Fig. 7.2), the position of all data points resulted from the determination of these two parameters. Fig. 7.3 shows the principle of calculation:
In the center of the class volume, those sources should be located that were most likely for this class. That is, the higher the value, the closer the sources are to the center (i.e. the smaller the radius). How is the direction in which the data point is drawn determined? Each source not only has a value for the most likely class membership, but one for each possible class. One could therefore assume that sources whose order of class affiliation is the same or similar also have comparable spectral properties and that this results in a correlation. The order of the data points should therefore be defined in such a way that the order of the classes is recognizable or similar objects are expressed by proximity. This means that the probability-weighted summation of the directions to the volume of the 2nd and 3rd most probable classes gave the position of the data point for an X-ray source within the volume of the most probable classes. Figure 7.4 shows how patterns develop naturally from the calculation. A change in the arrangement of the balls changes, due to the calculation logic, the corresponding distribution of the data points within a ball. The user has the possibility to examine the relationships between the classes. The color coding of the models makes it easy to differentiate between them and provides a picture of how the classifications change between the models.

### 7.1.2 Hierarchies of information

The data points in the room reveal the most important information from the classification of the X-ray sources: the class assignment and the underlying model. The user then has the option of moving through the room and specifically calling up more detailed information about individual X-ray sources. It was decided against the automatic rendering of all objects in the user's environment. The reason is that with the very high number of data points, firstly the frame rate is slowed down and secondly due to the sometimes very high density of the points only a small area is really covered, so that the rendered objects did not reveal any significant structure.

Instead, the data objects to be rendered are selected using the VR controller (see Section 7.3). The viewer can gradually reveal more and more information from a source. By approaching the controller, the one data point is activated and the model object is rendered in the associated color. The model object also has small signposts that indicate the direction to 1st, 2nd and 3rd class, in the order of the highest probability, and their numerical values. This makes it easier to understand how the position in the room comes about. At the same time, all other model objects from the same source are rendered and connected by a line. The connection visualizes the path that a single X-ray source takes through the classroom when it is classified by different models. Ideally, the entire path is within a class sphere, because then its properties can be clearly assigned to this class. The relationships between the model results should be visible through the connecting lines. Indirectly, the exploration of the classroom could lead to similar conclusions as the confusion matrix.

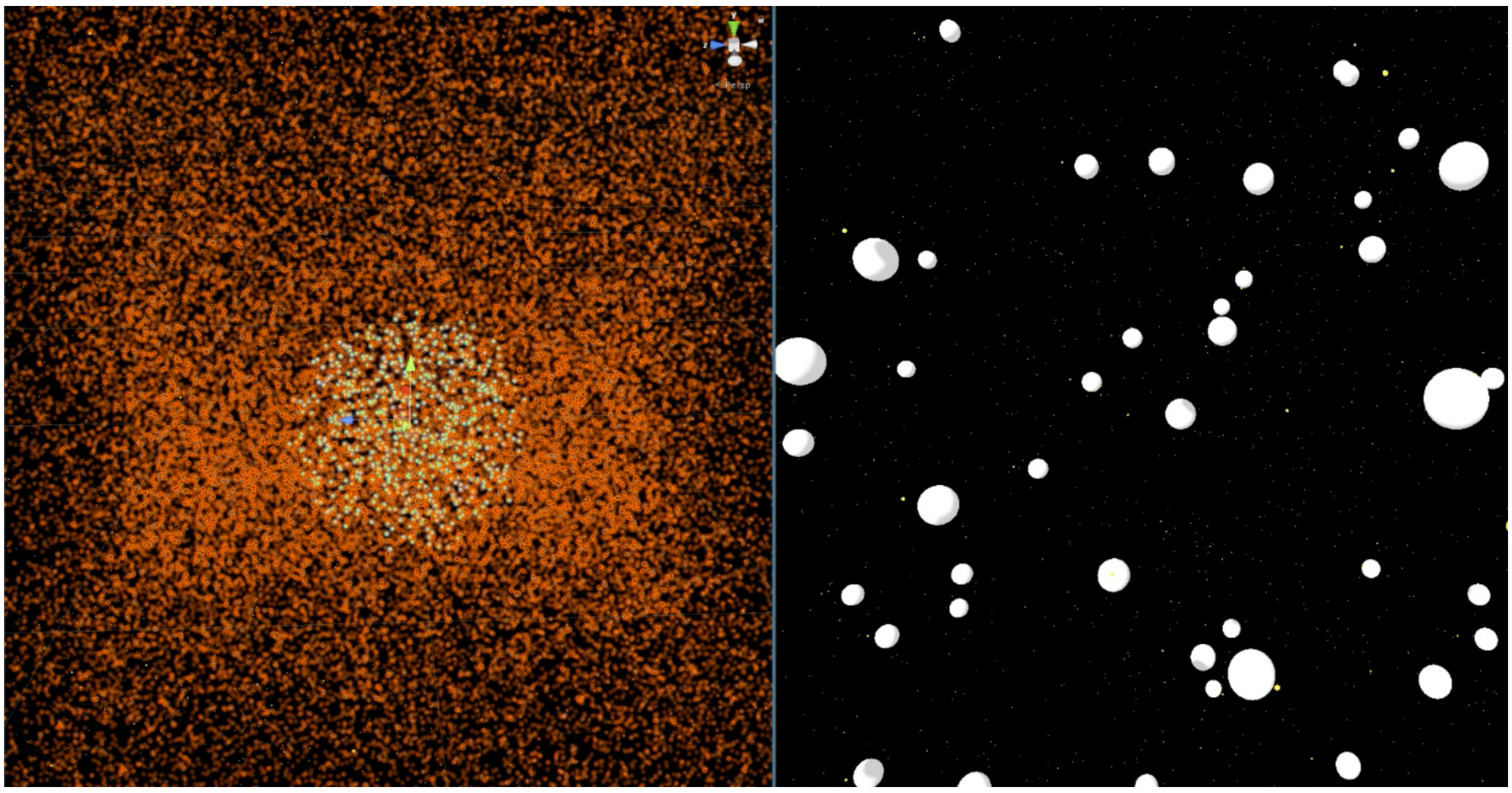

**Fig. 7.1:** Left: Representation of 150,000 objects (orange) as grid points, of which only 1000 objects (green) have been rendered in the vicinity of the user. Right: The user's view shows nearby rendered objects and distant grid points.

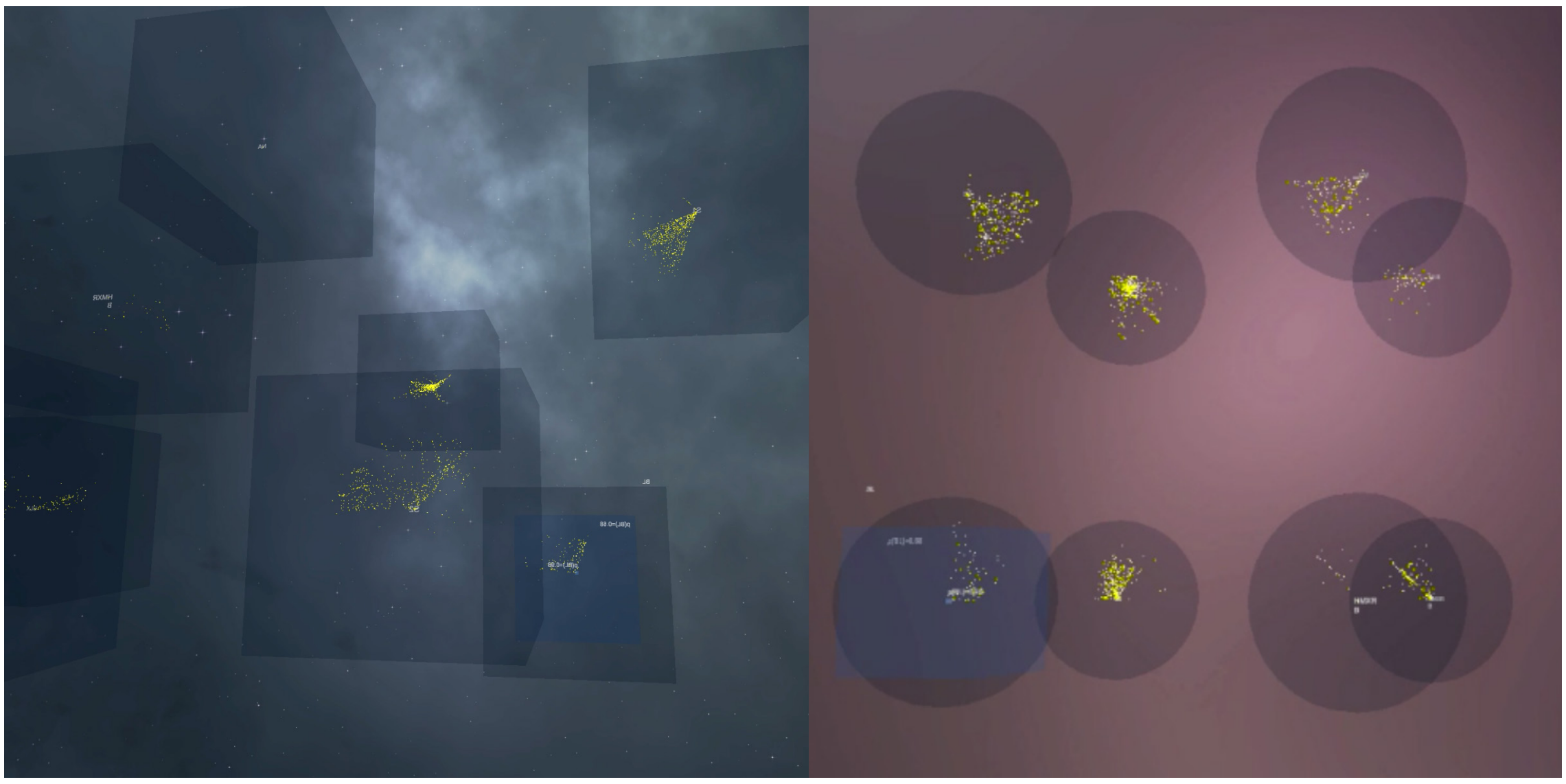

**Fig. 7.2:** Design of class volumes

For an activated data point, the viewer can delve deeper into the details of the data point by clicking on the object. Then detailed information about the model that created the data point appears. The most important distinguishing features between the models are the parameter lists and their error rate. In the ranking lists (see Fig. 7.6), the algorithm sorts the spectral parameters used in the order of their relevance for the classification. It is less the exact than the relative sizes of the ranking that are relevant. To represent this, a pie chart was created that shows the parmeter as pieces of cake (cf., Fig. 7.5). The higher the rank of a parameter, the larger its shape. This order helps to quickly identify dominant and important parameters. The shape of the pie chart was chosen because it is similar to the bar histogram, which is one of the most common statistical charts in astronomy. This makes it intuitively understandable and the information quickly visible.

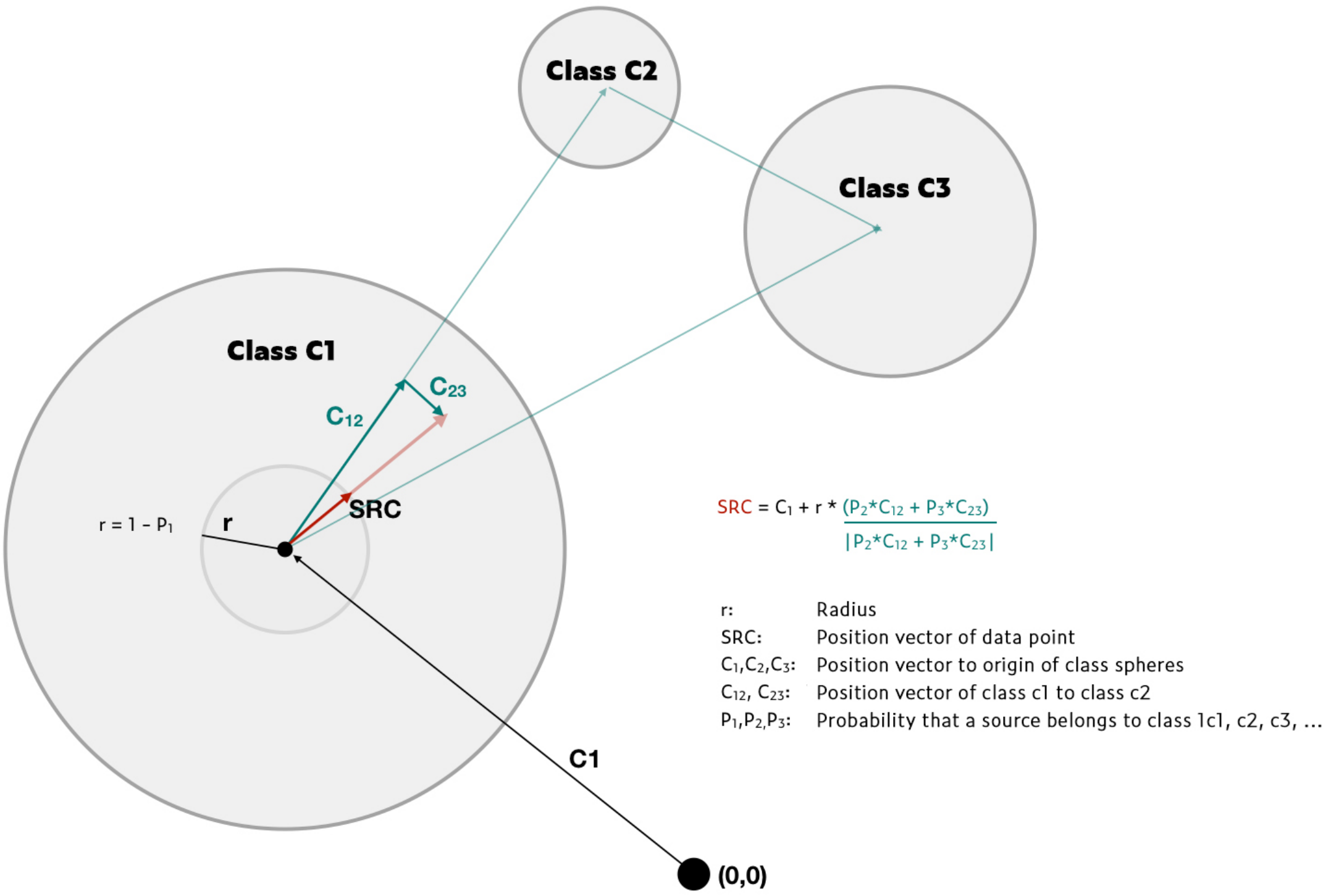

**Fig. 7.3:** Calculation of the position of data points in the classroom.

The third level of information is the spectral traces of the object. A renewed interaction with the activated model object leads the viewer into the parameter space. The transition must be fluid and intuitive, so that the two worlds are perceived as connected. This was implemented by a simultaneous, fluid enlargement of the classes and parameter space. The classroom grows until at some point the parameter space has grown to a perceptible size. Like zooming into a microscopic world. Conversely, when you leave the parameter space, both areas shrink and you are in the same place in the classroom as before.

## 7.2 The Parameter Space

### 7.2.1 Modular parameter cubes

In a first approach, the parameter space was constructed in a similar modular way to the classroom. According to this concept, the 3D coordinate systems common in astronomy spanned their own volumes in space, which were filled with the data points. The position of the data points resulted from the values of the parameters that were assigned to the axes of a coordinate cube. To link the position with numerical values here, the cubes showed orientation grids, which also showed the value ranges. A point in the parameter space was rendered by a colored ball and a line to the origin. The line again improved the perception of the relative position of the data point in space. The parameter cubes were initially regenerated each time the parameter space was called. This happened because the assignment of the spectral parameters to the axes of the cubes depended on the ranking of the selected model. However, in order to be able to compare several models with one another, the assignment was finally determined. Parameter groups of a model were in the same cube if possible. However, the data objects were connected with one line per model (see, Fig. 7.7 & 7.8). The order corresponded to the parameter ranking list of the model. This should make the trace of an X-ray source visible for a given model.

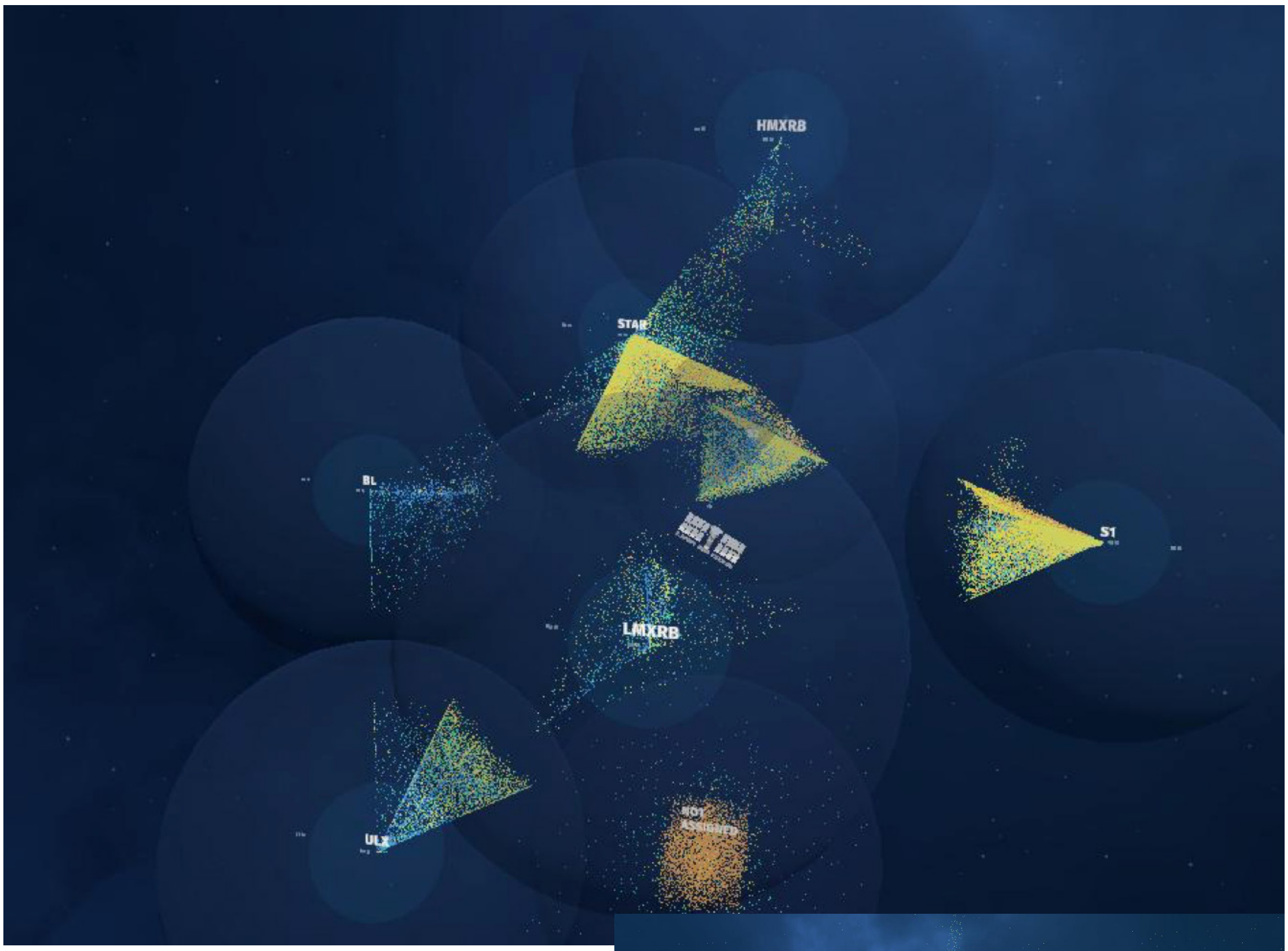

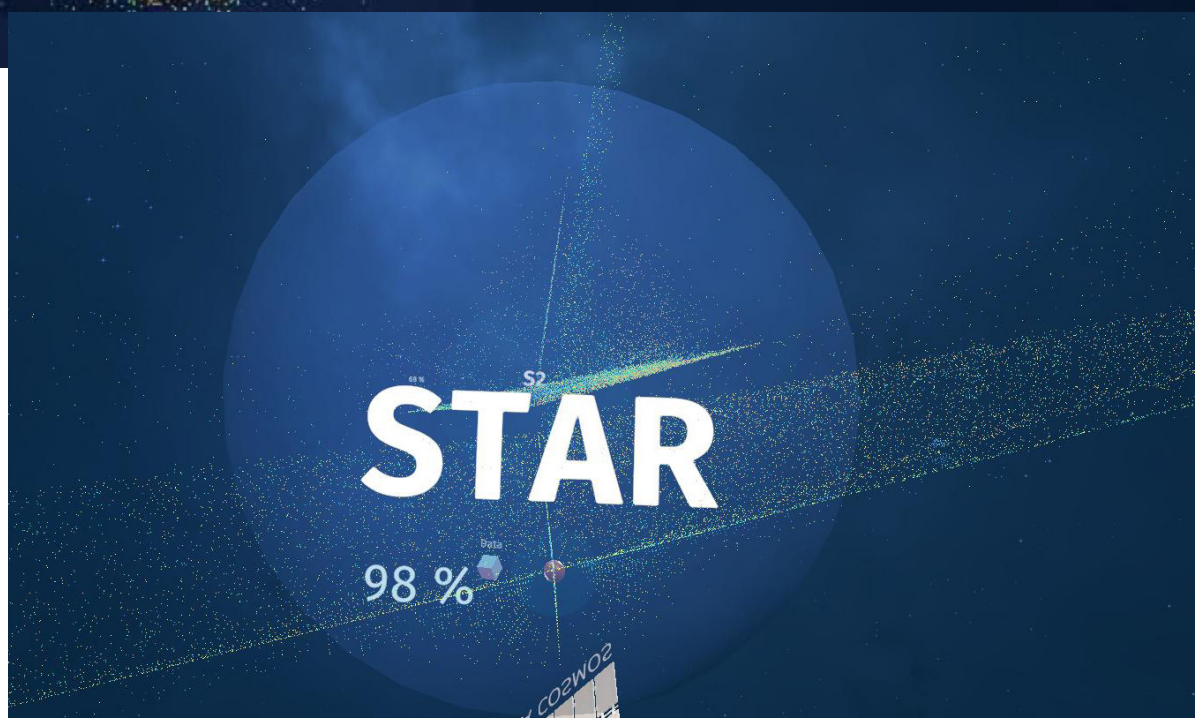

**Fig. 7.4:** Formation of different patterns from data point clouds through the arrangement of the class spheres.

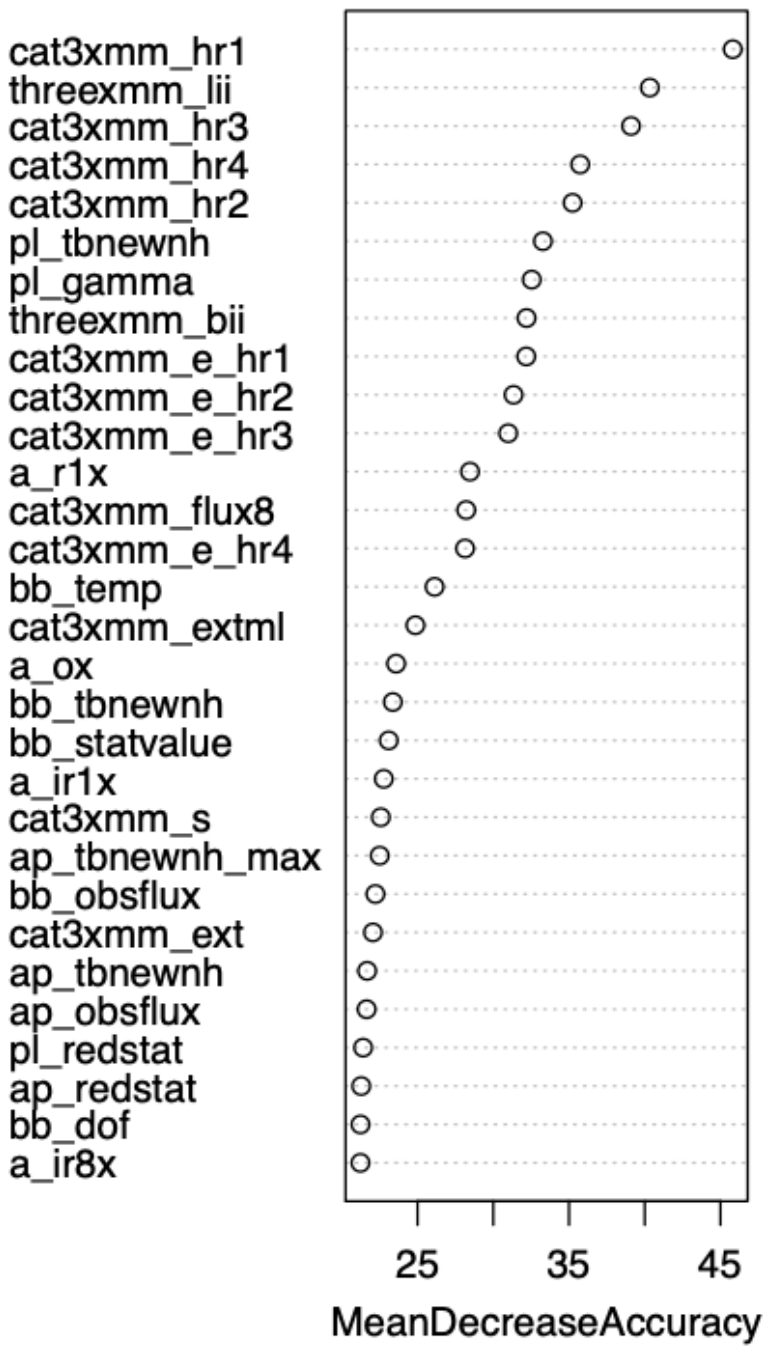

**Fig. 7.6:** Example of a conventional representation of parameter ranking of a classification model (left). The ranking was translated into a pie chart (right).

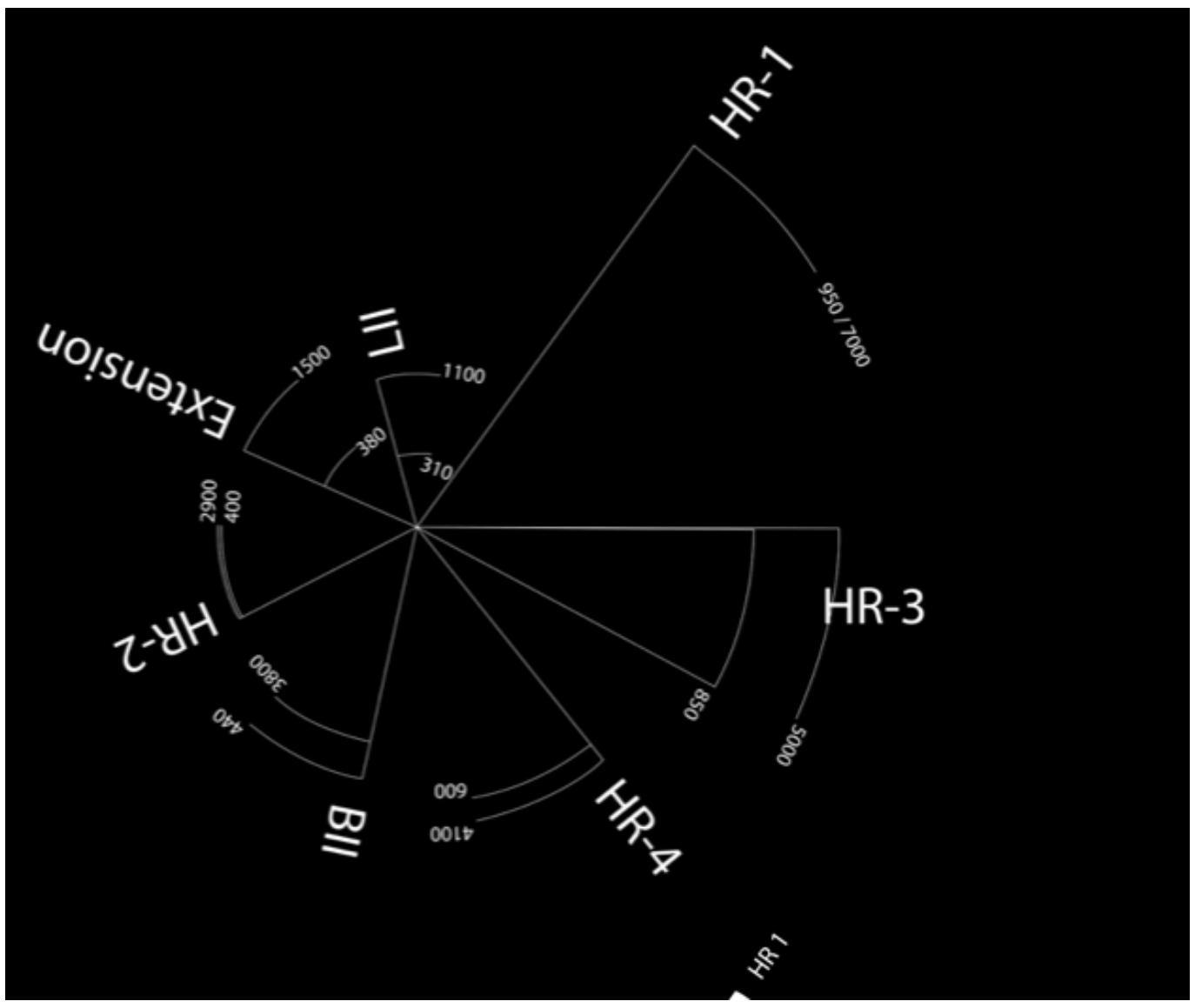

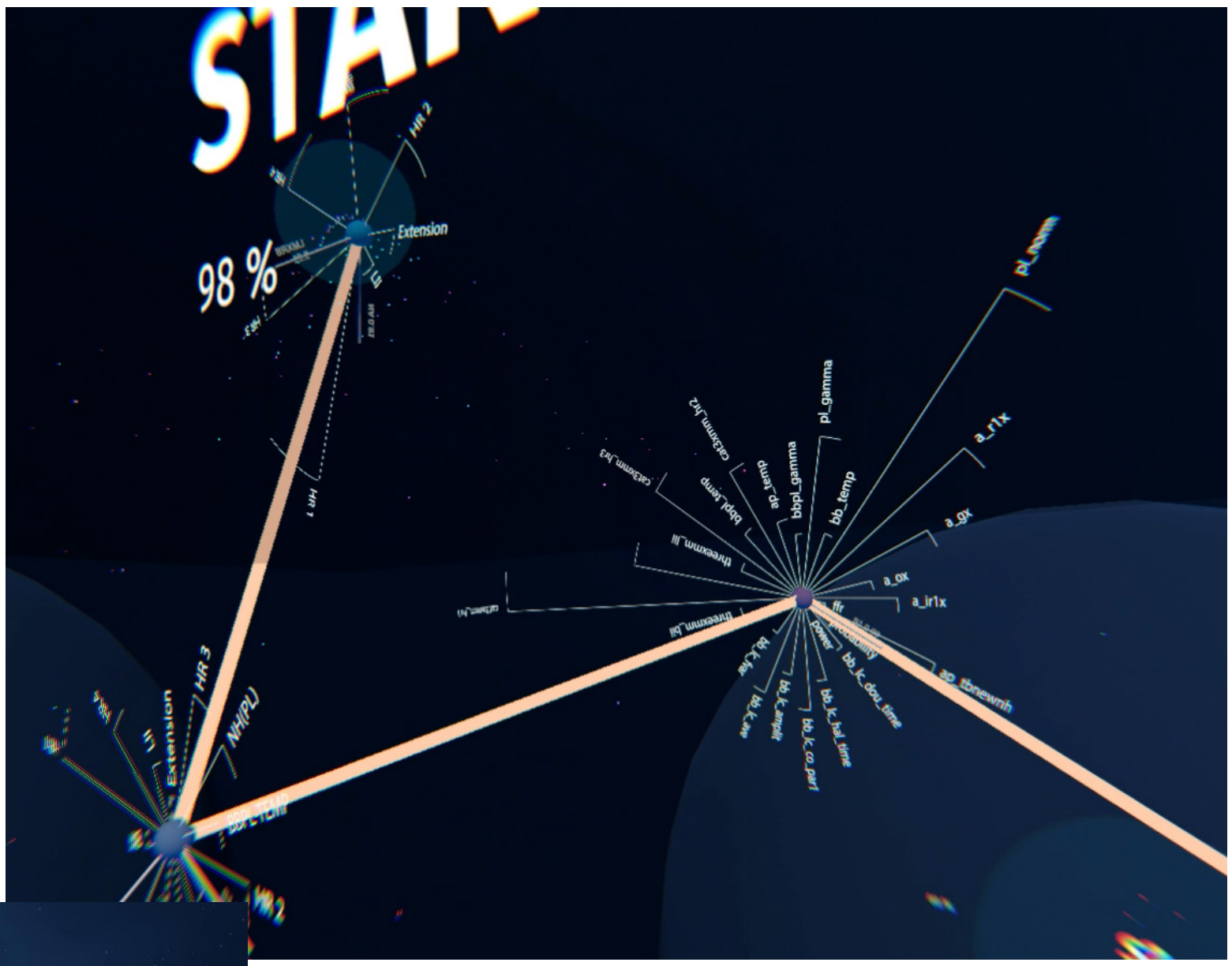

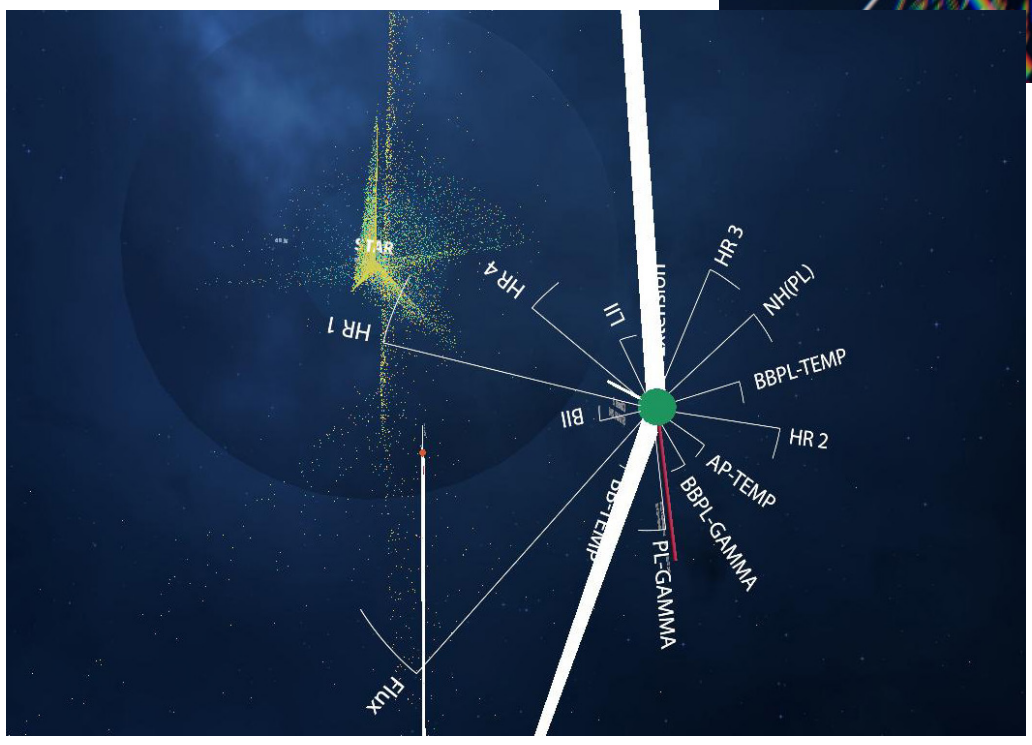

**Fig. 7.5:** Detailed model information for model objects: the pie charts shiws the parameters used by the classification model, as well as the parameter ranking that results from the training process.

For a small number of objects, the representation helped to recognize differences in their parameter values. However, for many objects that were to be compared, the lines finally became confusing and instead of supporting pattern recognition and perception of correlations, they made them more difficult (cf., Fig. 7.9). One of the reasons was that many lines crossed due to the different rankings and blocked the view.

Due to the large number of data points, which may be distributed over a small range of values, such representations inevitably overlap. As a viewer, you literally had the feeling that you wanted to pull the point cloud apart. The conventional concept of parameter cubes was therefore questioned for the representation of many data points.

**7.2.2 Parameter objects**
In a second approach, it was tested how the changes in a lattice shape due to the spectral parameters could serve as a representation. The parameters were assigned to the grid points of a generic object and shifted outwards or inwards relative to their difference from the respective parameter mean in the data record. Some of the resulting examples are shown in Fig.7.10. However, the disadvantage of this representation was that many objects were difficult to perceive or compare again. This approach could therefore not be used for the visualization of all data.

**Fig. 7.7:** Spectral trace in the parameter space for an object and different models.

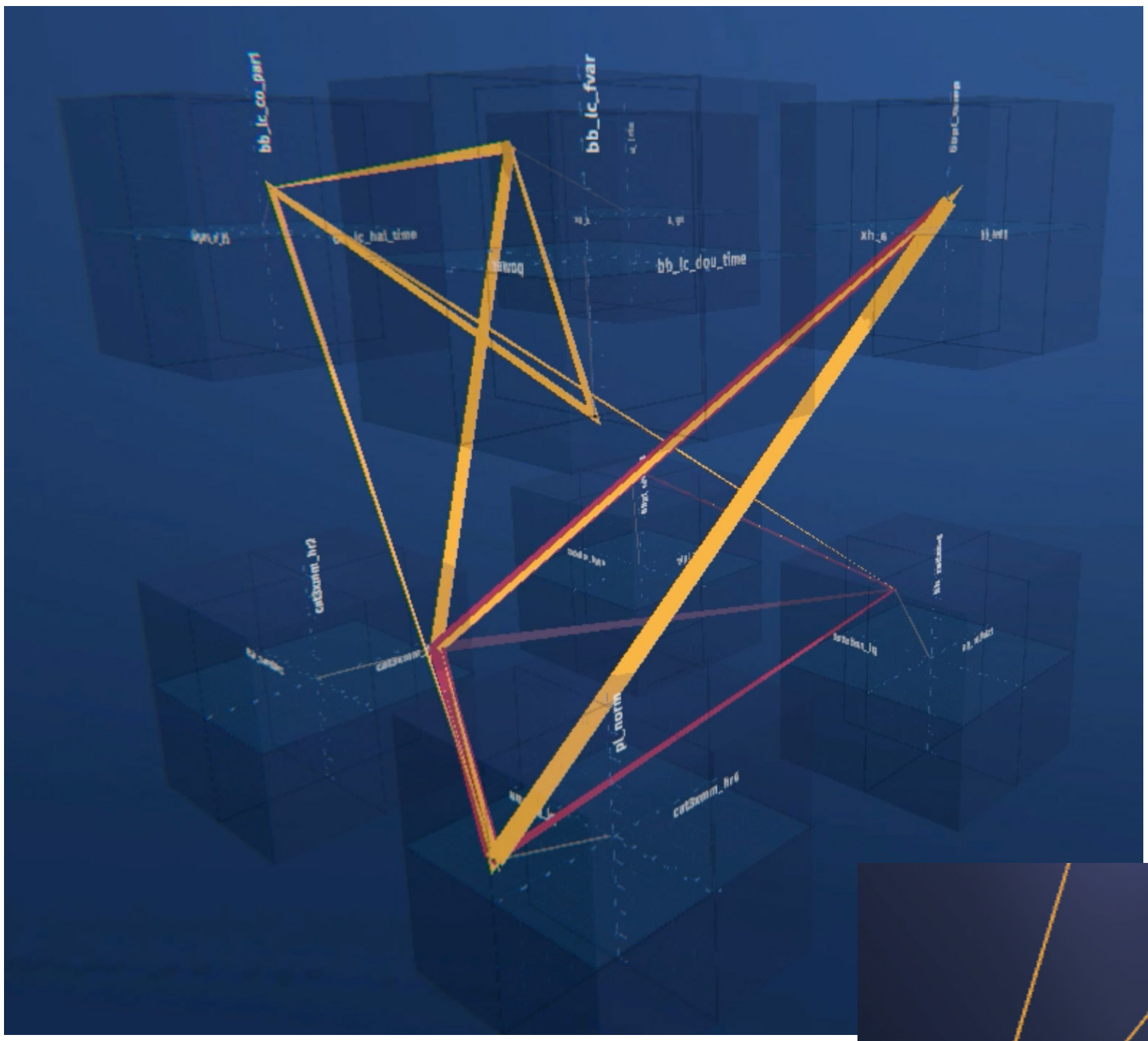

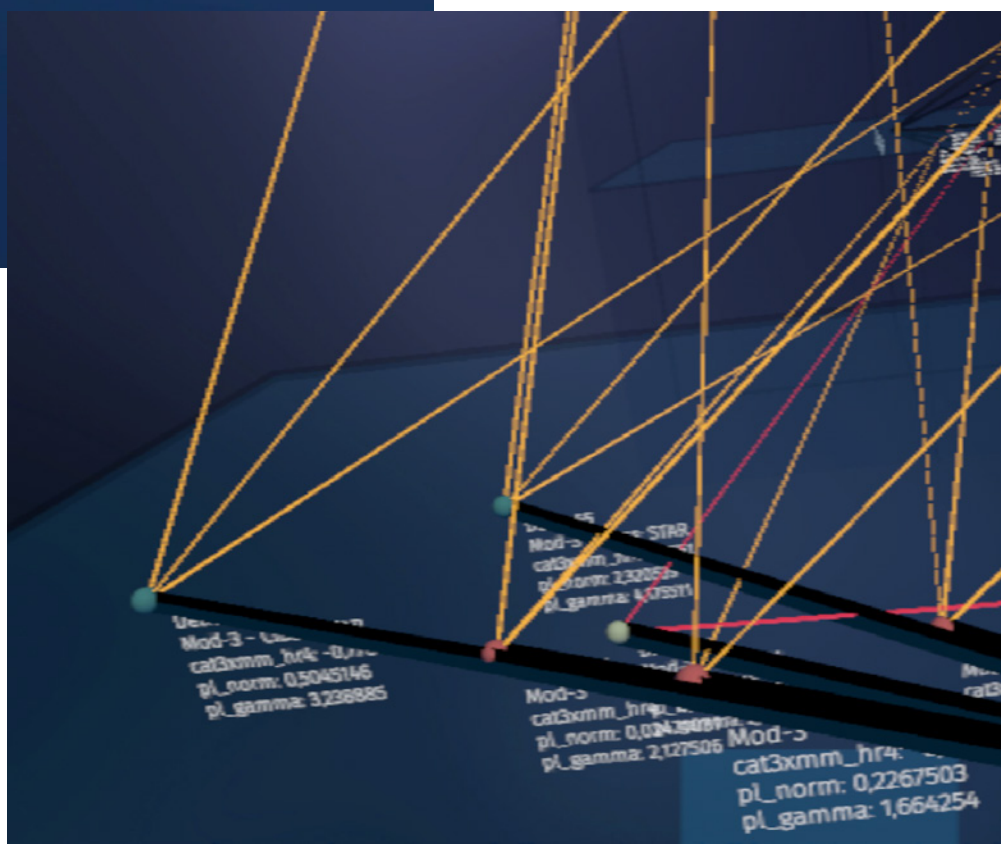

**Fig. 7.9:** Problematic line and information representation of spectral traces for several objects and models.

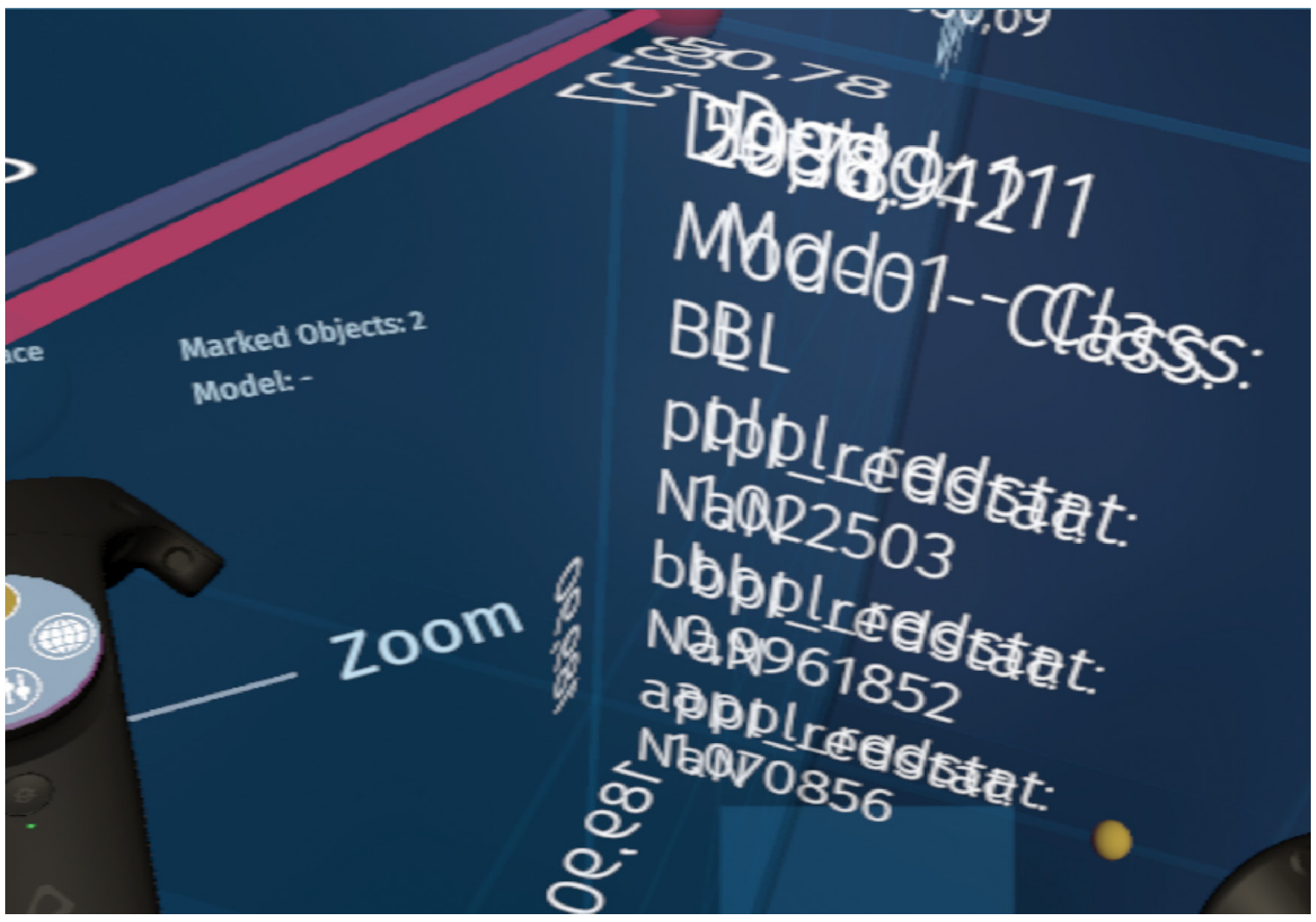

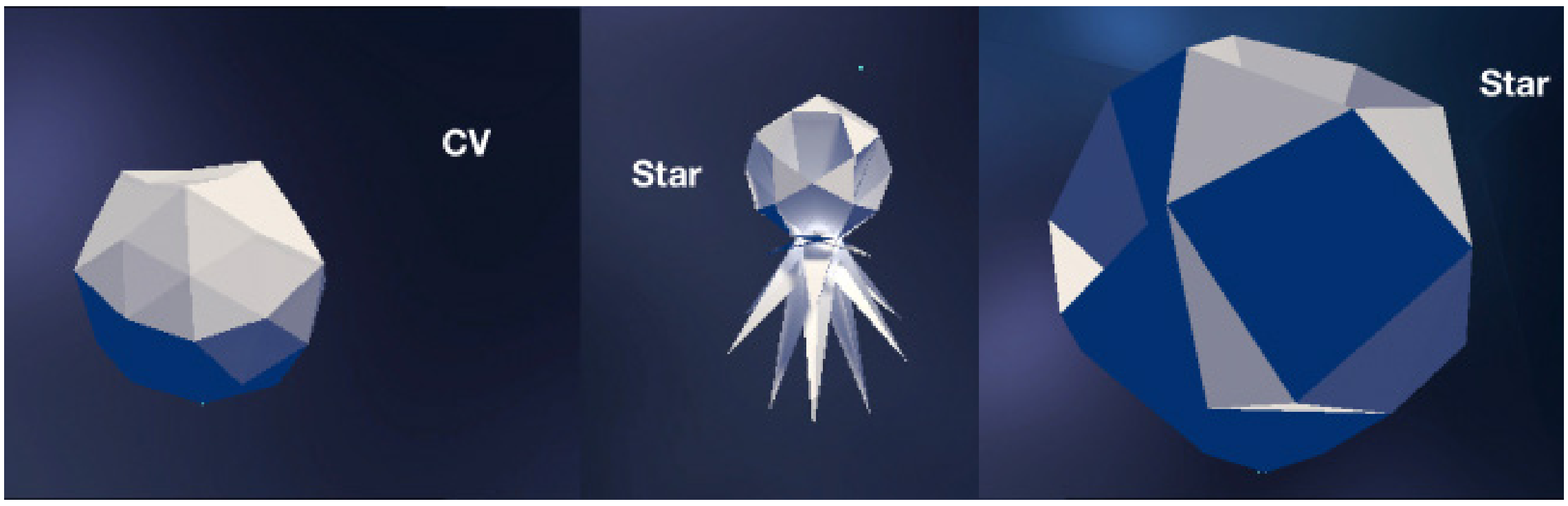

**Fig. 7.10:** Objects in the parameter space that have been deformed by spectral parameters.

### 7.2.3 Parameter cycles

The order of the parameters used by the model was used in this approach to connect the data points. However, the desire to pull the data points apart leads to a different approach. Can the ranking also be used to position the spectral data points?
With this approach in mind, different sketches were created, inspired by classic information graphics that map a lot of data to 2D. Finally, a calculation logic was derived which, like the logic in the classroom, is only determined by a distance and a direction. The difference is that all parameters of a ranking list are displayed one after the other and the range of values (i.e., the minimum and maximum values for a parameter) determines the direction. The logic is sketched in Fig.7.11. In retrospect, its basic features correspond to the visual method of epicyclic theory (see Chapter 2).

Starting from the parameter ranking lists, vectors are defined using circular coordinates. Circles or semicircles are drawn for each parameter according to the order in this list. The full scope (or half scope for semicircles) corresponds to the value range of the parameter in the data record. The parameter value of the source is now placed as a parameter object on the circumference of the circle and serves as the center for the next parameter circle. For the positioning on the circle, circle vectors are calculated for each parameter: The rank of a parameter in a model (i.e. the relevance for the classification) determines the radius of the circle. This also decreases with decreasing relevance. This means that the influence on the placement is reduced. The direction results from the parameter value, relative to the minimum and maximum value from the data set. Together, the parameter vectors result in a position in the (for example) XZ plane on which the user is moving.
One problem is the lack of values (NAN values) in the data records. To identify them, the parameter point is not drawn on the circumference of the circle, but at a fixed height above the origin of the respective circle. The data graphics are brought from the level into the room. In addition, the size of the final data point of the data cycle describes the probability with which the source is assigned to a class. To visualize the class, the data sets are color-coded differently. Figure 7.12 shows the first verison of three different data cycles for a) two objects of the same class, b) two objects that are very likely to belong to different classes, and c) two objects that are less likely to belong to two different classes. The effects of NAN values are shown in d). The final visual data cycles are shown in Fig. 7.13.

**Fig. 7.11:** Definition of parameter cycles

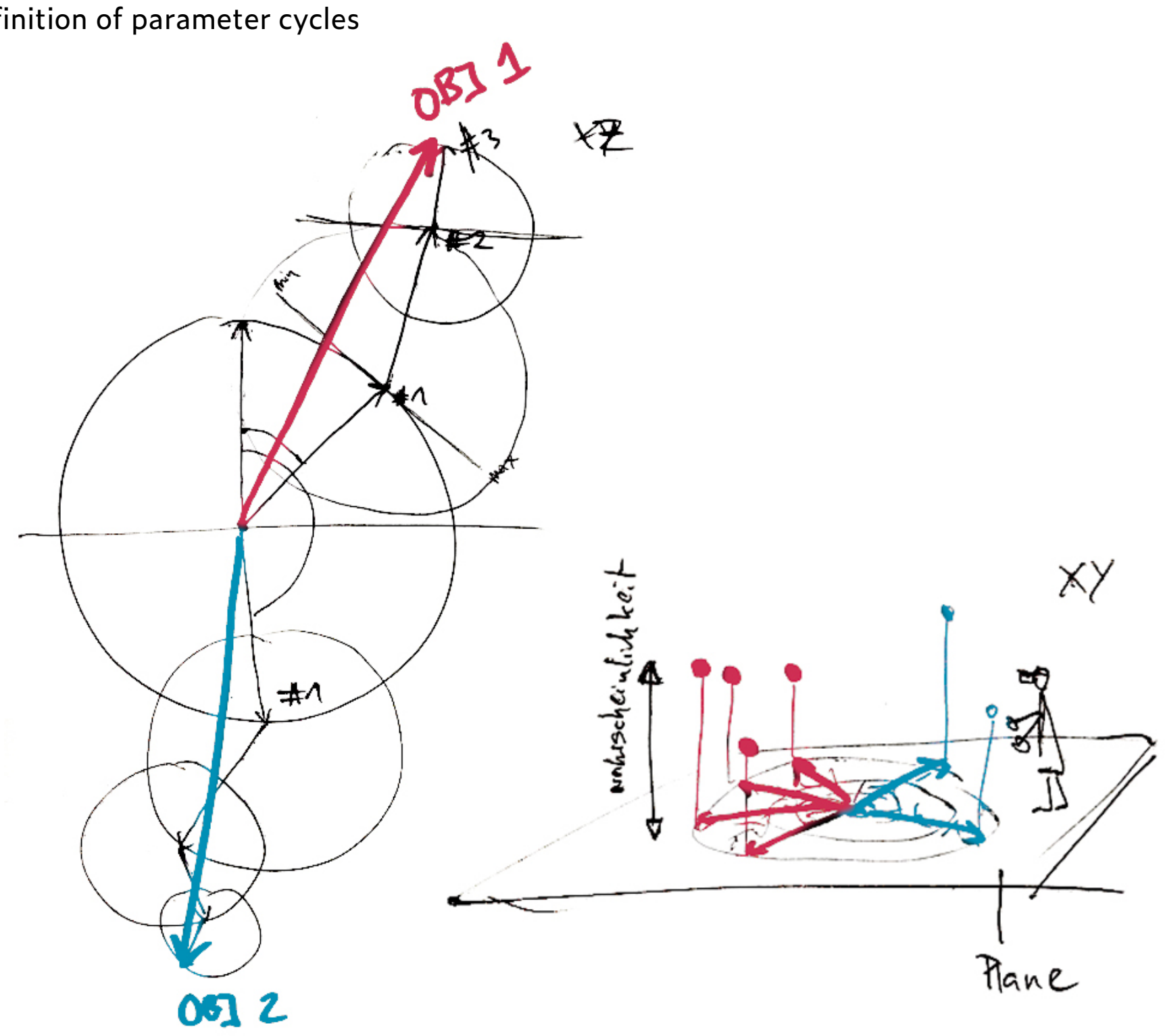

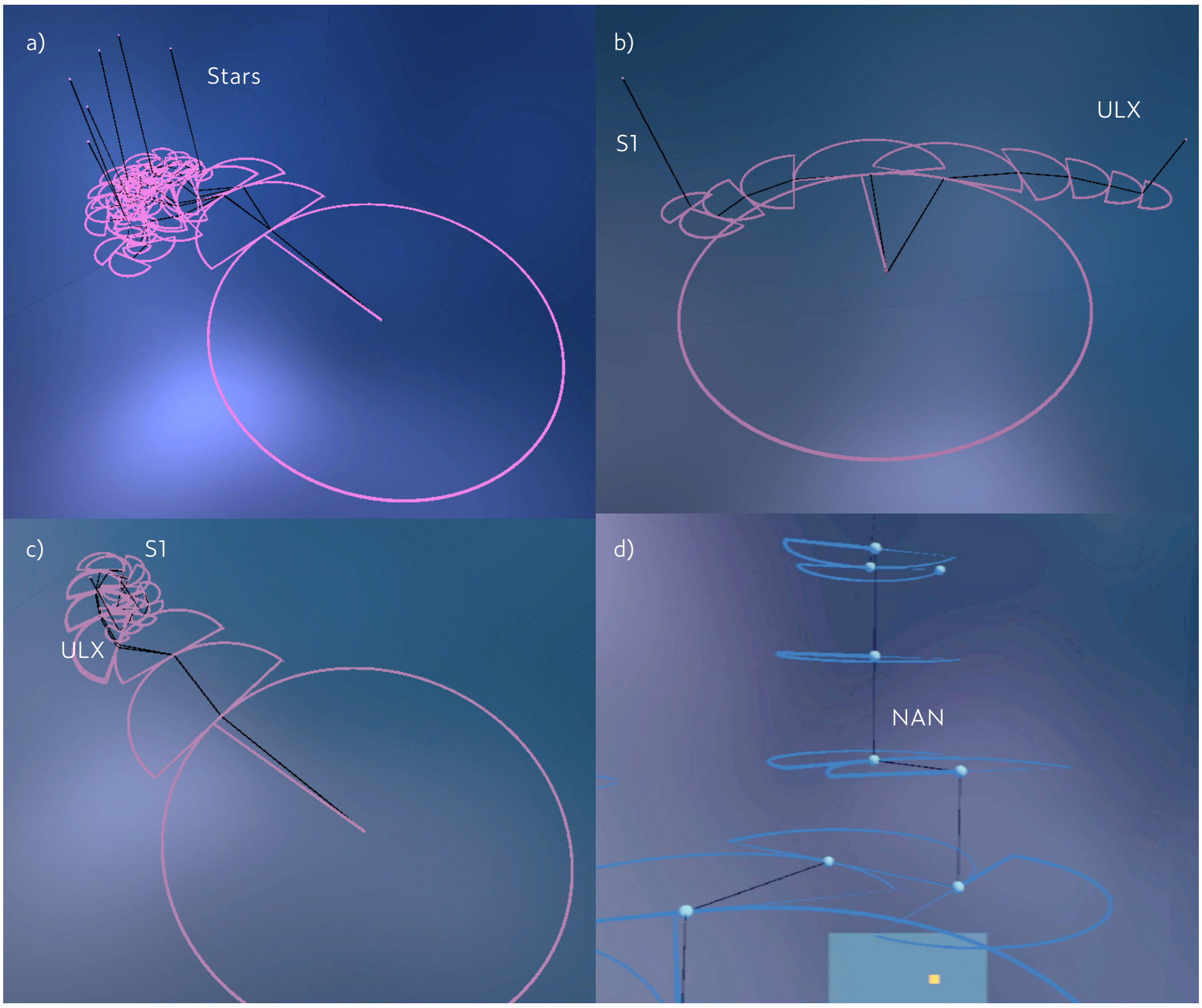

**Fig. 7.12:** Example of spectral traces using the parameter cycle a) two objects of the same class, b) two objects that are very likely to belong to different classes, c) two objects that are less likely to belong to two different classes, d) effects of NAN values.

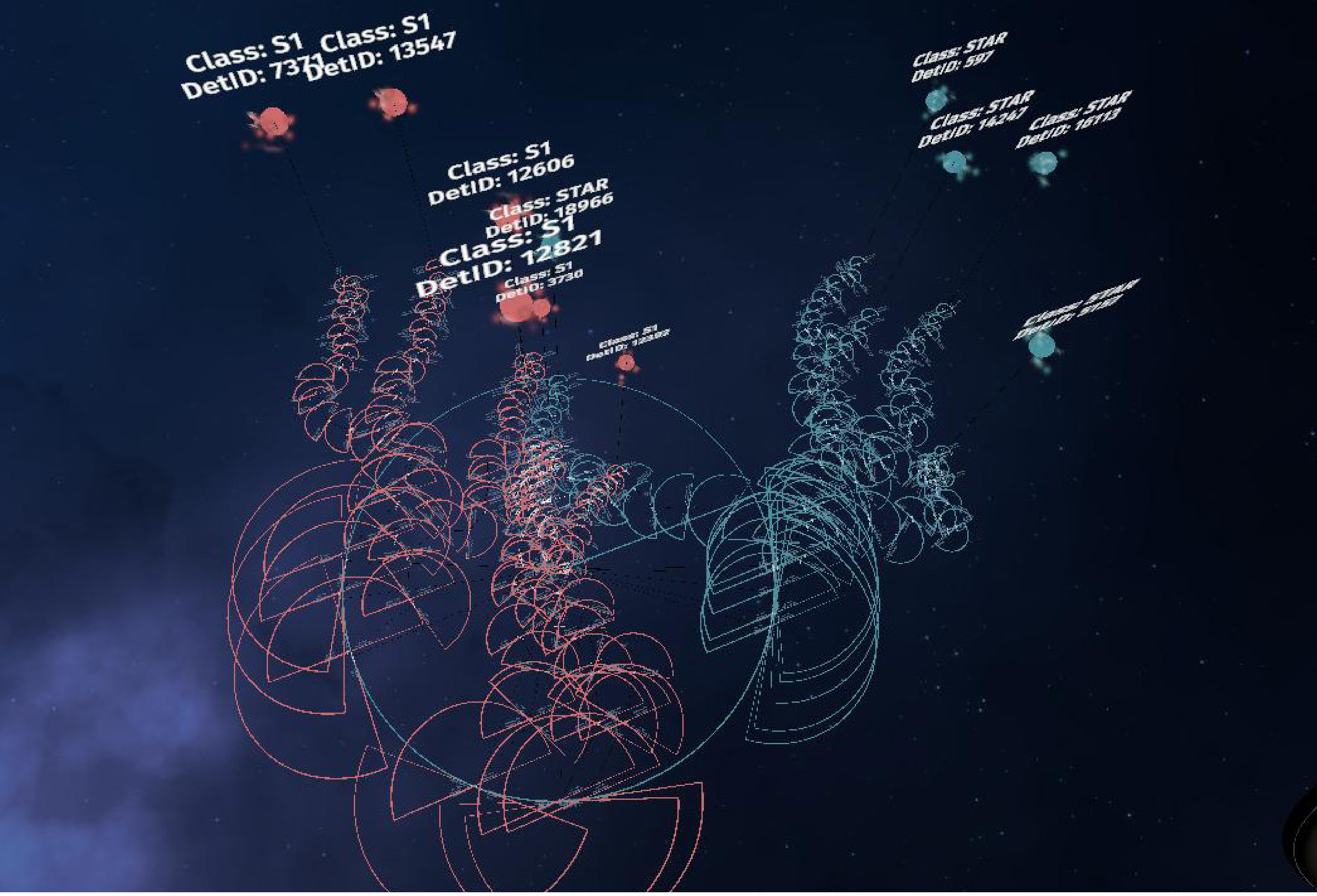

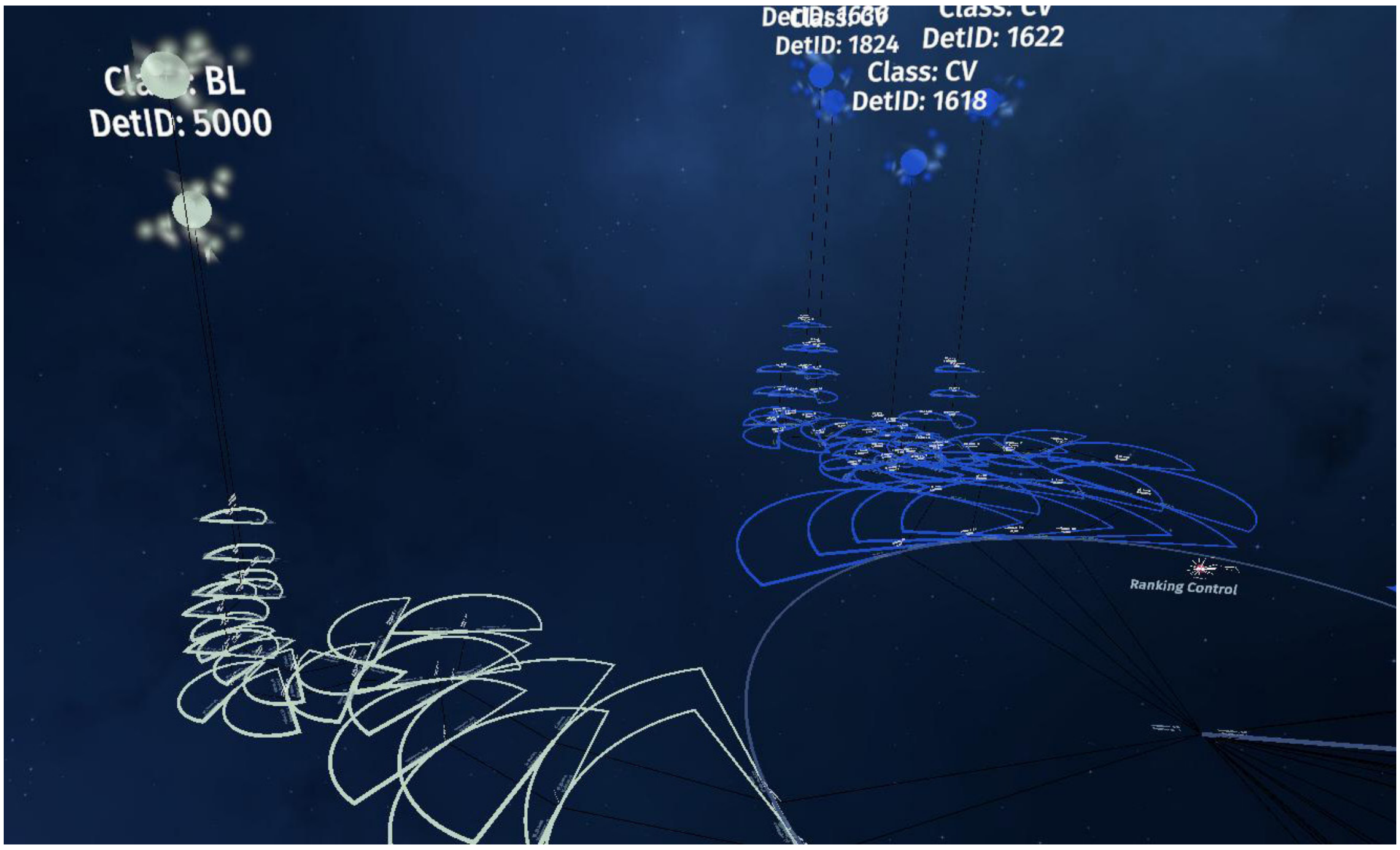

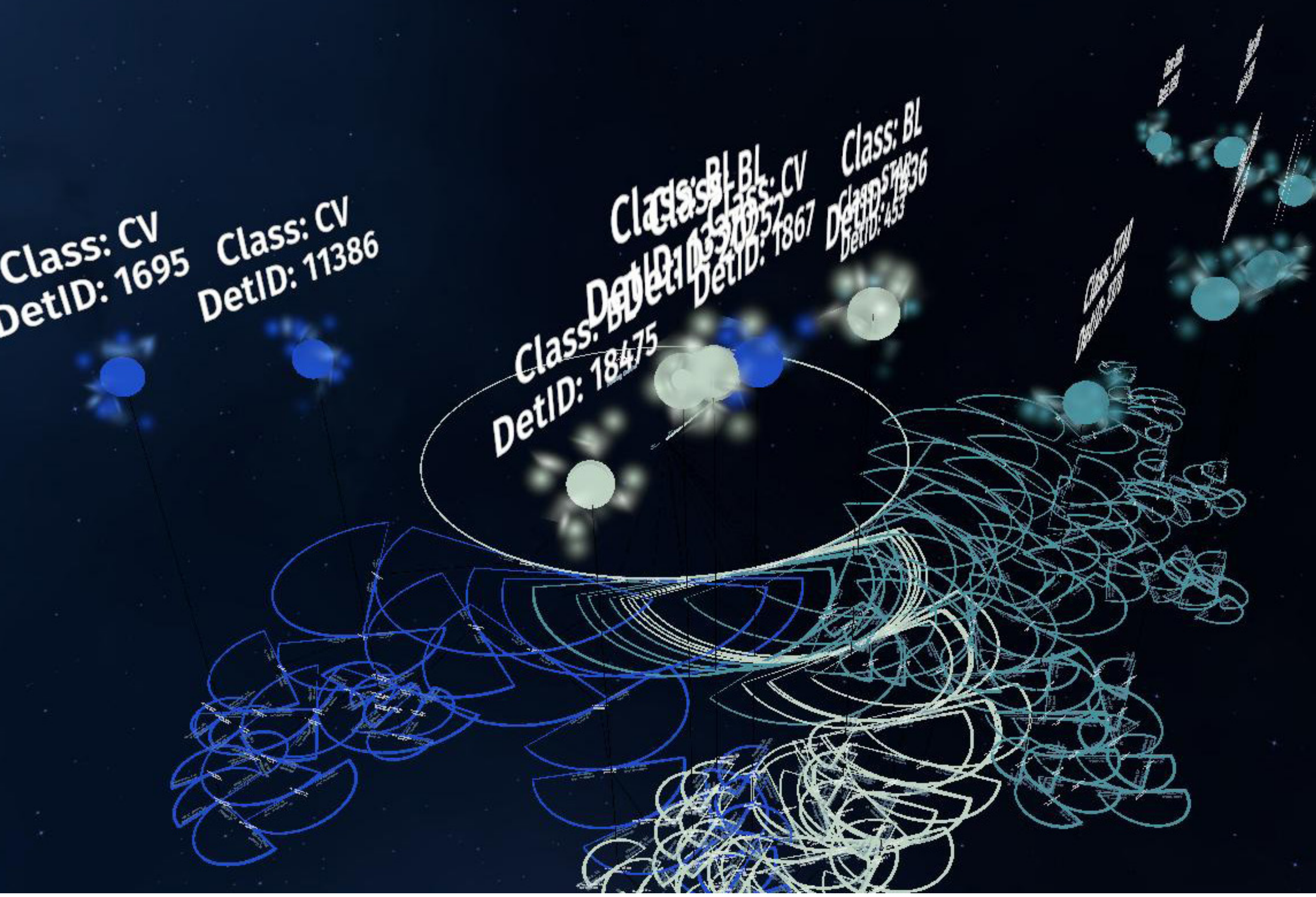

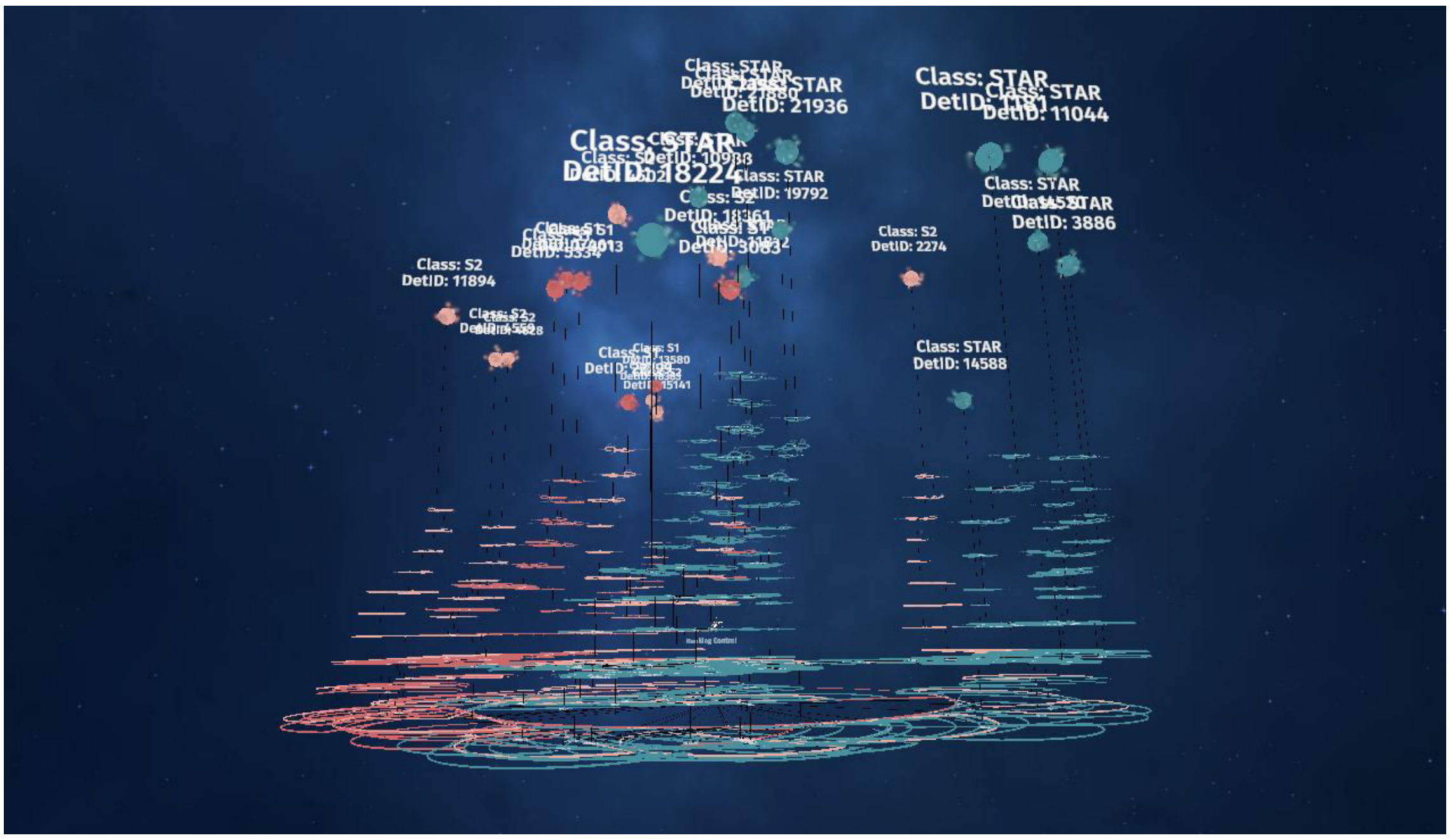

**Fig. 7.13:** Various examples of source traces within the parameter space. Different colors resemble different source classes. Each source describes a serious of epicylces, each of which visualize one parameter. The order of epicycles is given by the ranking for the selected model, respectively. The radius of each cycle is defined by the respective ranking of the parameter while its orientation is defined by the parameter values of an individual source. In case a parameter value is not available (NAN value), the cycles perform a step in the third dimension. Sources with similar properties will end up in the same area in space. Problematic sources with many NAN values are imediately detectable. Despite the final position of each source, the shape of the epicycles is also characteristic for similar sources.

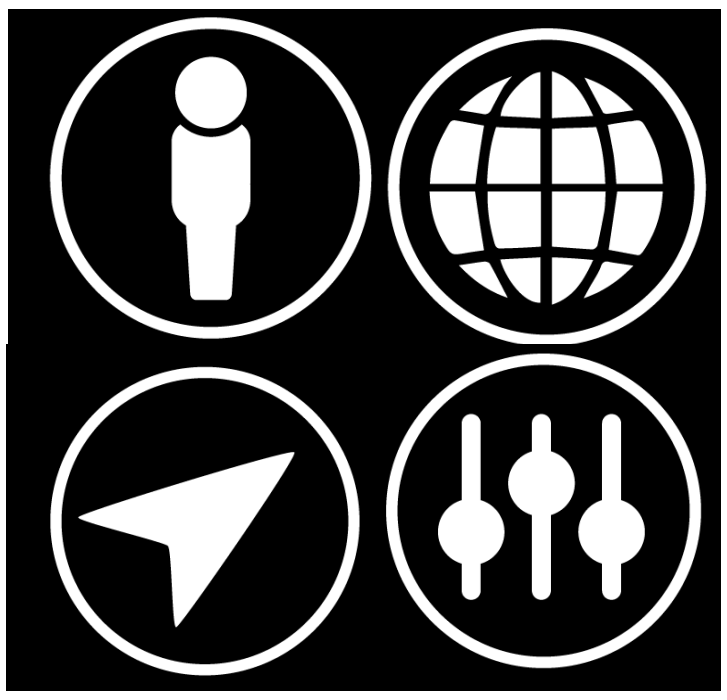

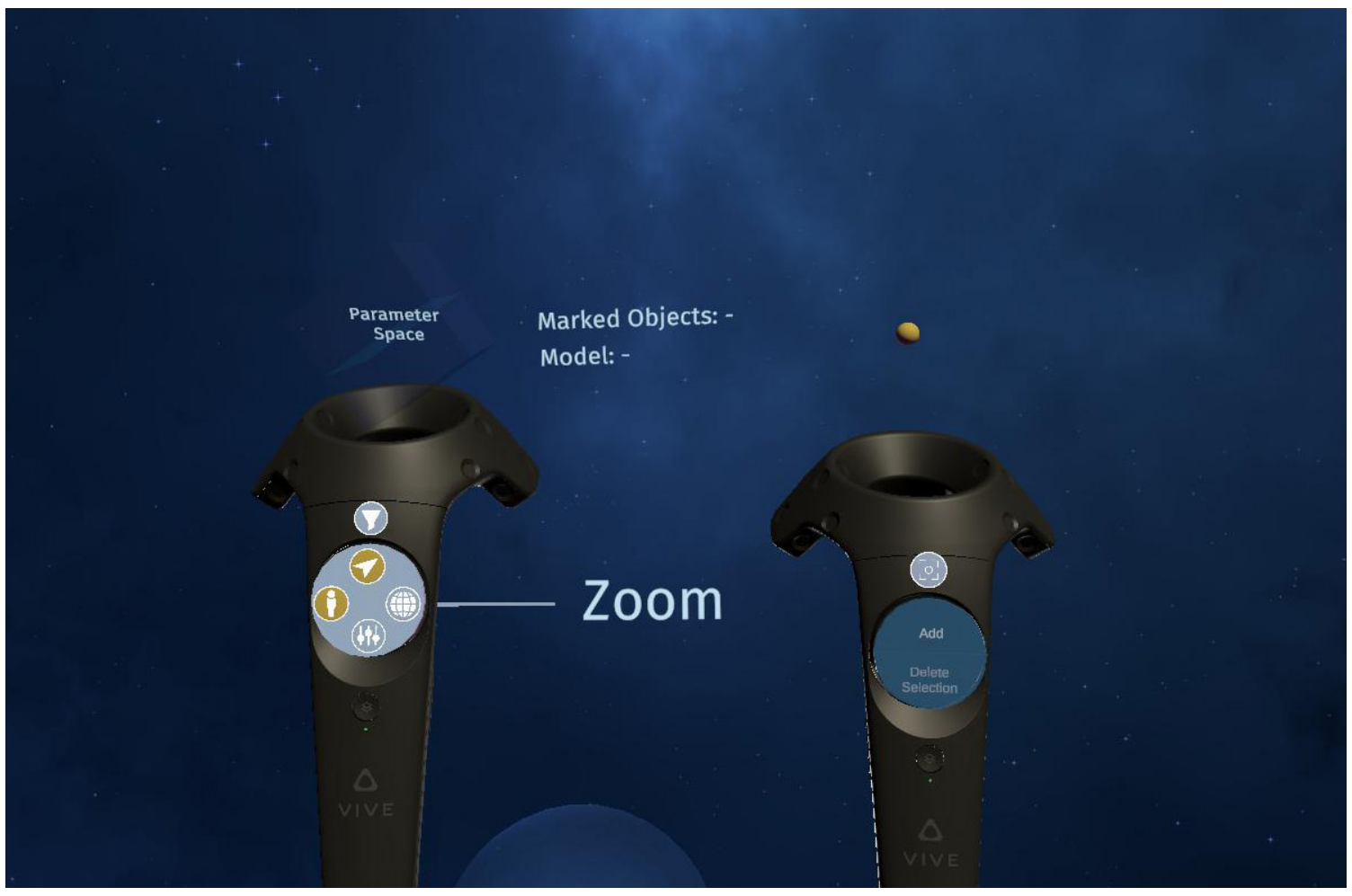

**Fig. 7.14:** Pictograms menu: Above: Player and World mode, below: Selection and manipulation tool

## 7.3. Interaction

So that the data visualization of this work can support the explorative analysis of the data, the user has various interaction options. These are regulated via the VR controller. The action options are divided into the left and right hand. Among other things, the left carries the selection menu, while the right usually becomes the selected tool. Exceptions are the actions for navigation in space and scaling / displacement of the same, as well as the spatial displacement of objects. The respective possible action or selection is marked either with text or pictograms.

### 7.3.1 Navigation & Scaling

In order to move in three-dimensional space, the user has two modes at all times, which are represented by pictograms
Marked are. In player mode (cf., Fig. 7.14, top left), the user can "swim" through the room by pressing both controller handles and the distance between their hands. He moves forwards or backwards along the line of sight. This mode is mainly used to overcome long distances. The second mode, the world mode (cf., Fig. 7.14, top right), allows the same action to scale and move the entire data visualization. This interaction allows two moments: on the one hand, the user can literally dive into the data and swim around in it. On the other hand, the display can be scaled smaller than the user himself, so that he can grasp, rotate and move the entire visualization with his hands, as is intuitively possible and common with a real object. Both modes are the only ones triggered by the buttons on the handles of the controllers to distinguish them from the tool interactions.

### 7.3.2 Manipulations

As mentioned in chapter 7.1, the distribution of the data clouds within the class volume depends on the relative directions of the classes. That is, by shifting the volume, the viewer is able to examine the influence of individual classes on the volume. To make this possible, the user can manipulate the volumes in the classroom. If this function is selected, a gripping zone appears on the controller and an indication of which keys trigger the gripping if the zone overlaps with a class volume (see Fig. 7.15). The data points are distributed in real time so that the user can see the consequences of the shift directly.
An equivalent action exists in the parameter space (see Fig. 7.16). The user can manipulate this sequence himself in order to research the influences of different parameter rankings. For this purpose, there is a model model object in the origin of the parameter space (and in the center of the first cycle) which, like the generated model objects in the classroom, represents a ranking of parameters across axes. With the gripping tool, the user can change the length of the axes and thus modify the order of the parameter cycles.

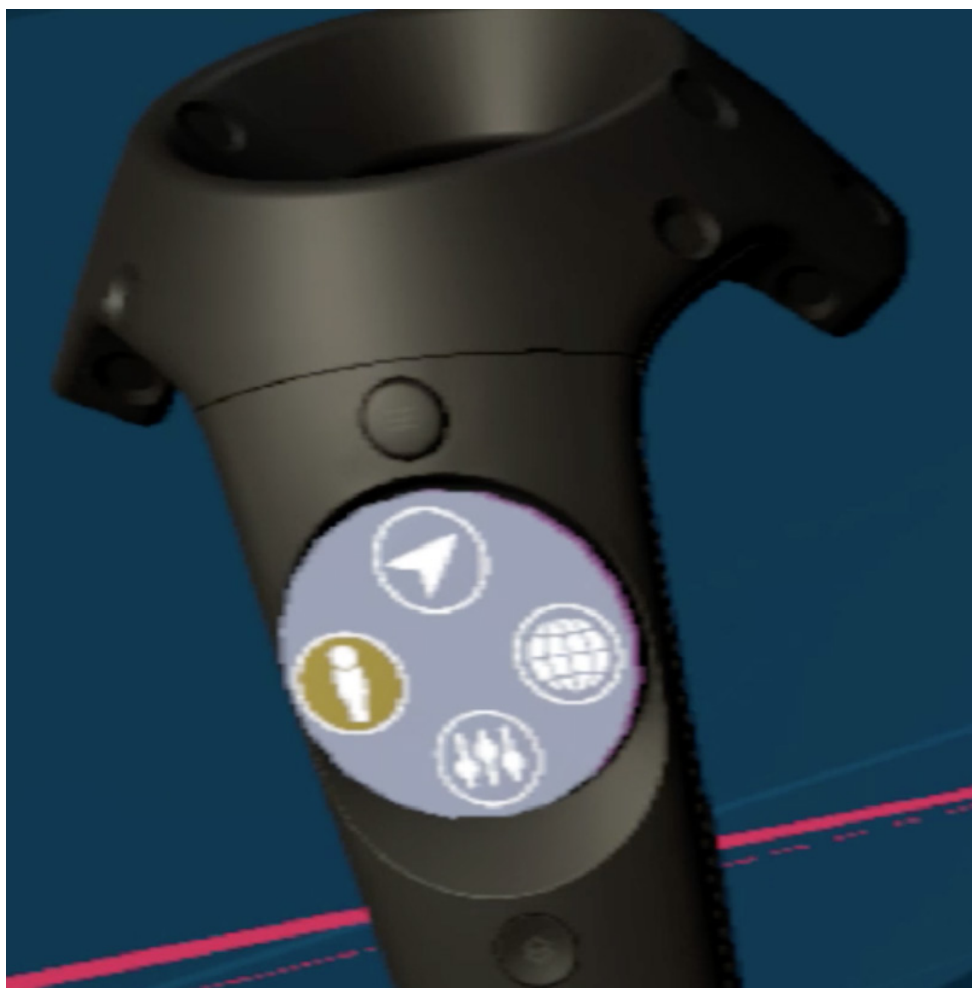

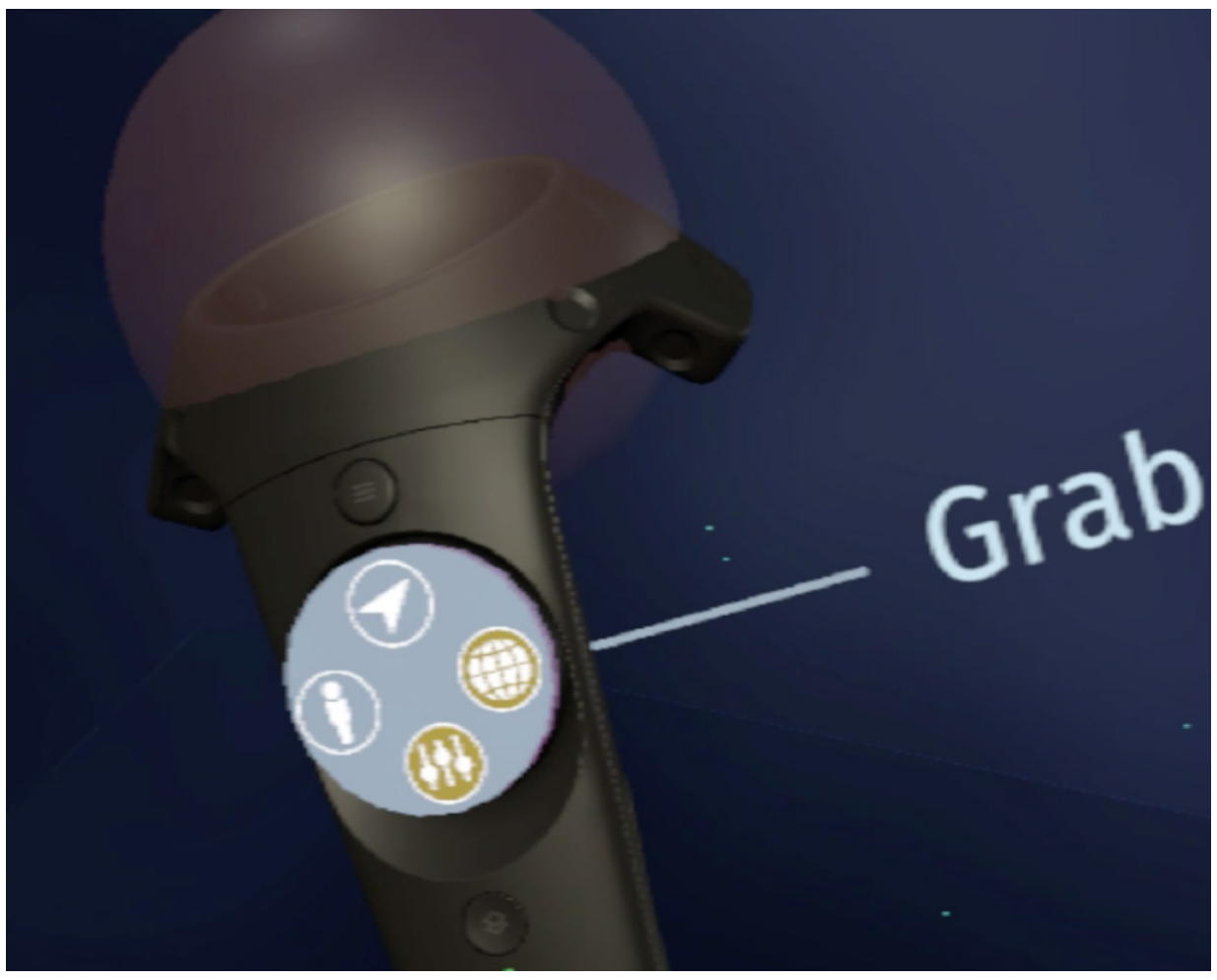

**Fig. 7.15:** Selection menu on the left VR controller.

**Fig. 7.16:** Selection of Grab tool

**7.3.3 Interacting with Data Points**
To interact with the data points in the class and parameter space, i.e. The selection tool is available to activate them and to mark them for the parameter space (cf., Fig. 7.17). If it is selected by the viewer, a small pointer appears on the right hand. This enables the first information levels of model objects to be activated by approximation. Additional levels can be opened using the trigger buttons on the index fingers, which are used for all interactions that are possible directly on the object. The activations of the data points can either be deactivated directly on the object, triggered again, or deleted for the entire classroom. The action button is displayed on the right controller.
If the selection tool is selected, the user can not only activate objects, but also mark them for the parameter space. In this way, spectral traces of several objects in the parameter space can be compared. There is also a button for this on the right controller. The number of marked objects and the selected model is shown on the left hand. Finally, you can switch to the parameter space using the trigger.

**7.3.4. Filter**
Filter options are available to select subspaces of the class spheres and thereby find interesting objects more easily. These are called up again via the left menu hand (cf., Fig. 7.18). At the time of writing, the user has two options: the selection of a minimum and maximum value for the probabilities shown. The sliders are moved over the right tool hand. The effects of the filters are visible in real time on the data clouds. For example, Figure 7.19 shows the influence when the maximum probability value is reduced.

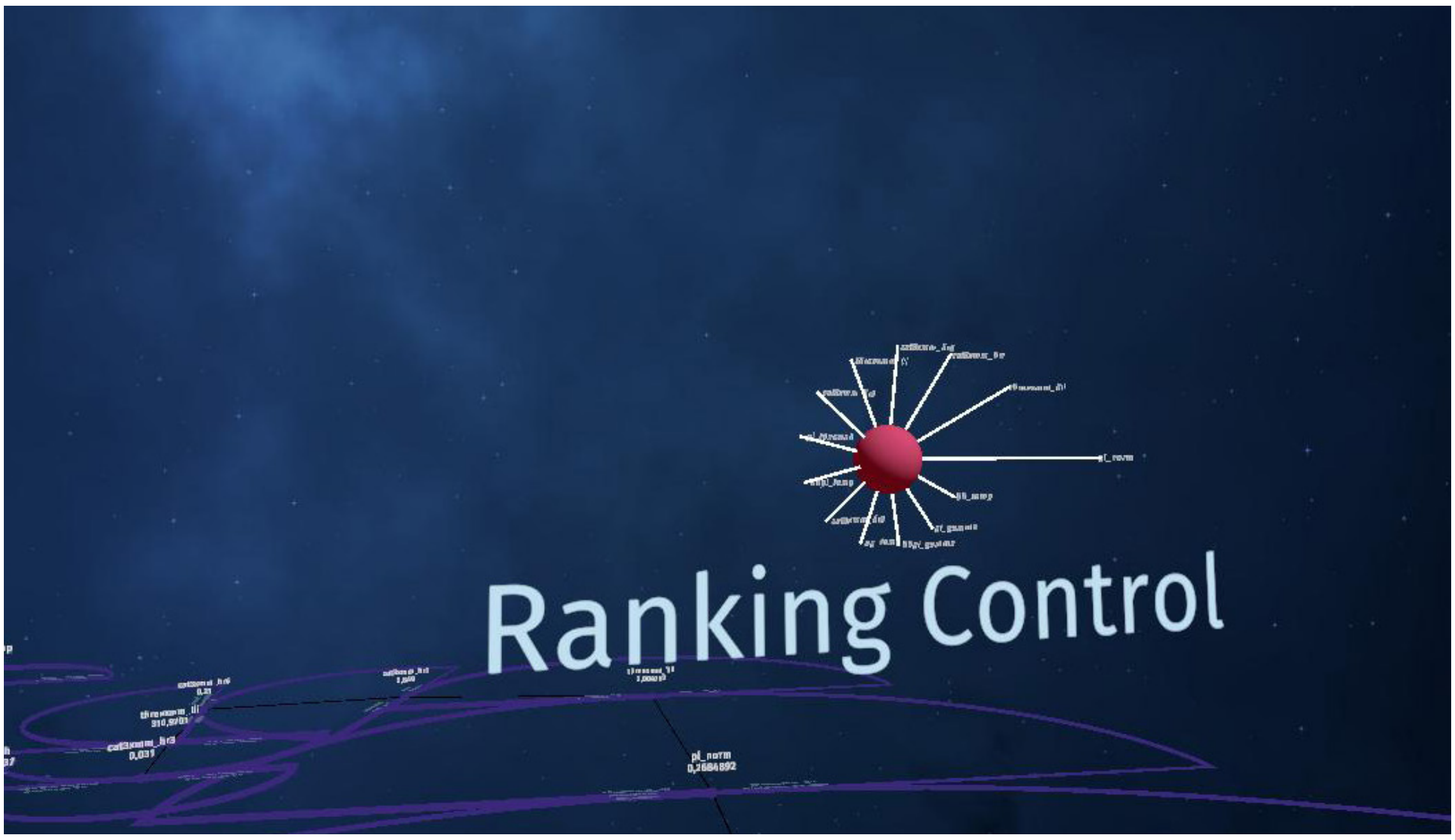

**Abb. 7.16:** Manipulation of parameter cycles via interaction with a sample object.

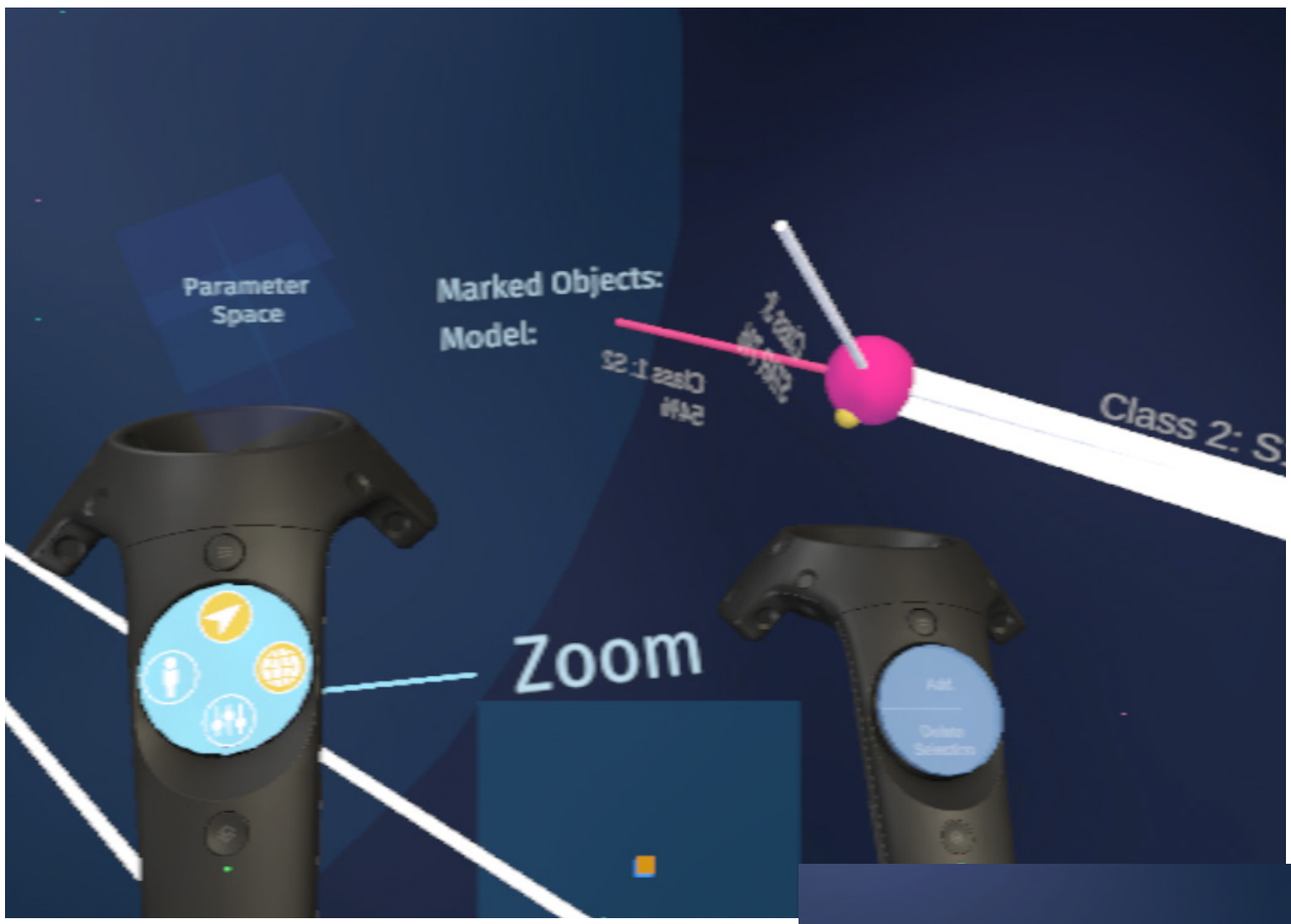

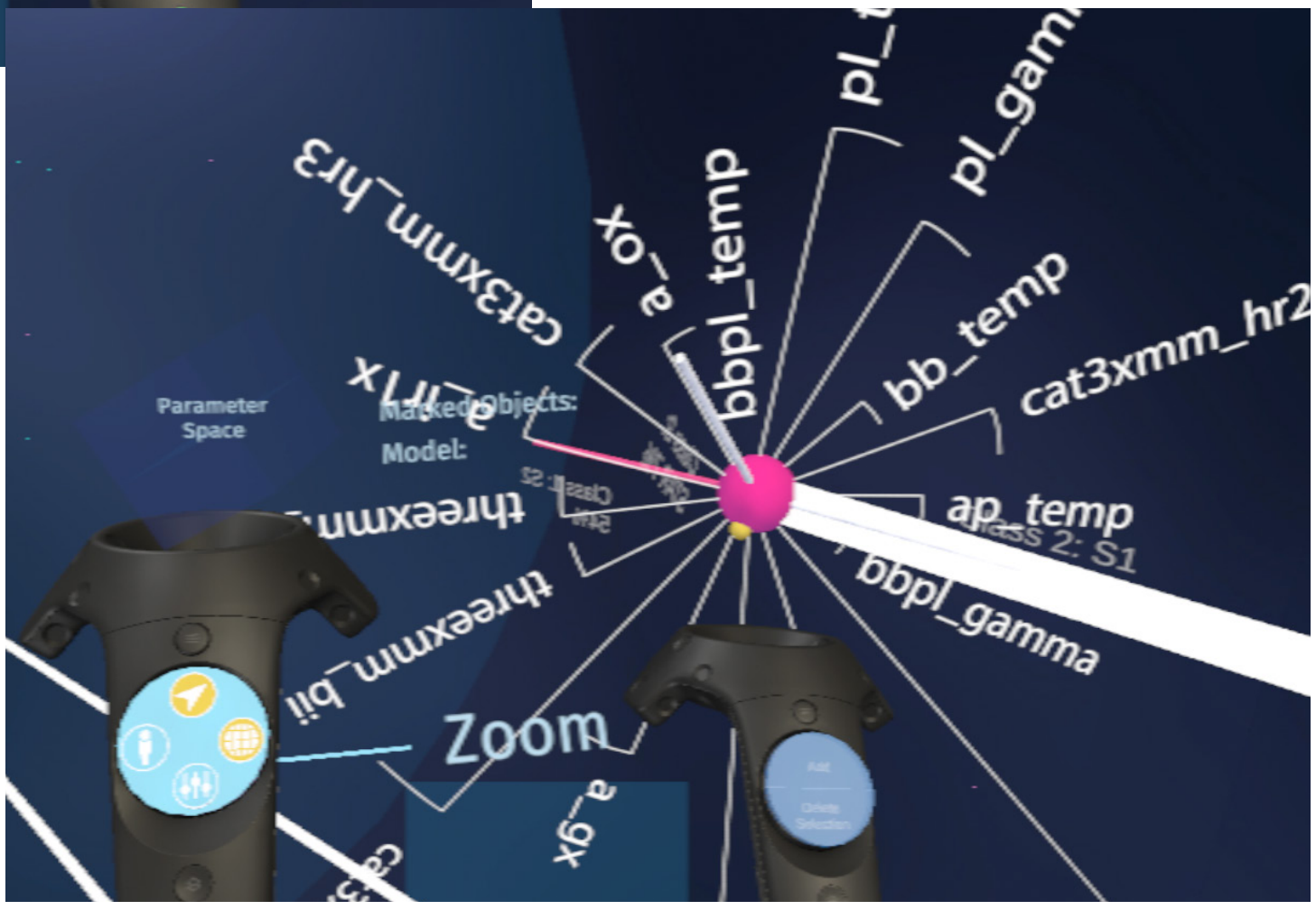

**Fig. 7.17:** With the selection tool selected, the user can interact with objects, call up details and mark them for display in the parameter space.

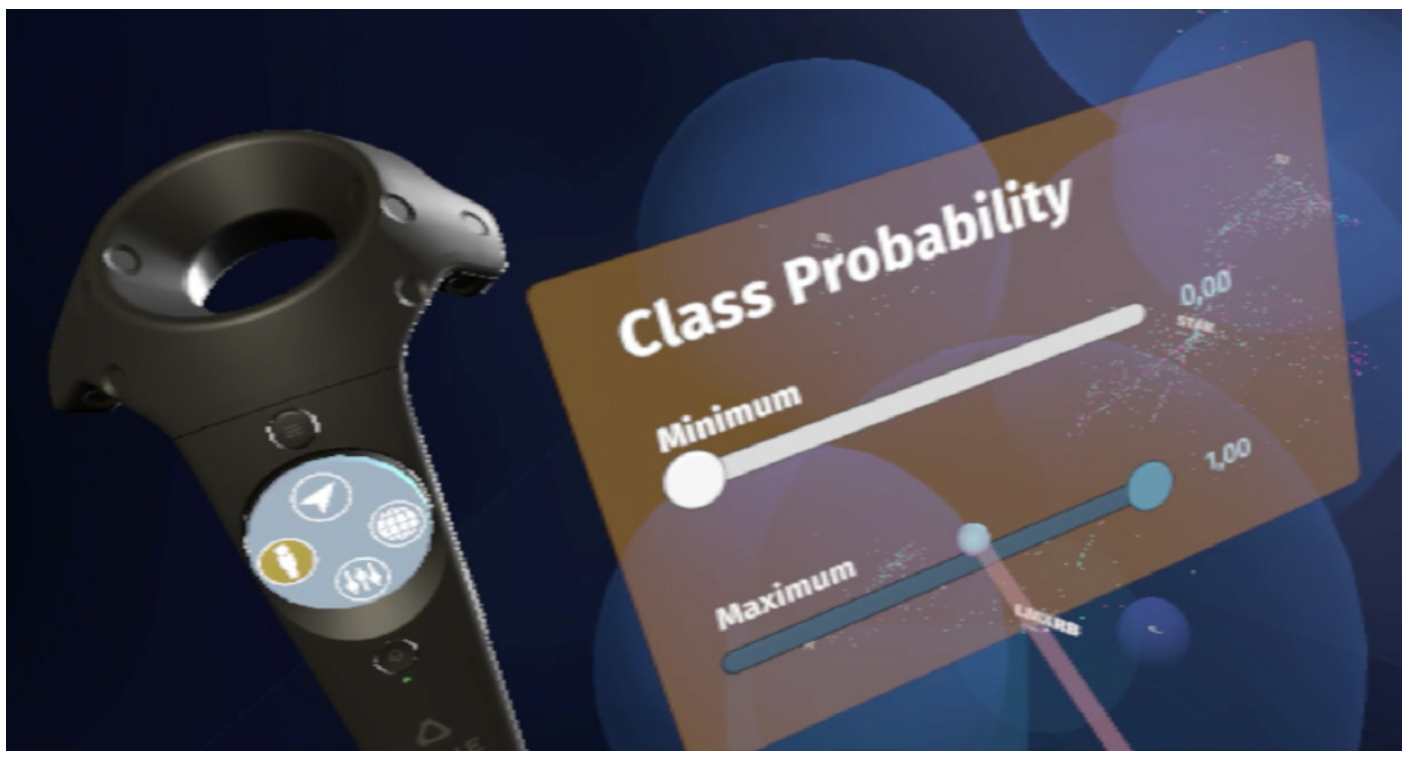

**Fig. 7.18:** Filter options for the display of data point clouds.

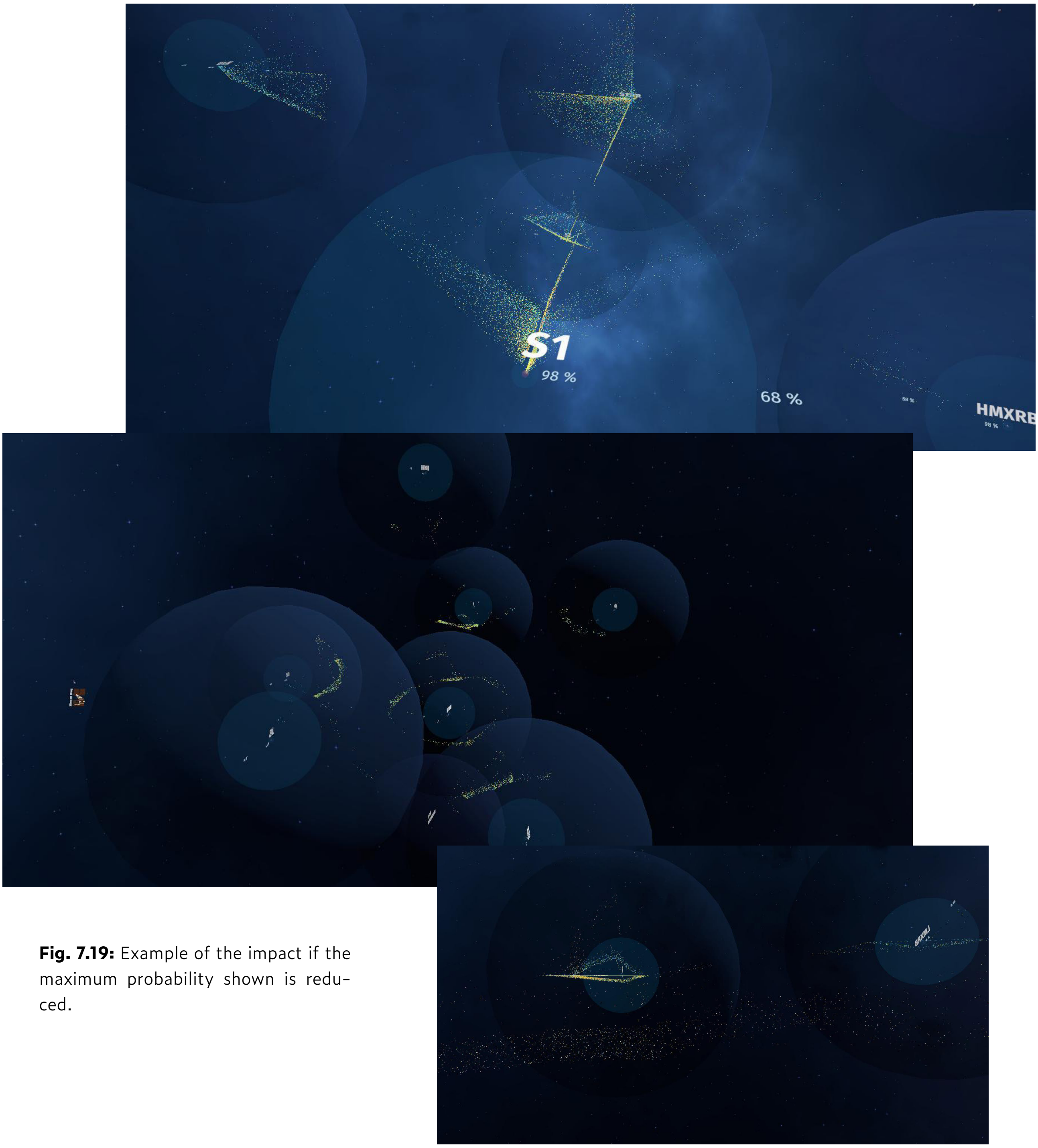

**Fig. 7.19:** Example of the impact if the maximum probability shown is reduced.

# 8 Summary and Outlook

Due to the ever increasing and faster growing data situation in modern astronomy, this discipline of the natural sciences faces new challenges. In order to encounter these mountains of data and to continue to be actively involved in the research of the internal structures and correlations of data sets as an astronomer, not only methodological solutions for statistical analysis have to be found, but also visual ones. The history of information graphics and the way in which people transform data into knowledge show that data visualization can also make a contribution. An interesting development in this area is the new media, i.e. the virtual and augmented reality. Compared to conventional data graphics on screens, these have a much larger field of view that can be used simultaneously and thus offer the possibility of visualizing more dimensions of data records. Especially in the field of astronomy, where visualizations always go hand in hand with the observations, it is therefore an exciting question to what extent large, multidimensional data sets can be visualized in VR and whether this not only adds value for the communication of results, but also arises for the analysis itself. Can the visualization in VR itself become an analysis tool?
This work dealt with this question and dealt with a data analysis that automatically classifies unknown X-ray sources into certain source classes based on their radiation properties. As more and more data is collected from satellites, their data archives are growing. It hides millions of radiation sources from which any unknown source could potentially produce new physical discoveries. The classification of unknown sources is therefore of great importance in modern astronomy and is increasingly adopted by intelligent algorithms due to the amount of data. The aim of this work was to find in various experiments, concepts and solutions that firstly represent a clear data visualization in VR and allow the user to understand the decisions of the algorithm. Second, it examines whether the presentation itself can serve the user to formulate and test new hypotheses.
In the development and implementation of the final concept, it was shown that the pure presentation of the data can create interesting virtual spaces, such as the classroom, whose aesthetics are determined by the parameters of the data. This generative approach implies that users interact with the data across this world. By exploring the virtual world, they are actually examining an abstract parameter space that is not necessarily visual in nature. Through this interaction with the virtual elements and their reactions based on the data, the visualization becomes a tool. The presentation that was developed for the classroom gives a picture of how the results of the different models differ with regard to the source distribution as a whole, but also for individual sources. The ability to interact with the classroom and manipulate its arrangement makes the display an analysis tool to better understand the structure of the data distribution and to find interesting objects.

In the course of the experiments on the spatial ordering of data in the parameter space, it was shown that it can make sense for the display in this area to leave the usual mapping of data behind in coordinate systems. Instead, a type of parameter representation via so-called parameter cycles, which is unusual for statistically dominated, modern astronomy, was chosen. This method tried to distribute the ra-diation source in three-dimensional space based on the results of the algorithms and the parameters themselves. In the first experiments with different selections of classified sources, very interesting patterns emerged, which separated the sources surprisingly well in space. These representations give the viewer a feeling of how the intelligent algorithm, which does not need an idea to recognize patterns in multidimensional spaces, may work. This allows the results to be better assessed in terms of their quality and susceptibility to errors. The astronomer no longer has to trust the values alone, but can control and understand them. An important part of this presentation, which allows the step to a real exploratory visual analysis, is in turn the possibility to manipulate the order of the parameter cycles and to test your own hypotheses.

The visualizations developed in this project provide clues that the data visualization in VR can be used for visual, explorative data analysis and contains exciting possibilities. The VR application of this work represents only one of many possible visualizations. Once solved from conventional representations, the design of the parameter space with the radiation properties of the sources raises many new questions. How can parameters be separated more precisely? Are there better representations that allow the parameter correlations to be analyzed even more clearly? These questions could be clarified in further studies and experiments. Another point is the generation of new statistical representations from the analysis. If, for example, the user has defined a meaningful ranking list for the classification in the parameter space, it would be exciting to make its effects visible for the classifications in the classroom. For this feedback, the control data set on which the algorithms were trained would have to be built in, but was not dealt with in this work.

Especially in the context of the new x-ray satellite eROSITA6, which started in 2019 and will carry out even more extensive surveys of the x-ray sky, the VR application created in this work is very exciting. The extent to which it can become part of an analysis chain depends on the classification algorithms. Due to the generative approach and the modular class and parameter spaces, the application is not limited to the sample data set. Any classification that relates a range of probabilities to underlying source parameters could be represented in this application.

In conclusion it can be said that this work demonstrates the applicability but also the limitations of VR data visualization as a solution to the visual challenges in modern astronomy. At the same time, some questions were raised so that the work can serve as a starting point for further investigations in this area.

[6] eROSITA Satellite – www.mpe.mpg.de/800290/eROSITA

# Image References

01|2020